\documentclass[twocolumn]{aastex631} 

\newcommand{\vlos}{$v_{\text{los}}$}
\newcommand{\sigmav}{$\sigma_{v, \text{los}}$}

\begin{document}

\title{X-ray emission signatures of galactic feedback in the hot circumgalactic medium: predictions from cosmological hydrodynamical simulations}

\author[0000-0002-1616-5649]{Emily M. Silich}
\affiliation{Cahill Center for Astronomy and Astrophysics, California Institute of Technology, Pasadena, CA 91125, USA}
\affiliation{Center for Astrophysics $\vert$ Harvard \& Smithsonian, 60 Garden St., Cambridge, MA 02138, USA}

\author[0000-0003-3175-2347]{John ZuHone}
\affiliation{Center for Astrophysics $\vert$ Harvard \& Smithsonian, 60 Garden St., Cambridge, MA 02138, USA}

\author[0000-0001-6411-3686]{Elena Bellomi}
\affiliation{Center for Astrophysics $\vert$ Harvard \& Smithsonian, 60 Garden St., Cambridge, MA 02138, USA}

\author[0000-0002-3817-8133]{Cameron Hummels}
\affiliation{Cahill Center for Astronomy and Astrophysics, California Institute of Technology, Pasadena, CA 91125, USA}

\author[0000-0002-3391-2116]{Benjamin Oppenheimer}
\affiliation{University of Colorado, Center for Astrophysics and Space Astronomy, 389 UCB, Boulder, CO 80309, USA}

\author[0000-0003-3729-1684]{Philip F. Hopkins}
\affiliation{Cahill Center for Astronomy and Astrophysics, California Institute of Technology, Pasadena, CA 91125, USA}

\author[0000-0003-1785-8022]{Cassandra Lochhaas}
\affiliation{Center for Astrophysics $\vert$ Harvard \& Smithsonian, 60 Garden St., Cambridge, MA 02138, USA}
\affiliation{Hubble Fellow}

\author[0000-0002-7484-2695]{Sam B. Ponnada}
\affiliation{Cahill Center for Astronomy and Astrophysics, California Institute of Technology, Pasadena, CA 91125, USA}

\author[0000-0001-8121-0234]{Alexey Vikhlinin}
\affiliation{Center for Astrophysics $\vert$ Harvard \& Smithsonian, 60 Garden St., Cambridge, MA 02138, USA}

\begin{abstract}
Little is currently known about the physical properties of the hot circumgalactic medium (CGM) surrounding massive galaxies. Next-generation X-ray observatories will enable detailed studies of the hot CGM in emission. To support these future efforts, we make predictions of the X-ray emission from the hot CGM using a sample of 28 $\sim$Milky Way-mass disk galaxies at $z=0$ from seven cosmological hydrodynamical simulation suites incorporating a wide range of galactic feedback prescriptions. The X-ray surface brightness (XSB) morphology of the hot CGM varies significantly across simulations. XSB-enhanced outflows and bubble-like structures are predicted in many galaxies simulated with AGN feedback and in some stellar-feedback-only galaxies, while other galaxies exhibit more isotropic XSB distributions at varying brightnesses. Galaxies simulated without cosmic ray physics exhibit radial XSB profiles with similar shapes ($\propto r^{-3}$ within $20-200$~kpc), with scatter about this slope likely due to underlying feedback physics. The hot CGM kinematics also differ substantially: velocity maps reveal signatures of bulk CGM rotation and high-velocity biconical outflows, particularly in simulations incorporating AGN feedback. Some stellar-feedback-only models also generate similar AGN-like outflows, which we postulate is due to centrally-concentrated star formation. Simulations featuring AGN feedback frequently produce extended temperature enhancements in large-scale galactic outflows, while simulations incorporating cosmic ray physics predict the coolest CGM due to pressure support being provided by cosmic rays rather than hot CGM. Individually-resolved X-ray emission lines further distinguish hot CGM phases, with lower-energy lines (e.g., O VII) largely tracing volume-filling gas, and higher-energy lines (e.g., Fe XVII) highlighting high-velocity feedback-driven outflows.
\end{abstract}

\section{Introduction} \label{sec:intro}

The circumgalactic medium (CGM) is a diffuse, gaseous atmosphere that surrounds galaxies and permeates the space out to their virial radii and beyond (see \citealt{Tumlinson2017} and \citealt{FGrev2023} for recent reviews). Both observations and numerical simulations have indicated that the CGM is a multi-phase gas \citep[e.g.,][]{Tumlinson2011, Werk2016, Rupke2019} that is generally classified into cool ($T \lesssim 10^5$ K), warm ($10 ^5 \lesssim T \lesssim 10^6$ K), and hot ($T \gtrsim 10^6$ K) components. These phases host a complex combination of velocity structures, such as inflows of accreting matter onto galactic disks to fuel star formation and outflows of metal-enriched material driven by supernovae, stellar winds, and active galactic nuclei (AGN) \citep{Tumlinson2017}. Such processes comprise a galactic feedback cycle that regulates the growth and quenching of galaxies, and are therefore central to determining the thermodynamic, kinematic, and chemical states of the CGM.  

Existing observational studies of the CGM have predominantly focused on characterizing the cool/warm CGM gas phases by employing absorption and emission line techniques at ultraviolet (UV) wavelengths \citep[e.g.,][]{Tumlinson2011, Tumlinson2013, Werk2013, Stocke2013, Werk2014, Burchett2016, Hayes2016, Werk2016, Burchett2019} or in the optical (rest-frame UV) for higher-$z$ objects \citep[e.g.,][]{steidel2000, Nielsen2013, Churchill2013, cai2018, Burchett2021}. These observations are largely consistent with theoretical predictions, where the majority of diffuse CGM gas cooling is expected to occur via continuum and line emission \citep{Bertone2013}, although some discrepancies remain in the detailed abundances of specific ions according to different feedback physics \citep[e.g.,][]{Hummels2013,Suresh2015,Ji2019,vandeVoort2019}. UV absorption line methods dominate the existing literature due to the extremely low surface brightness of the diffuse CGM gas in UV emission \citep{Hayes2016}. While UV absorption studies have provided information on the mass and metal content of the CGM, these methods rely on the presence of suitable background sources (e.g., high-$z$ quasars). Generally, only one sight-line will be available to probe an individual galaxy's CGM \citep{Tumlinson2017}, and these absorption line studies only yield one-dimensional measures of gas surface density, making contributions from spatial variations in the CGM difficult to isolate and interpret.

For galaxies of $\sim$Milky Way (MW)-mass and above, the hot phase of the CGM at and above the virial temperature is expected to be substantial. Emission from the hot CGM is expected to peak in the soft X-ray waveband, though measurements of the hot CGM are also possible in stacks of galaxies via the thermal Sunyaev-Zeldovich (tSZ) effect \citep{Bregman2022}, and as excesses in dispersion measure derived from extragalactic fast radio burst (FRB) observations \citep[e.g.,][]{Wu2023,Connor2022,Ravi2019}. Studies of the hot CGM with emission and absorption lines in the soft X-ray waveband are exceptionally challenging due to the very low surface brightness of the gas at these energies, particularly at large galactic radii \citep[see e.g.,][]{Anderson2011}. To date, hot CGM emission has been observed in X-rays predominantly in stacking analyses of galaxies with XMM-Newton \citep[][]{Li2018}, ROSAT \citep[][]{Anderson2013, Anderson2015}, and eROSITA \citep{Chadayammuri2022, Comparat2022, Zhang2024}, and only in a small number of individual systems \citep[e.g.,][]{Anderson2011, Humphrey2011, Bogdan2013, bogdan2017, Li2017, Das2019, Das2020}. Since galactic feedback processes should imprint distinct signatures in the hot CGM \citep[e.g.,][]{Pillepich2021, ZuHone2024, Schellenberger2024}, mapping large-scale emission from the hot CGM is vital to characterizing the nature of the feedback physics that shapes the evolution of MW-like galaxies.

In order to comprehensively map the physical properties of the hot CGM out to galactic virial radii using X-ray observations, new facilities with e.g., improved field of view, spatial and spectral resolutions, and throughput, are imperative. High spectral resolution microcalorimeter capabilities are currently in use by the Resolve instrument on the XRISM observatory \citep[see e.g.,][]{Porter2024}, but the gate valve for Resolve has failed to open since the spacecraft launch, severely compromising analyses below $\sim2$ keV (the energy range of interest for studies of the hot CGM). In addition, the spatial resolution of XRISM is large ($\gtrsim 1'$), making disentangling small scale features in the hot CGM, even for nearby galaxies, difficult. Future mission concepts such as Lynx \citep{Bandler2019} and Athena \citep{athena2018} will leverage microcalorimeters to achieve the spectral resolution required to probe the hot CGM velocity field via emission line shifts and broadening, as well as maintaining the spatial resolution needed to resolve small scale features in the hot CGM. 

However, there are several key gaps between our theoretical understanding of the CGM and how these quantitatively map to X-ray observables. For example, while the velocity field of a galaxy's hot CGM may be physically determined by the nature of feedback in the system, it may also be systematically dependent on, e.g., the specific spectral line(s) used to probe the gas velocity via observational line shifts since the different line energies may probe different gas temperatures with different velocities. Therefore, a detailed parameter space exploration of the relationships between galactic feedback processes (e.g., the injection of mass, energy, momentum, and metals from supernovae and AGN outflows, or the acceleration of cosmic rays by supernovae shocks) and the observational signatures that the next generation of X-ray observatories could probe is necessary in order to inform the design of mission architectures with optimized capabilities to study galaxy evolution via the hot CGM. 

In the absence of a representative inventory of existing observations, numerical simulations provide a means to explore how the complex processes that govern galaxy evolution can be traced by the hot CGM. Galactic feedback implementation choices in these simulations can significantly impact the physical properties of the hot CGM in various ways, leading to vastly different predictions for the density, temperature, composition, and velocity structures. Some simulations, such as the fiducial ``Feedback In Realistic Environments'' (FIRE-2) simulations \citep{FIRE1, FIRE2}, focus on explicitly modeling stellar feedback via supernovae, stellar winds, photoionization, and radiation pressure on the ISM/CGM, but neglect contributions from AGN. Numerical and physical implementations of stellar feedback can further vary significantly from one simulation to another. For example, the ``Figuring Out Gas \& Galaxies in Enzo'' \citep[FOGGIE;][]{Peeples2019} and TEMPEST \citep{Hummels2019} simulations both implement a simple thermal feedback model from supernovae as the sole source of stellar feedback, and these simulations therefore produce galaxies with very different global properties than FIRE2 \citep[e.g., stellar mass; see][]{Wright2024,FIRE2}. Additional simulations include the effects of cosmic rays in their feedback models. For instance, certain runs of the FIRE2 simulations \citep{FIRE2CR} build on the FIRE2 feedback prescription by incorporating cosmic ray physics, and therefore can estimate how high-energy particles injected by supernovae influence the hot CGM, though the cosmic ray transport models are not as observationally well constrained in the CGM of MW- and lower-mass galaxies. 

\begin{deluxetable*}{lccccc}[hbt!]  
\caption{Overview of the cosmological hydrodynamical simulations analyzed in this study \label{tab:sims_table}}
\tablehead{\multicolumn{1}{l}{Simulation} & \colhead{Run}  & \colhead{Code} & \colhead{$N_{\text{gal}}$} & \colhead{Feedback physics} & \colhead{Reference(s)}}
\startdata
Illustris-TNG & 50 & Arepo & 8 & stellar + AGN & \citet{tng1, tng2} \\
FIRE2 & fiducial & Gizmo & 3 & stellar & \citet{FIRE2} \\
FIRE2 & CR-700 & Gizmo & 3 & stellar + cosmic ray & \citet{FIRE2CR}\\
EAGLE & NEQ zooms & Gadget-3 & 5 & stellar + AGN & \citet{Schaye2015, Oppenheimer2016} \\
FOGGIE & fiducial & Enzo & 3 & stellar & \citet{Peeples2019, Wright2024} \\
TEMPEST & 10 & Enzo & 1 & stellar & \citet{Hummels2019} \\
Simba & high-res & Gizmo & 5 & stellar + AGN & \citet{SIMBA}
\enddata
\end{deluxetable*}

Other simulations, such as the Illustris-TNG (hereafter, TNG) project \citep{tng1, tng2}, the ``Evolution and Assembly of GaLaxies and their Environments'' (EAGLE) simulations \citep{Schaye2015}, or the Simba simulations \citep{SIMBA} incorporate both stellar feedback and contributions from AGN. While these implementations model the combined influence of AGN activity and stellar processes on the ISM/CGM, in addition to using different stellar feedback prescriptions, the numerical and physical implementations of AGN feedback differ drastically. Notably, the EAGLE simulations use a single-mode form of AGN feedback where thermal energy injection at the location of the black hole is determined by the gas accretion rate. TNG, however, implements a dual-mode AGN feedback prescription that, in addition to having a high accretion rate mode of thermal feedback similar to EAGLE, also implements a low accretion rate, purely kinetic mode in which momentum is applied to the gas cells surrounding the SMBH in an isotropic fashion, which has been shown to produce winds and cavities \citep{Pillepich2021}. Alternatively, Simba implements two forms of AGN kinetic feedback as bi-directional winds and jets, as well as a thermal energy injection mode. These numerical models can drastically alter the physical state of the CGM; for instance, the median CGM mass fractions predicted by TNG and EAGLE galaxies at a given halo mass are discrepant \citep{Davies2020}, especially at low halo masses ($M_{halo} \lesssim 10^{12.5}~M_\odot$), where halos are more gas-poor in EAGLE and gas-rich in TNG \citep[see also][Figure 9]{Crain2023}.

In this work, we utilize a suite of  Milky Way-mass disk galaxies formed in cosmological hydrodynamical simulations with varying feedback prescriptions explore the resulting X-ray signatures that the next generation of X-ray observatories could use to distinguish between different physical models. In Section \ref{sec:sims}, we provide an overview of the simulated galaxies studied in this work. Section \ref{sec:obsgen} details our methods for generating mock X-ray observables, and Section \ref{sec:results} describes the main results predicted by these mock observables, including comparisons to other observational and computational works. In Section \ref{sec:outflows}, we discuss the predicted manifestation of large-scale biconically-oriented galactic outflows driven by galactic feedback processes across the simulated galaxy sample. Section \ref{sec:summary} provides an overview of our findings.

\begin{deluxetable*}{lcccccr}[hbt!]  
\caption{Physical properties of simulated galaxies used in this analysis \label{tab:features_table}}
\tablehead{\colhead{Galaxy ID} & \colhead{$r_{200c}$} & \colhead{$M_{200c}$} & \colhead{$M_{*}$} & \colhead{SFR} & \colhead{$m_{\text{gas}}$} & \colhead{Notes} \\ \colhead{} & \colhead{[kpc]} & \colhead{[$10^{12} M_{\odot}$]} & \colhead{[$10^{10} M_{\odot}$]} & \colhead{[$M_{\odot}$ yr$^{-1}$]} & \colhead{[$M_{\odot}$]} & \colhead{}}
\startdata
\multicolumn{7}{c}{TNG50} \\
\hline
    TNG50-31 & 360 & 5.0 & 17 & 4.5 & $8.6_{-1.6}^{+1.8} \times 10^{4}$ & \\
    TNG50-37 & 350 & 4.6 & 15 & 1.6 & $8.9_{-1.5}^{+1.6} \times 10^{4}$ & \\
    TNG50-53 & 344 & 4.3 & 9.8 & 1.0 & $8.7_{-1.6}^{+1.8} \times 10^{4}$ & Relaxed satellite criterion; \\
     & & & & & & X-ray bright nearby subhalo \\
    TNG50-80 & 311 & 3.2 & 4.6 & 1.2 & $8.7_{-1.7}^{+1.9} \times 10^{4}$ & Relaxed satellite criterion; \\  & & & & & & X-ray bright nearby subhalo \\
    TNG50-101 & 268 & 2.1 & 11 & 12 & $8.4_{-1.7}^{+1.9} \times 10^{4}$ & \\ 
    TNG50-211 & 216 & 1.1 & 6.2 & 1.1 & $8.5_{-1.6}^{+1.8} \times 10^{4}$ & \\
    TNG50-318 & 195 & 0.79 & 4.3 & 0.0 & $8.5_{-2.0}^{+2.4} \times 10^{4}$ & \\
    TNG50-322 & 190 & 0.73 & 4.2 & 0.0 & $8.2_{-1.8}^{+2.1} \times 10^{4}$ & ISM recently disrupted by AGN \\
\hline
\multicolumn{7}{c}{FIRE2} \\
\hline
    FIRE2-m12f & 226 & 1.3 & 9.3 & 11 & $7.1_{-0.0}^{+0.09} \times 10^{3}$ & \\
    FIRE2-m12m & 223 & 1.3 & 13 & 11 & $7.1_{-0.0}^{+0.09} \times 10^{3}$ & \\
    FIRE2-m12i & 202 & 0.94 & 7.1 & 6.2 & $7.1_{-0.0}^{+0.04} \times 10^{3}$ & \\
\hline
\multicolumn{7}{c}{FIRE2CR} \\
\hline
    FIRE2CR-m12f & 220 & 1.2 & 3.8 & 5.7 & $7.1_{-0.0}^{+0.02} \times 10^{3}$ & \\
    FIRE2CR-m12m & 210 & 1.1 & 3.5 & 3.3 & $7.1_{-0.0}^{+0.03} \times 10^{3}$ & \\ 
    FIRE2CR-m12i & 196 & 0.86 & 2.7 & 2.4 & $7.1_{-0.0}^{+0.02} \times 10^{3}$ & \\
\hline
\multicolumn{7}{c}{EAGLE zooms} \\
\hline
    EAGLEz-002 & 266 & 2.0 & 1.8 & 0.96 & $2.9_{-0.0}^{+0.10} \times 10^{4}$ & \\
    EAGLEz-003 & 243 & 1.5 & 1.3 & 0.59 & $2.8_{-0.0}^{+0.08} \times 10^{4}$ & Recent merger?\\
    EAGLEz-001 & 238 & 1.4 & 1.2 & 1.1 & $2.8_{-0.0}^{+0.05} \times 10^{4}$ & \\
    EAGLEz-004 & 220 & 1.1 & 1.0 & 0.89 & $2.8_{-0.0}^{+0.06} \times 10^{4}$ & \\
    EAGLEz-007 & 189 & 0.72 & 0.73 & 1.3 & $2.8_{-0.0}^{+0.06} \times 10^{4}$ & \\ 
\hline
\multicolumn{7}{c}{FOGGIE} \\
\hline
    FOGGIE-H & 247 & 1.7 & 24 & 4.7 & $5.4_{-4.2}^{+33} \times 10^{1}$ & Polar disk galaxy \\
    FOGGIE-B & 216 & 1.1 & 14 & 2.1 &  $1.4_{-1.2}^{+4.1} \times 10^{2}$ & \\
    FOGGIE-M & 208 & 1.0 & 11 & 2.0 & $2.9_{-2.6}^{+4.0} \times 10^{2}$ & \\
\hline
\multicolumn{7}{c}{TEMPEST} \\
\hline
    TEMPEST-10 & 154 & 0.41 & 2.7 & 0.20 & $2.9_{-2.4}^{+7.5} \times 10^{1}$ & \\
\hline
\multicolumn{7}{c}{Simba} \\
\hline
    SIMBA-38 & 287 & 2.5 & 10 & 0.0 & $2.3_{-0.0}^{+0.3} \times 10^{6}$ & X-ray bright nearby subhalo \\
    SIMBA-40 & 282 & 2.4 & 13 & 6.1 & $2.3_{-0.0}^{+0.1} \times 10^{6}$ & \\
    SIMBA-49 & 245 & 1.6 & 8.6 & 4.2 & $2.3_{-0.0}^{+0.1} \times 10^{6}$ & \\
    SIMBA-59 & 237 & 1.4 & 6.5 & 1.7 & $2.3_{-0.05}^{+0.6} \times 10^{6}$ & \\
    SIMBA-81 & 214 & 1.1 & 4.7 & 12 & $2.3_{-0.0}^{+0.1} \times 10^{6}$ & \\ 
\enddata
\tablecomments{Column descriptions: (1) galaxy identifier; (2) radius within which the mean enclosed density is 200 times the critical density at $z=0$; (3) total enclosed mass at $r_{200c}$; (4) stellar mass within a radius of 30 kpc; (5) star formation rate (SFR) averaged over 10 Myr within a radius of 30 kpc; (6) median hot gas ($T > 3 \times 10^5$ K and $\rho < 3 \times 10^{-25}$ g cm$^{-3}$) resolution element mass and 1-$\sigma$ range within $r_{200c}$.}
\end{deluxetable*}
    
\section{Simulated galaxies} \label{sec:sims}
We examine 28 $\sim$Milky Way-mass disk galaxies at $z = 0$ from seven cosmological hydrodynamical simulation suites that incorporate substantial variations in feedback processes. While the precise parameter values can vary slightly between simulations, each simulation assumes a flat $\Lambda$CDM cosmology with $H_0 \simeq 70$ km s$^{-1}$ Mpc$^{-1}$ and $\Omega_M$ = $1 - \Omega_{\Lambda} \simeq 0.3$. Brief overviews of each simulation are detailed below. 

\subsection{TNG50} \label{subsec:TNG}
The TNG50 simulation \citep{tng1, tng2} is the highest resolution volume of the IllustrisTNG project. It was performed with the Arepo code \citep{arepo} and spans a volume of $\sim 50^3$ cMpc$^3$, adopting the \citet{Planck2016} cosmology. Among the eight TNG50 galaxies studied in this work (see below and Table \ref{tab:features_table}), the median resolution element mass of a hot, X-ray emitting gas cell (defined in Section \ref{sec:obsgen} and Table \ref{tab:features_table}) is $\langle m_{\text{gas}} \rangle \simeq 8.6 \times 10^4~M_{\odot}$. Throughout, we refer to galaxies from TNG50 by the group number for which each galaxy is the most massive subhalo. 

The TNG50 physics model is outlined in \citet{tng1} and described in detail by \citet{Pillepich2018}, though we highlight the details of its implementation relevant to this study here. TNG50 models galactic feedback physics associated with stellar formation, evolution, and feedback via supernovae (Types Ia and II) and asymptotic giant branch (AGB) stellar winds, in addition to supermassive black hole (SMBH)/AGN formation, growth, and isotropic feedback via both thermal and kinetic modes. Treatments for metal-line cooling and heating from background radiation fields are included in TNG50, though prescriptions for radiation and cosmic ray physics are not. 

The kinetic AGN feedback mode in TNG50, which takes the form of a wind-like energy injection, naturally produces large-scale galactic outflows which manifest as X-ray emitting bubbles as they propagate outwards into the CGM for $\sim$MW/M31-like galaxies at $z \sim 0$ \citep[see][]{Pillepich2021}. In this work, we select six galaxies from TNG50 satisfying the fiducial MW/M31 selection criteria outlined in \citet{Pillepich2021, Pillepich2024}, namely with stellar masses $M_{*}(<30~\text{kpc}) = 10^{10.5-11.2}~M_{\odot}$, exhibiting disc-like stellar morphologies, having no massive ($M_{*} > 10^{10.5}~M_{\odot}$) galaxies within a 500 kpc radius, and being embedded within a host halo with $M_{200c} < 10^{13}~M_{\odot}$. We further include two additional TNG50 galaxies that were not members of the \citet{Pillepich2021, Pillepich2024} sample. These galaxies were identified in the TNG50 volume with identical selection criteria as in the aforementioned studies, except we relaxed the criterion excluding galaxies with nearby massive satellite galaxies in order to match the often non-isolated environments of galaxies selected from other simulation suites in this analysis (e.g., in Simba or FOGGIE). As noted in an analysis of a similar TNG50 MW/M31 sample \citep{ZuHone2024}, one galaxy—TNG50-322—has undergone a recent AGN feedback-induced disruption of the galactic disk, which has drastically altered its morphology and kinematics.

\subsection{FIRE} \label{subsec:FIRE}

\subsubsection{Fiducial run}
The fiducial FIRE2 simulations \citep{FIRE2} are cosmological zoom-in simulations run in the Gizmo\footnote{\url{http://www.tapir.caltech.edu/~phopkins/Site/GIZMO.html}} code \citep{gizmo} for a range of halo masses ($10^9 \lesssim M_{\text{halo}} \lesssim 10^{12}$ M$_{\odot}$). FIRE2 is an update to the original FIRE1 simulations, notably leveraging a more accurate hydrodynamics solver \citep[see][]{FIRE1, FIRE2}. The fiducial FIRE2 simulations assume a cosmology consistent with \citet{planck2014}, and among all the FIRE2 galaxies studied in this work (both fiducial and those with cosmic rays; see below and Table \ref{tab:features_table}), the median resolution element mass of a hot, X-ray emitting gas cell is $ \langle m_{\text{gas}} \rangle \simeq 7.1 \times 10^3~M_{\odot}$.

The fiducial FIRE2 physics model is outlined in detail by \citet{FIRE2}; here we highlight processes relevant to this study. FIRE2 models galactic feedback physics associated with gas cooling, star formation, evolution, and feedback via supernovae (Types Ia and II), O/B and AGB stellar winds, photoionization, photoelectric heating, and radiation pressure in an explicit manner. The fiducial FIRE2 model does not include a prescription for AGN formation, growth, or feedback, nor does it include a prescription for cosmic ray physics. 

We select three $\sim$MW-mass galaxies from FIRE2 for this study. In the literature, these are commonly referred to as m12i, m12m, and m12f. These galaxies all host unambiguous disks and are well-studied representatives of their mass regime \citep[e.g.,][]{FIRE2, Garrison-Kimmel2018, El-Badry2018}.

\subsubsection{CR+ run}
The FIRE2CR simulations \citep{FIRE2CR} are modified runs of the fiducial FIRE2 simulations. In addition to the physics incorporated in the fiducial FIRE2 simulations, they additionally model magnetic fields, in the ideal MHD limit, with anisotropic Spitzer-Braginskii conduction and viscosity \citep[see][for details on the numerical implementation]{Su2017, hopkins2016}, and cosmic ray physics \citep[described in][]{Chan2019,FIRE2CR}. 

In the FIRE2CR model, a ``single-bin'' of $\sim$GeV cosmic rays are injected as a relativistic fluid with $\gamma_{\rm CR} = 4/3$ at supernova shocks and undergo transport including anisotropic cosmic ray streaming along local gas magnetic field lines, diffusion, and advection, with hadronic, Coulomb/collisional, and ionization losses tracked self-consistently \citep{FIRE2CR}. Each run uses an empirically-motivated diffusion coefficient of $\kappa_{\rm \|} = 3 \times 10^{29}~\text{cm}^2~\text{s}^{-1}$ \citep{Chan2019,FIRE2CR}, and streaming at the local gas Alfv\`en speed, and have been shown to produce galaxy properties consistent with observational constraints on cosmic rays and magnetic fields \citep{FIRE2CR,ponnada_magnetic_2022}.

At larger halo masses ($M_{200} \gtrsim 10^{11} M_{\odot}$) and lower redshifts ($z \lesssim 1–2$), contributions from cosmic rays can become a significant source of pressure support in the CGM \citep{FIRE2CR, Ji2020, Hopkins2021}, dominating thermal CGM pressure. This results in a cooler CGM than in the fiducial FIRE2 simulations, with the cosmic ray pressure supporting cooler gas against gravity \citep{FIRE2CR}. We thus select the FIRE2CR analogs of the three fiducial FIRE2 galaxies (i.e., run with the same initial conditions as m12i, m12m, and m12f) in order to make a direct comparison of the X-ray signatures of each halo with and without cosmic ray contributions (though, as detailed in Section \ref{subsec:emissgen}, we exclusively model the thermal emission from the hot CGM, neglecting possible additional contributions to the X-ray emissivity in these simulations from, e.g., cosmic ray inverse-Compton scattering processes).

\subsection{EAGLE Zooms} \label{subsec:EAGLE}
The EAGLE non-equilibrium (NEQ) zoom simulation suite \citep{Oppenheimer2016} consists of cosmological zoom-in simulations run using a version of the Gadget-3 code and the subgrid models introduced in \citet{Schaye2015,Crain2015} for 10 MW-mass and 10 group-sized halos, the former being of interest to this work. The MW-mass NEQ runs are an update to the original EAGLE volume simulations \citep{Schaye2015}, but incorporate non-equilibrium ionization and cooling in the presence of background radiation fields \citep{Oppenheimer2013}. Like the original EAGLE simulations, the EAGLE NEQ zoom simulations adopt the \citet{planck2014} cosmology. We select five EAGLE NEQ galaxies for this study, all run at the highest ``M4.4'' resolution described in \citet{Oppenheimer2016}. These zooms are $8\times$ higher resolution than the EAGLE-Recal simulation \citep{Schaye2015} and main M5.3 NEQ zooms from \citet{Oppenheimer2016}. Among these galaxies (see below and Table \ref{tab:features_table}), the median resolution element mass of a hot, X-ray emitting gas cell is $ \langle m_{\text{gas}} \rangle \simeq 2.8 \times 10^4~M_{\odot}$. 

The EAGLE physics model is described in \citet{Schaye2015}, and the NEQ updates are outlined in \citet{Oppenheimer2016}, though we include a summary of the relevant physical processes here. The EAGLE NEQ zoom simulations include radiative cooling, photoheating from the cosmic UV/X-ray background, as well as star formation, evolution, and feedback via supernovae (Types Ia and II) and stellar winds from O/B and AGB stars. They further include prescriptions for SMBH/AGN formation, growth, and isotropic AGN feedback injected as thermal energy in an amount proportional to the AGN accretion rate \citep{Schaye2015}. The specific prescriptions use the ``Recal'' parameters listed in Table 3 of \citet{Schaye2015}. These simulations do not incorporate cosmic ray physics. 

We note that while the CGM properties of EAGLE NEQ zooms are generally reproduced at the $8\times$ lower ``M5.3'' resolution suite, the stellar masses of the M4.4 galaxies are lower than the M5.3 galaxies. At a given halo mass ($M_{200c}$), the EAGLE M4.4 galaxies generally produce the lowest stellar masses of any galaxies across the sample. 

\begin{deluxetable*}{lcccl}[t]  
\caption{Soft X-ray emission lines used as probes of the hot CGM in this study \label{tab:lines_table}}
\tablehead{\colhead{Line} & \colhead{$E$ ($z = 0$)} & \colhead{$E$ ($z = 0.01$)} & \colhead{$T_{i}$} & \colhead{Transition type} \\ \colhead{}  & \colhead{[keV]} & \colhead{[keV]} & \colhead{[$10^6$ K]} & \colhead{}}
\startdata
     O VII & 0.561 & 0.555 & $0.25 - 3.1$ & magnetic dipole (forbidden) \\
     O VIII & 0.654 & 0.648 & $1.4 - 4.7$ & electric dipole (allowed) \\
     Fe XVII & 0.727 & 0.720 & $2.2 - 9.3$ & electric dipole (allowed) 
\enddata 
\tablecomments{Column descriptions: (1) emission line name; (2) line energy at $z=0$; (3) observed-frame line energy when emitted from $z=0.01$; (4) $T_{i}$: range of temperatures within which the ion fraction is over 10\% for a plasma in collisional ionization equilibrium; (5) transition type for the associated line.}
\end{deluxetable*}

\subsection{TEMPEST} \label{subsec:TEMPEST}
The TEMPEST simulations \citep{Hummels2019} are a suite of cosmological zoom-in simulations of one $\sim$MW-mass halo run in Enzo \citep{enzo} at four different resolutions. The TEMPEST simulations assume a cosmology consistent with \citet{planck2014}. These simulations employ the Enhanced Halo Resolution technique \citep[EHR;][]{Hummels2019}, which forces a higher level of spatial refinement than traditional refinement algorithms that act on mass density alone (and therefore under-resolve the hot CGM relative to, e.g., the much denser galactic disk). For this study, we choose the highest resolution TEMPEST halo, TEMPEST-10, which has been run to $z=0$. TEMPEST-10 is forced to have a spatial resolution of 0.5 comoving kpc throughout most of its virial radius, which translates to a median mass resolution for a hot, X-ray emitting gas cell of $ \langle m_{\text{gas}} \rangle \simeq 2.9 \times 10^1~M_{\odot}$ (see Table \ref{tab:features_table}). This makes the hot CGM of the TEMPEST-10 halo the best resolved of all galaxies in our sample.

The TEMPEST galactic physics model is provided in \citet{Hummels2019}, though we highlight relevant processes here. TEMPEST models metal-line cooling and non-equilibrium heating from background radiation fields, as well as galactic feedback physics associated with stellar formation, evolution, and feedback solely via Type II supernovae. The TEMPEST feedback model is much simpler than the stellar feedback schemes of other cosmological simulations, since it injects energy from supernovae only as thermal energy and mass, and it neglects contributions from any other form of stellar feedback (e.g., Type Ia supernova or O/B and AGN stars). TEMPEST does not implement cosmic ray physics or AGN formation, growth, and feedback. While it is the highest resolution simulation studied here, we note that the TEMPEST-10 halo is also the least massive galaxy in our sample. 

\subsection{FOGGIE} \label{subsec:FOGGIE}
The FOGGIE simulations \citep{Peeples2019} are a suite of cosmological zoom-in simulations of six $\sim$MW-mass halos run with Enzo \citep{enzo}. The FOGGIE simulations assume a cosmology consistent with \citet{planck2014}. We select three FOGGIE galaxies that have been run to $z=0$ for this study: FOGGIE-H, FOGGIE-B, and FOGGIE-M (which are called Hurricane, Blizzard, and Maelstrom, respectively, in FOGGIE papers; see Table \ref{tab:features_table}). Among these galaxies, the median mass resolution of a hot, X-ray emitting gas cell is $ \langle m_{\text{gas}} \rangle \simeq 1.2 \times 10^2~M_{\odot}$. Like TEMPEST, this high mass resolution in the hot CGM is due primarily to the additional implementation of EHR in the central galaxy of a given zoom-in simulation (see \citet{Hummels2019} for an introduction to EHR methods and \citet{Peeples2019} for a first application of EHR in the FOGGIE simulations). Additional refinement on the gas cooling time enhances the resolution in fast-cooling regions of the CGM, particularly close to the galaxy center. 

The physical processes incorporated in the FOGGIE simulations are outlined in \citet{Wright2024}, and we highlight the relevant details here. FOGGIE models galactic feedback physics associated with stellar formation, evolution, and feedback via Type II supernovae exclusively, while accounting for metal-line cooling and non-equilibrium heating from background radiation fields. As in TEMPEST, the FOGGIE feedback model deposits energy as an exclusively thermal component in the gas surrounding stellar particles. The FOGGIE simulations do not include prescriptions for cosmic rays or AGN formation, growth, and feedback. 

One of the FOGGIE galaxies, FOGGIE-H, was previously identified as a polar ring galaxy at $z = 0$, containing both a central and a polar disk \citep{Wright2024} populated with young stars. We further note that at a given halo mass, the FOGGIE galaxies generally produce the highest stellar masses of any galaxies across the sample. 

\subsection{Simba} \label{subsec:SIMBA}
The Simba simulations used in this study are from the highest resolution publicly-available Simba run, spanning a volume of $\sim 37^3$ (25 $h^{-1}$)$^3$ cMpc$^3$ \citep{SIMBA}. Simba is run with a modified version of the Gizmo code \citep{gizmo, gizmoPR}, and adopts a cosmology consistent with \citet{Planck2016}. Among the five Simba galaxies used in this analysis (see below and Table \ref{tab:features_table}), the median resolution element mass of a hot, X-ray emitting gas cell is $ \langle m_{\text{gas}} \rangle \simeq 2.3 \times 10^6~M_{\odot}$. 

The Simba physics model is given by \citet{SIMBA}, and we include a summary of the relevant processes as follows. Simba models galactic feedback physics associated with stellar formation, evolution, and feedback via supernovae (Types Ia and II) and AGB stellar winds. Simba models AGN formation, growth, and feedback via two kinetic modes and a thermal mode. The kinetic modes of the Simba AGN feedback scheme are explicitly anisotropic, injecting mass, energy, and momentum into bipolar outflows directed perpendicular to the galactic disk, in contrast to many other AGN feedback implementations in cosmological simulations. At high black hole/AGN accretion rates, this energy injection takes the form of a radiative wind, and at low accretion rates, a jet is launched. A thermal (X-ray) energy deposition is also applied during this jet mode. Simba includes a prescription for non-equilibrium radiative cooling and photoionization heating from background radiation fields, though it does not include a prescription for cosmic ray physics/feedback.

\section{Generating X-ray observables} \label{sec:obsgen}
In this study, we utilize the thermodynamic state of the hot CGM in each of the simulated galaxies to predict characteristic X-ray emission signatures. In this section, we describe the methods used to realize these predictions. For all simulated galaxies, we utilize snapshots at $z = 0$. For each snapshot, we generate projected maps of physical quantities of the hot CGM that could be accessed in the soft X-ray band by a generalized high-resolution future X-ray observatory to probe the morphological and kinematic state of the hot CGM. These quantities include the 0.5-1 keV band X-ray emissivity (as calculated in Section \ref{subsec:emissgen}), temperature (\textit{kT}), line-of-sight (LOS) velocity ({\vlos}), and LOS velocity dispersion ({\sigmav}) of the hot CGM. To isolate gas cells in the simulations that represent the hot, X-ray emitting CGM, we apply both a temperature criterion ($T > 3 \times 10^5$ K) and a density criterion ($\rho < 3 \times 10^{-25}$ g cm$^{-3}$). These cuts act to exclude cool gas in the CGM and ISM that would emit at energies below the soft X-ray band, as well as gas within a reasonable star formation threshold in the simulations. This definition of hot, X-ray emitting gas cells is identical to that used in \citet{ZuHone2024}.

\begin{figure*}[thb!]
    \centering 
        \includegraphics[width=1\textwidth]{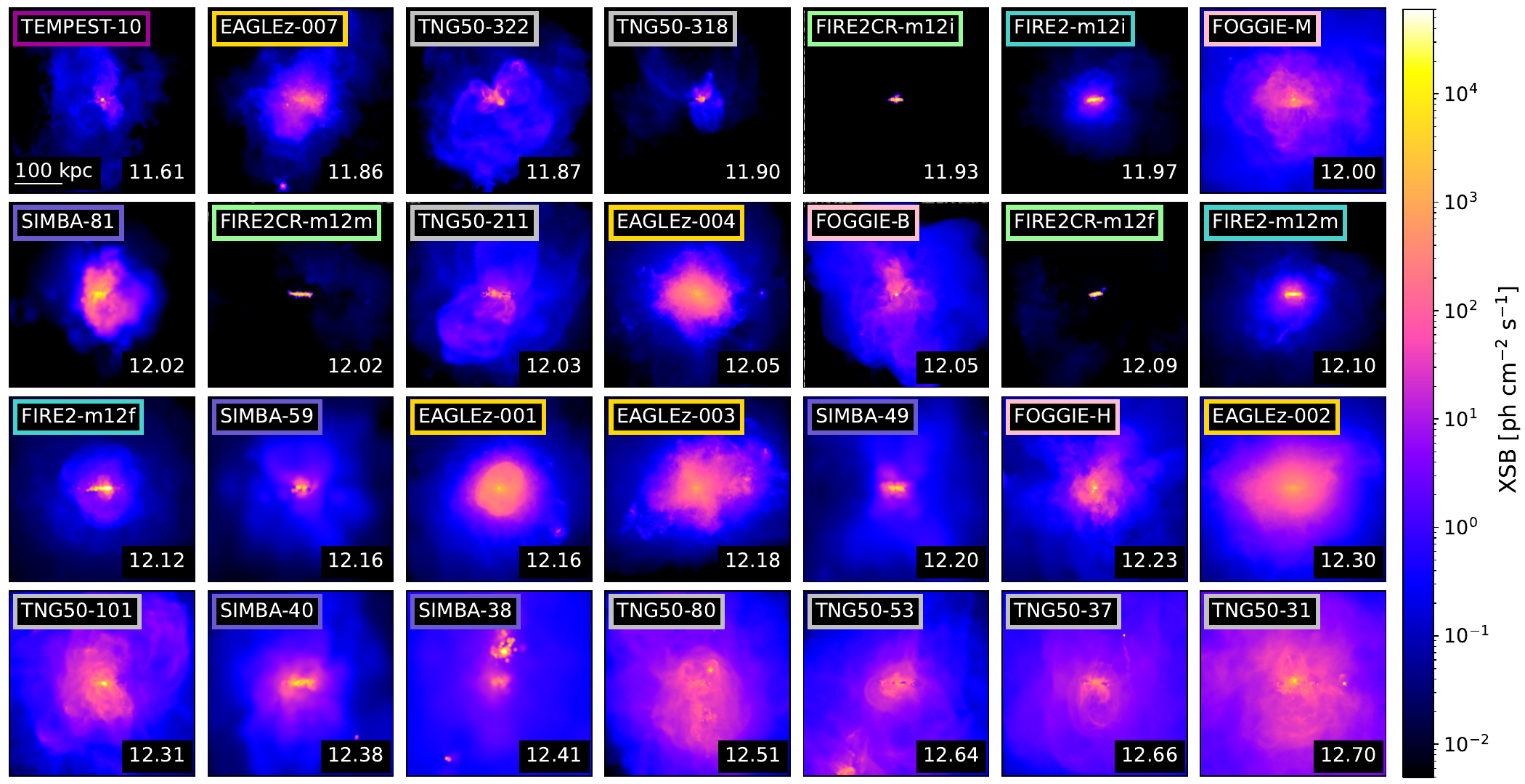}
        \caption{Edge-on projections of broadband thermal 0.5-1~keV X-ray surface brightness (XSB) for all galaxies in the sample sorted from lowest to highest halo mass (with $\log(M\;[M_{\odot}])$ indicated in each bottom right corner). The XSB morphology in each galaxy is strongly dependent on the underlying galactic feedback physics and environment. \vspace{0.5em}}  \label{fig:xsb_galsamp}
\end{figure*} 

\begin{figure}[th] 
    \centering 
        \includegraphics[width=0.47\textwidth]{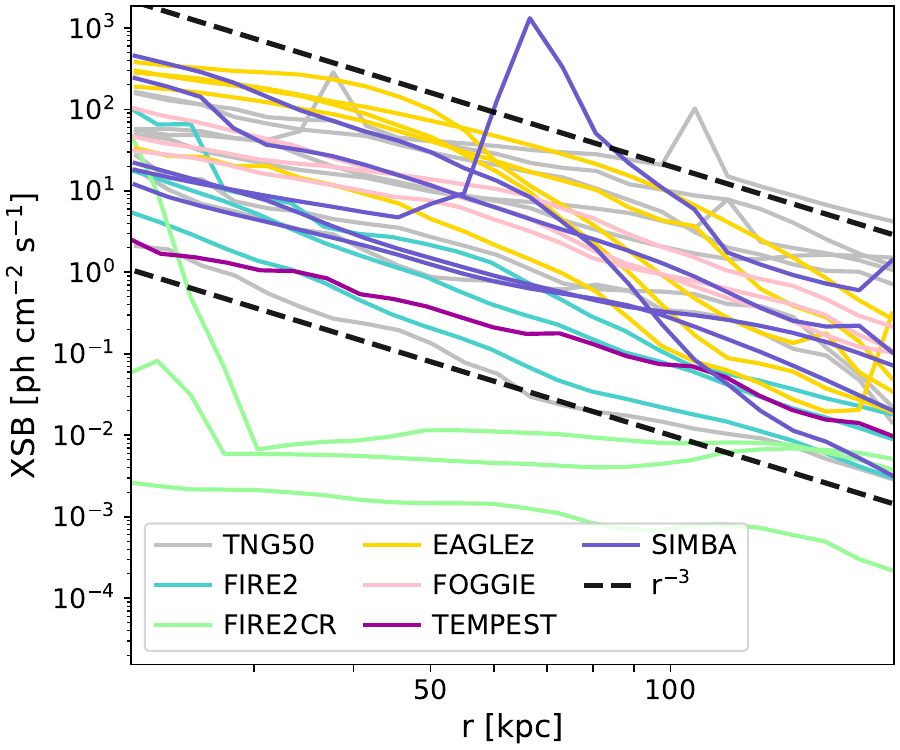}
        \caption{Radial profiles calculated from broadband thermal 0.5-1~keV XSB edge-on projections of all galaxies in the sample. Profiles are calculated in concentric circular annuli centered on the galaxy out to a radius of $\simeq200$ kpc. For all galaxies except those incorporating cosmic ray physics, the mean XSB profile slope is $\propto r^{-3}$ (illustrated with multiple amplitudes for reference)).} \label{fig:xsb_edge_1d}
\end{figure} 

To mimic maps derived from X-ray observations, we first project \textit{kT}, {\vlos}, and {\sigmav} of the hot gas cells along a given LOS weighted by the broadband X-ray emissivity, while unweighted projections of the broadband X-ray emissivity yield the X-ray surface brightness (XSB) maps. We place each simulated central galaxy nearby (at $z = 0.01$; see discussion below), and we generate projected maps (``mocks''), along three primary lines of sight: oriented relative to the galactic disk angular momentum vector in edge-on, face-on, and midway between the former two orientations. We calculate this angular momentum vector for each galaxy using the population of young stars (which we define to have ages $t_{\text{age}} < 5$ Gyr). 

In practice, continuum measurements of the hot CGM of external galaxies will not be possible due to the bright MW foreground emission which will dominate. Therefore, in addition to the broadband quantities, we identify three soft X-ray emission lines as probes of the hot CGM for use in this study (see Table \ref{tab:lines_table}). These emission lines (O VII, O VIII, and Fe XVII) are expected to be prominent in the $\sim 0.1 - 1$~keV thermal plasma of the hot CGM in the soft X-ray band \citep{Schellenberger2024}. These lines will be brighter than the MW foreground, and importantly, at $z=0.01$, they are redshifted sufficiently far from the Milky Way CGM foreground emission lines in the observer frame \cite[see, e.g., Figure 7 in][]{Kraft2022} such as to be distinguished from them, assuming a spectral resolution $\lesssim$~3~eV (such measurements are impossible for CCD instruments on current X-ray observatories). We therefore generate analogous mocks of the aforementioned physical quantities in the hot CGM weighted by the X-ray emissivity of each of the O VII, O VIII, and Fe XVII lines. The X-ray emissivity of each line is derived from the broadband X-ray spectrum by taking a small (3 eV) spectral window centered on the rest-frame line energy.

\subsection{X-ray emissivity calculation} \label{subsec:emissgen}
At soft X-ray energies below $\sim$1 keV, photoionization of the low-density warm/hot CGM by the cosmic UV/X-ray background radiation fields becomes important relative to collisional ionization, which is the dominant ionization process in the warm/hot CGM above this energy. Photoionization acts to shift ions to higher ionization states than would be expected from collisional ionization alone (see \citealt{Churazov2001, Wijers2020} and Appendix \ref{appendix:photoion}). We therefore calculate soft X-ray emissivities of lines and continuum with a model for an optically-thin thermal plasma including photoionization by the cosmic UV/X-ray background radiation fields \citep{Churazov2001,Khabibullin2019}. At temperatures above $\sim 1$~keV, this reduces to a plasma in collisional ionization equilibrium (CIE). The spectra are calculated with a \texttt{cloudy}-based CIE model based on the Cloudy radiative transfer package \citep{Ferland2017}. 

One simulation suite in this study, FIRE2CR, includes cosmic rays in the CGM model. In this work, we do not consider additional potential contributions to the X-ray emissivity from the inverse-Compton scattering of cosmic microwave background (CMB) photons with GeV cosmic ray electrons, which may contribute non-negligibly to the broadband X-ray emissivity for $\sim$MW-mass halos \citep{Hopkins2025, Lu2025}. We instead save a more detailed quantitative analysis of these effects on the X-ray spectra of MW-mass galaxies for future work. We also do not include contributions from resonant scattering from the cosmic X-ray background (CXB) given the small expected emission contribution of this component, which we expect to be insignificant except in low-density regions (i.e., large radii). Finally, we verified that there are no significant differences between mock X-ray observables generated with a \texttt{cloudy}-based CIE model relative to those generated with the commonly used Astrophysical Plasma Emission Code \citep[APEC;][]{Smith2001,Foster2012} model. 

\section{Results: hot CGM properties across galaxy simulations} \label{sec:results}
\subsection{X-ray surface brightness} \label{subsec:xsb}
The simulated galaxies we include in our analysis make very different predictions for the morphology of the hot CGM in their respective galaxies. To illustrate this, in Figure \ref{fig:xsb_galsamp}, we show projections of the broadband (again, 0.5-1~keV) XSB for all simulated galaxies in the sample, as viewed by an observer oriented edge-on to the galactic disk. In general, most of the simulations predict a volume-filling diffuse X-ray emitting CGM out to large ($r \gtrsim 100$~kpc) radii, though the azimuthal anisotropy of the CGM morphology in each simulation varies significantly, especially when all galaxies are viewed edge-on. Enhancements in XSB along the galaxy's minor axis due to AGN outflows are apparent in many of the galaxies. For example, TNG50-211 and TNG50-101 clearly exhibit a series of bubble-like shells preferentially oriented above and below the disk plane, driven by the AGN kinetic feedback mode. Many Simba galaxies, most obviously SIMBA-49 and SIMBA-81, also indicate XSB enhancements perpendicular to the disk plane, though the coarser resolution of these simulations compared to TNG50 makes identifying edge morphology or substructures more difficult. 

\begin{figure*}[t]
    \centering 
        \includegraphics[width=0.85\textwidth]{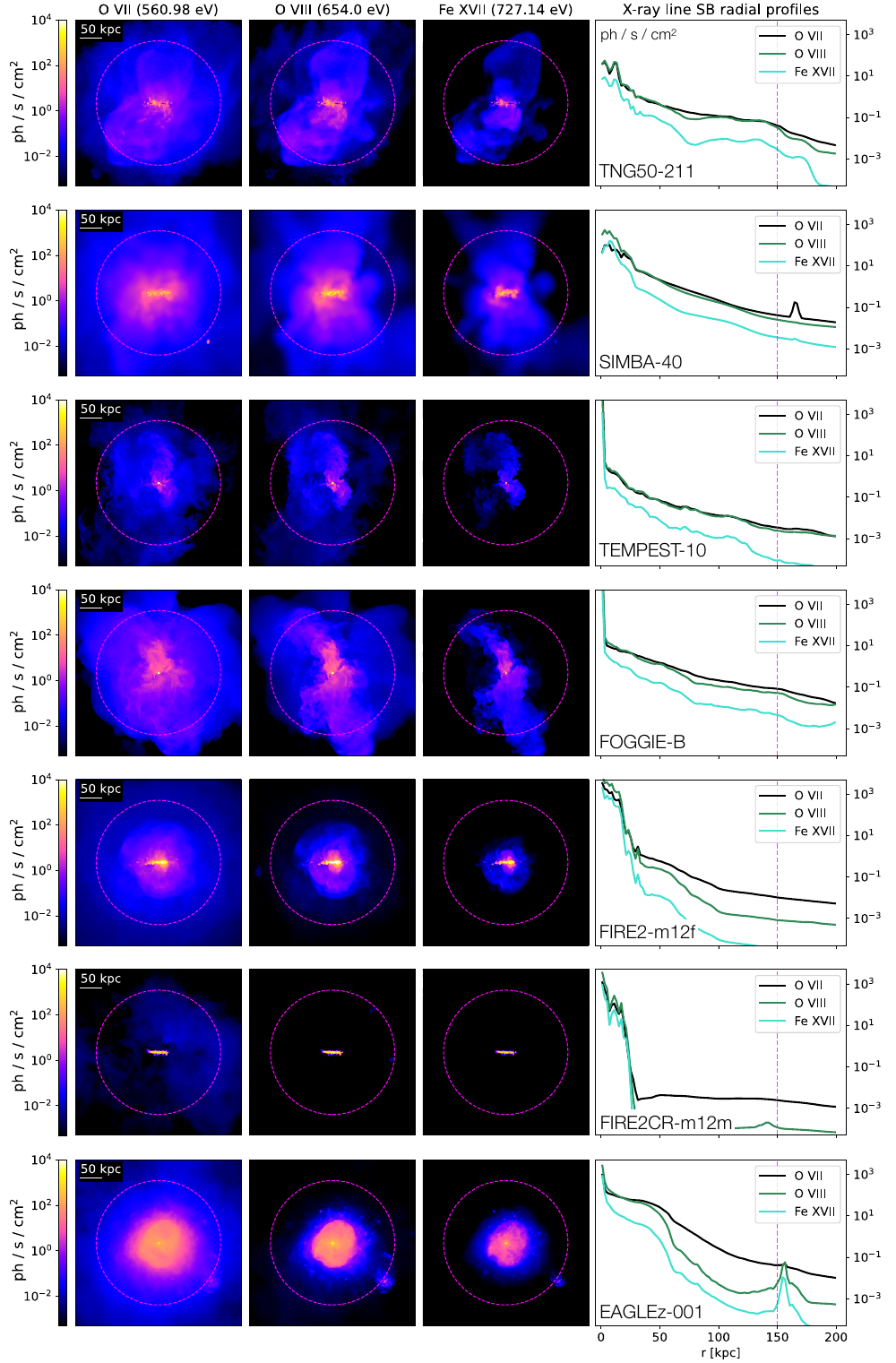}
        \caption{X-ray emission line surface brightness maps (from left to right: O VII, O VIII, Fe XVII) for one galaxy from each simulation suite (from top to bottom: TNG50, Simba, TEMPEST, FOGGIE, FIRE2, FIRE2CR, EAGLEz) with radial profiles derived from each map.} \label{fig:xsb_lines_maps_profiles}
\end{figure*} 

\begin{figure}[h]
    \begin{interactive}{animation}{XSB_paper_4123.mov}
        \includegraphics[width=0.43\textwidth]{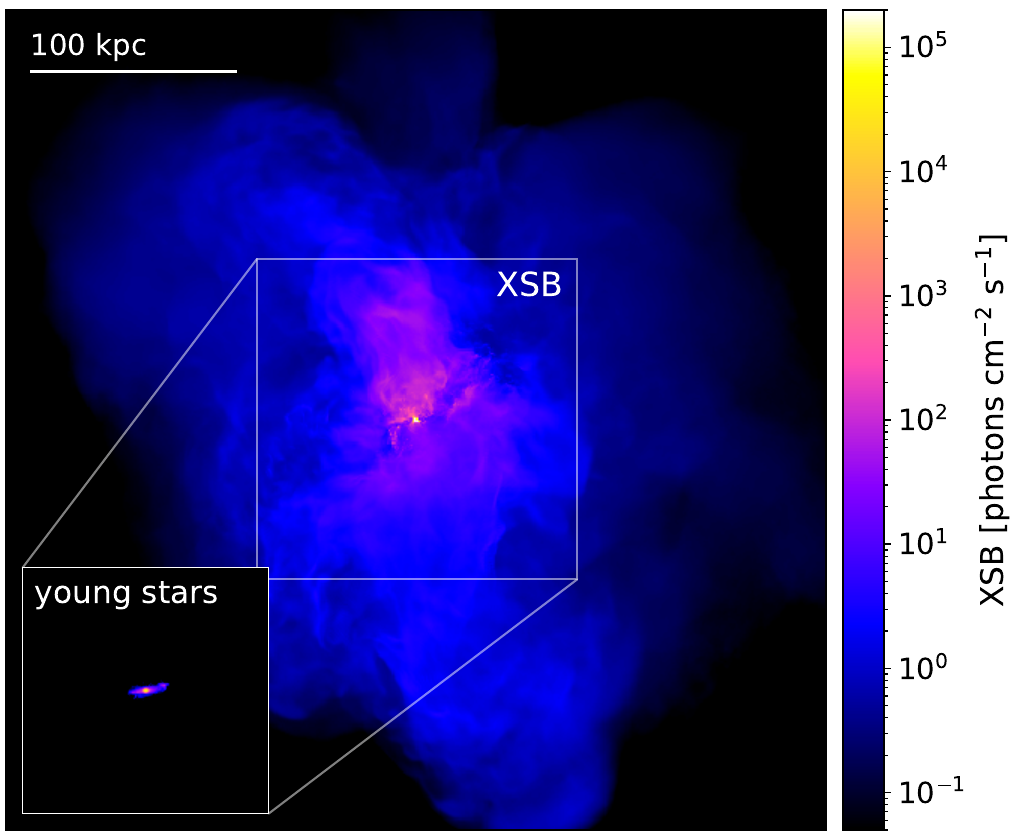}
    \end{interactive}
    
    \begin{interactive}{animation}{XSB_paper_f.mov}
        \includegraphics[width=0.43\textwidth]{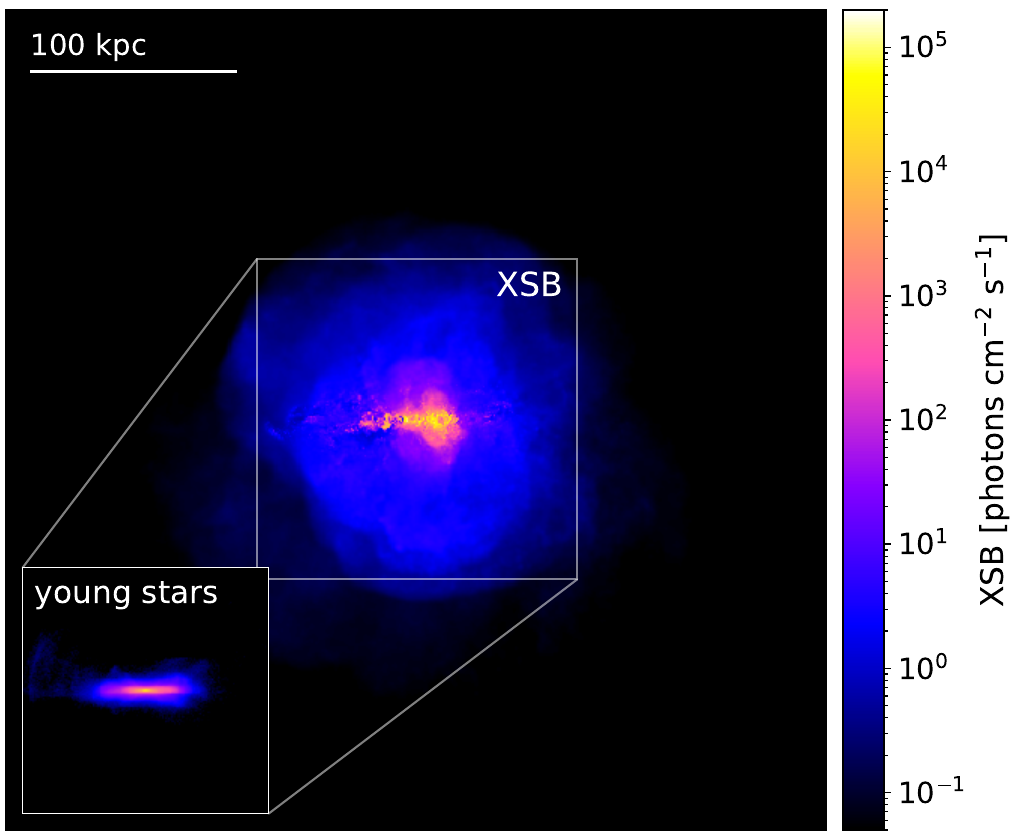}
    \end{interactive}
    
    \caption{Example animations of the broadband thermal XSB and the young stellar disk mass density as viewed by observers at locations sweeping out $360^{\circ}$ in viewing angle from an edge-on orientation with the galaxies FOGGIE-B and FIRE2-m12f. The XSB morphology predictions vary significantly from one simulation to another.} \label{fig:xsb_animation}
\end{figure}

Most of the fiducial FIRE2 galaxies \citep[all of which do not have a central AGN, but rather have more extended stellar feedback in the disk, see][]{Orr2020} show largely azimuthally symmetric XSB distributions on large spatial scales. However, in other simulation suites some of the galaxies simulated with only stellar feedback prescriptions can host anisotropic XSB enhancements. For example, the TEMPEST-10 and FOGGIE-B galaxies show clear regions of bright XSB preferentially above and below the disk planes associated with strong central outflows, which we discuss in detail in the following sections. In contrast to all other simulation suites neglecting cosmic ray feedback, there is very little thermal X-ray emission from the hot CGM in the FIRE2CR galaxies outside a couple tens of kpc from the halo centers, in agreement with \citet{FIRE2CR}. This prediction is also qualitatively consistent with predictions of a cooler CGM supported by cosmic ray pressure in simulations of MW-like galaxies run in other codes with a variety of cosmic ray transport prescriptions \citep[e.g.,][]{Buck2020,Butsky2018,Salem2016}.

Most of the galaxies in our sample are reasonably isolated (in most cases by selection), but three galaxies (SIMBA-38, TNG50-80, and TNG50-53) host visibly X-ray bright nearby subhalos. We note by visual examination that the morphological anisotropy in XSB is maximized when galaxies are viewed edge-on, while face-on projections of XSB are more azimuthally symmetric. Population-averaged stacking analyses from cosmological simulations confirm this trend, having broadband XSB distributions enhanced along the galactic disk angular momentum vector (minor axis) of $\sim$MW-mass galactic disks in TNG and EAGLE \citep{Truong2021}, likely due to hot outflows from AGN feedback propagating outward above and below the galactic disk plane. In contrast to other simulation suites that implement AGN feedback schemes, some of the EAGLE NEQ zoom galaxies (notably EAGLEz-002 and EAGLEz-003) exhibit broadband XSB distributions that are elongated along the major axis, which is not in agreement with the large minor-to-major axis XSB ratios typically found in EAGLE \citep[Ref-L0100N1504; see][]{Truong2021} when evaluated using a similar energy band. The source of this discrepancy is not clear, though it could be influenced by, e.g., differences in the physics models, simulation resolutions, or environments of individual galaxies. 

The distribution of hot CGM can also be characterized using azimuthally averaged radial profiles of the broadband XSB. In Figure \ref{fig:xsb_edge_1d}, we plot XSB radial profiles for each simulated galaxy in the sample when observed from an edge-on orientation relative to the galactic disk calculated in concentric circular annuli centered on the galaxy out to a radius of $\simeq200$ kpc. We note that nearby massive halos can result in spikes in the XSB radial profiles (e.g., the large spike in one of the Simba XSB radial profiles is associated with an X-ray bright nearby subhalo in the SIMBA-38 galaxy; see Figs. \ref{fig:xsb_edge_1d} and \ref{fig:xsb_galsamp}). 

Interestingly, for all galaxies simulated without cosmic ray physics, even given the wide range of halo masses, hot CGM spatial/mass resolutions, feedback schemes, and projected XSB morphologies spanned by the simulated galaxies, the azimuthally averaged radial XSB profiles generally follow similar shapes. The mean slope of these XSB profiles over the range of $20-200$ kpc is $\propto r^{-3}$ (overplotted at two representative amplitudes for reference in Figure \ref{fig:xsb_edge_1d}). We outline a very basic argument highlighting the plausibility of this shape for a simple hot CGM as follows. In an approximation of a roughly isothermal halo where the hot CGM is in CIE and the temperature and metallicity do not vary by orders of magnitude (the former being explicitly shown to be the case for our simulated galaxies in Section \ref{subsec:kT}), the X-ray emissivity will be strongly dependent on density ($\epsilon \propto n^2$). For a roughly isothermal halo, the gas density profile of the hot CGM should follow the isothermal sphere solution ($n \propto r^{-2}$). Projecting the X-ray emissivity along the LOS, the projected radial profile of the XSB would thus be roughly $\propto r^{-3}$. Therefore, the mean slope of our XSB radial profiles ($r^{-3}$) represents a plausible shape for a simple isothermal hot halo with roughly constant metallicity. 

However, the scatter in these XSB radial profile slopes is non-negligible between simulations. The slopes ($\alpha$) for galaxies in each simulation are within the following ranges: TNG50: $-1.4 \leq\alpha\leq -3.1$, FIRE2: $-3.4 \leq \alpha \leq -3.8$, EAGLEz: $-3.0 \leq \alpha \leq-5.0$, FOGGIE/TEMPEST: $-2.4 \leq \alpha \leq-3.3$, and Simba: $-1.1 \leq \alpha \leq-5.8$. On average, TNG50 exhibits the shallowest broadband XSB radial profile slopes, and EAGLEz the steepest. This trend is not strongly correlated with the spatial/mass resolution of the simulations. 

Individually, the profiles can be biased in the presence of an X-ray bright nearby subhalo. In the case of TNG50, the slopes of galaxies with identified X-ray bright nearby subhalos (TNG50-53 and TNG50-80) are not outliers relative to the radial XSB profile slopes of isolated halos in TNG50 of similar mass (e.g., TNG50-37 or TNG50-101). However, in Simba, the galaxy with an X-ray bright nearby subhalo (SIMBA-38) has an extremely shallow slope that is likely affected by the companion halo, so if the presence of these subhalos are not accounted for in observational analyses, biases in the XSB radial profile slopes are possible. Overall, it is likely that the primary differences in XSB slope are dominated by the effects of the feedback prescription in each simulation suite, i.e., how and on what spatial/mass scales the stars and/or AGN in each simulation can redistribute the hot CGM gas and metal content. 

For comparison, observed broadband stacked XSB radial profiles for a sample of $\sim$MW-mass central galaxies in the eROSITA All-Sky Survey \citep{Zhang2024} indicate a much shallower slope at large radii $\propto r^{-1.6}$ (with $\beta \simeq 0.43$ from $\beta$-model fits). A robust comparison of the broadband XSB radial profiles from our simulated galaxy sample to the \citet{Zhang2024} result calls for realistically matching the observed galaxy sample selection criteria, stacking technique, and profile generation conventions (e.g., X-ray energy band and radial range utilized). While these efforts are beyond the scope of this study, they will be described in a future paper. 

We find that the broadband XSB emission from the hot CGM at nearly all radii in the three FIRE2CR galaxies is much fainter than that of any other simulation suite, often by $\gtrsim$ an order of magnitude. These galaxies show a noticeable lack of hot, X-ray emitting CGM at any radius beyond the galactic disk. The broadband XSB maps (Figure \ref{fig:xsb_galsamp}) and profiles (Figure \ref{fig:xsb_edge_1d}) predominantly trace only the hot gas within the galactic disks, and the XSB out to a radius of $\sim$200 kpc is the lowest of any simulation suite. This is consistent with the analyses of \citet{Ji2020, chan2022}, who both note a decreased fraction of hot CGM in FIRE2 runs with cosmic ray physics relative to the fiducial FIRE2 runs for a sample of the ``m12'' galaxies, especially at larger heights from the galactic disks. In these FIRE2CR runs, and as is generally predicted by other simulations incorporating cosmic ray physics \citep[e.g.,][]{Buck2020,Butsky2018,Salem2016}, the cosmic rays can provide the necessary pressure support against gravity in the halo, so hydrostatic equilibrium can be maintained at lower gas temperatures than halos that are supported predominantly by thermal pressure (which therefore have higher amounts of hot, X-ray emitting CGM dominated by line emission throughout the halos). 

In these cosmic ray pressure-dominated galaxies, the broadband XSB could potentially be boosted by X-ray emission generated by the inverse-Compton scattering of CMB photons with GeV cosmic ray electrons. As motivated in \citet{Hopkins2025} and explored in \citet{Lu2025}, an inverse-Compton continuum component from an extended distribution of cosmic rays in a galaxy could contribute non-negligibly to the broadband X-ray emissivity for $\sim$MW-mass halos and further alter the slope of the radial XSB profiles.  

While broadband images of the XSB of the hot CGM show its distribution, narrow-band images in individual emission lines can reveal its different sub-phases. To illustrate this, in Figure \ref{fig:xsb_lines_maps_profiles}, we show maps of the XSB of the O VII, O VIII, and Fe XVII lines (Table \ref{tab:lines_table}), respectively, and the associated radial profiles for one galaxy from each simulation suite for an observer viewing the galaxy edge-on. Across all of the galaxies, which are $\sim$MW-mass, the volume-filling hot CGM is best traced with the O VII emission line, while feedback-driven outflows from the galactic disks are traced by the higher energy emission lines. Fe XVII nearly exclusively traces hot, high-velocity outflows (see, e.g., the FOGGIE-B XSB maps in Figure \ref{fig:xsb_lines_maps_profiles}) or shocks (e.g., the bipolar outflow lobes of EAGLEz-001 in Figure \ref{fig:xsb_lines_maps_profiles}), whereas O VIII can often simultaneously be a good tracer of both outflows and the volume-filling gas to large galactic radii (e.g., TNG50-211 in Figure \ref{fig:xsb_lines_maps_profiles}). 

These maps are consistent with trends observed by \citet{Truong2023} in an analysis of several individual Milky Way-like galaxies in TNG50. In particular, \citet{Truong2023} found that bubble structures due to strong outflows reminiscent of the Milky Way eROSITA/Fermi-like bubbles \citep[as originally explored in][]{Pillepich2021} were progressively better traced by higher energy X-ray emission lines. At the population level, \citet{Truong2023} similarly note that XSB profiles of low energy X-ray lines, including O VII, are largely isotropic for galaxies selected from TNG100, EAGLE, and Simba at $z = 0$ (within a similar stellar mass range as this study). Higher energy emission lines, including O VIII and Fe XVII, were found to exhibit a high degree of anisotropy, enhanced along the angular momentum vector of the galactic disks, which is in general agreement with our findings for the higher resolution TNG50, EAGLE NEQ zoom, and Simba halos in this study (see Figure \ref{fig:xsb_lines_maps_profiles}). However, two EAGLE NEQ zoom galaxies provide a strong counter-example of this, as EAGLEz-002 and EAGLEz-003 exhibit broadband XSB and emission line emissivities elongated along the major axis. 

Of the three emission lines analyzed here, the O VII and O VIII X-ray line SB maps generally trace the brightest gas within 200~kpc of each halo center. The Fe XVII line, tracing more anisotropic CGM features associated with feedback at super-virial temperatures, is typically the faintest in total SB out to 200 kpc (see Figure \ref{fig:xsb_lines_maps_profiles}). These trends are in general agreement with mock microcalorimeter observations of similar mass galaxies to those studied here from TNG100, EAGLE, and Simba produced by \citet{Schellenberger2024}, who found the same qualitative contributions of O VII, O VIII, and Fe XVII as above. Finally, when we exclude regions in the inner halo in/near the galactic disk region (enforcing $20 < r < 200$ kpc), the relative XSB contributions from the O VII and O VIII lines are similar, and both are dominant to the Fe XVII line.

\begin{figure*}[thb!]
    \centering 
        \includegraphics[width=1\textwidth]{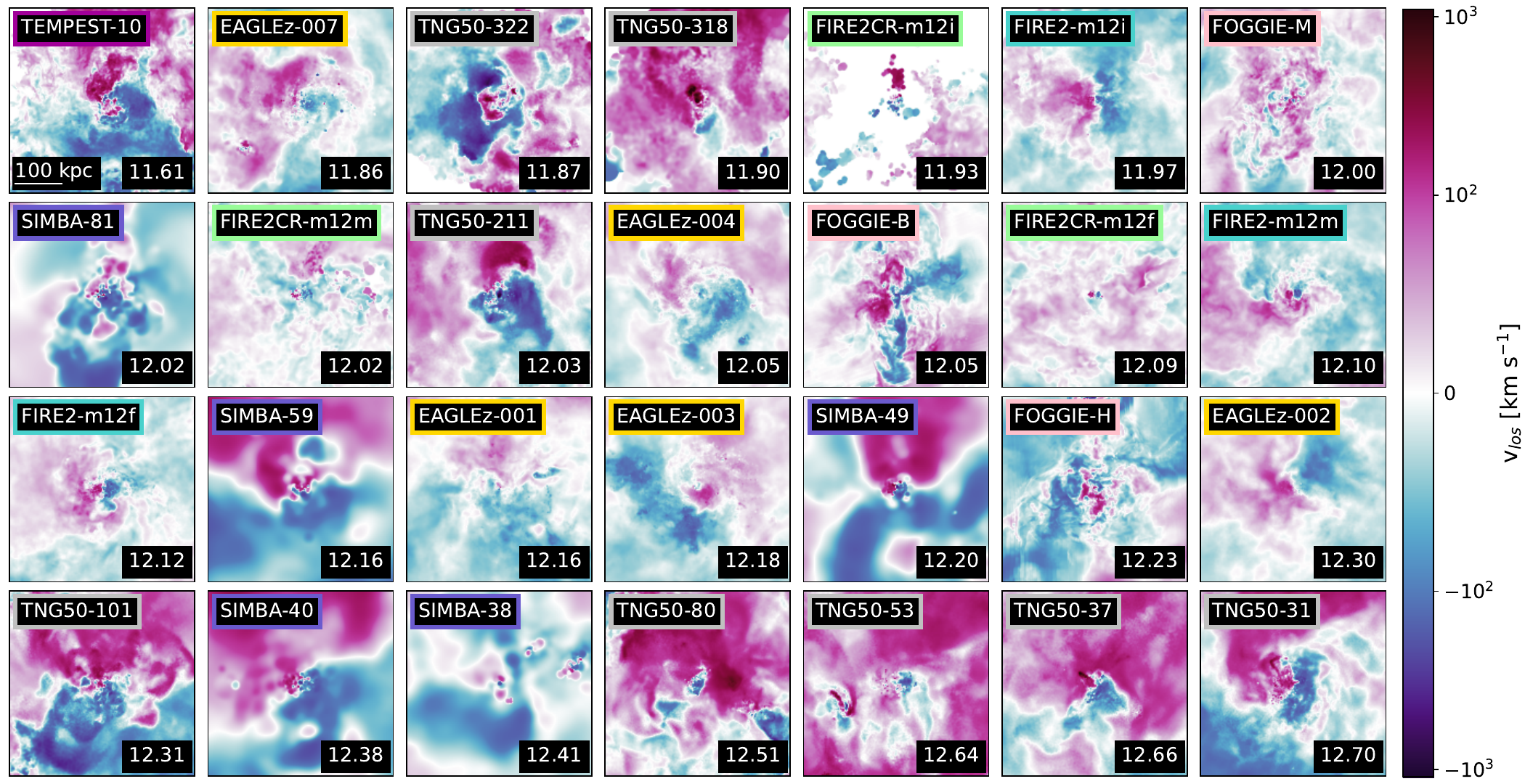}
        \caption{Intermediate (midway between edge- and face-on)  projections of LOS velocity ({\vlos}) weighted by the broadband X-ray emissivity for all galaxies in the sample sorted from lowest to highest halo mass (with $\log(M\;[M_{\odot}])$ indicated in each bottom right corner). The {\vlos} structure varies significantly, even between halos of similar masses, as a function of simulation and environment. \vspace{1em}}  \label{fig:vlos_xsb_galsamp}
\end{figure*}

\begin{figure*}[thb!]
    \centering 
        \includegraphics[width=1\textwidth]{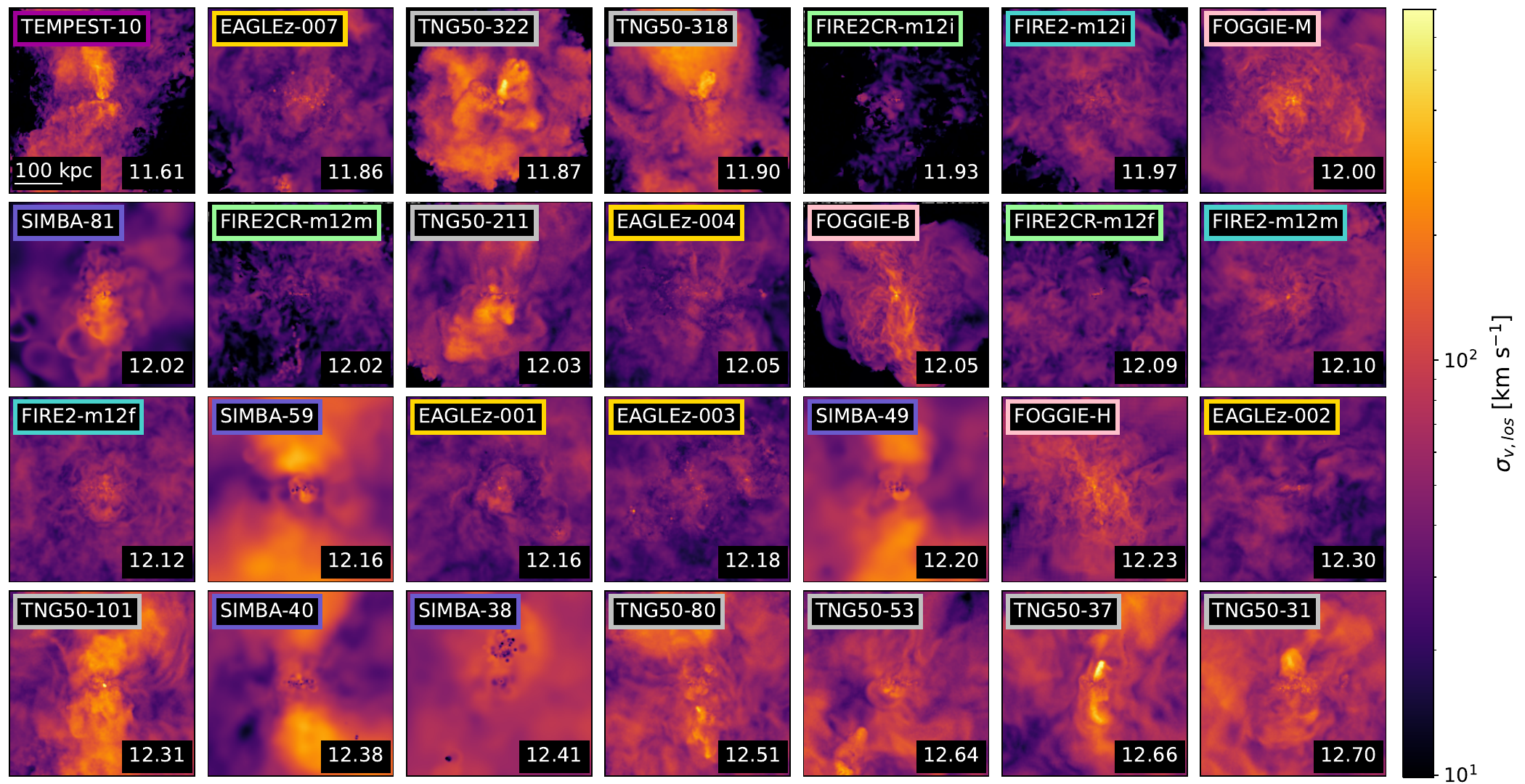}
        \caption{Edge-on projections of LOS velocity dispersion ({\sigmav}) weighted by the broadband X-ray emissivity for all galaxies in the sample sorted from lowest to highest halo mass with $\log(M\;[M_{\odot}]$ indicated in each bottom right corner). Galaxies with strong large-scale feedback-driven outflows generally exhibit corresponding regions of high {\sigmav}.} \label{fig:vlos_sigma_xsb_galsamp} \vspace{1em}
\end{figure*} 

\begin{figure}[htb!] 
    \centering 
        \includegraphics[width=0.47\textwidth]{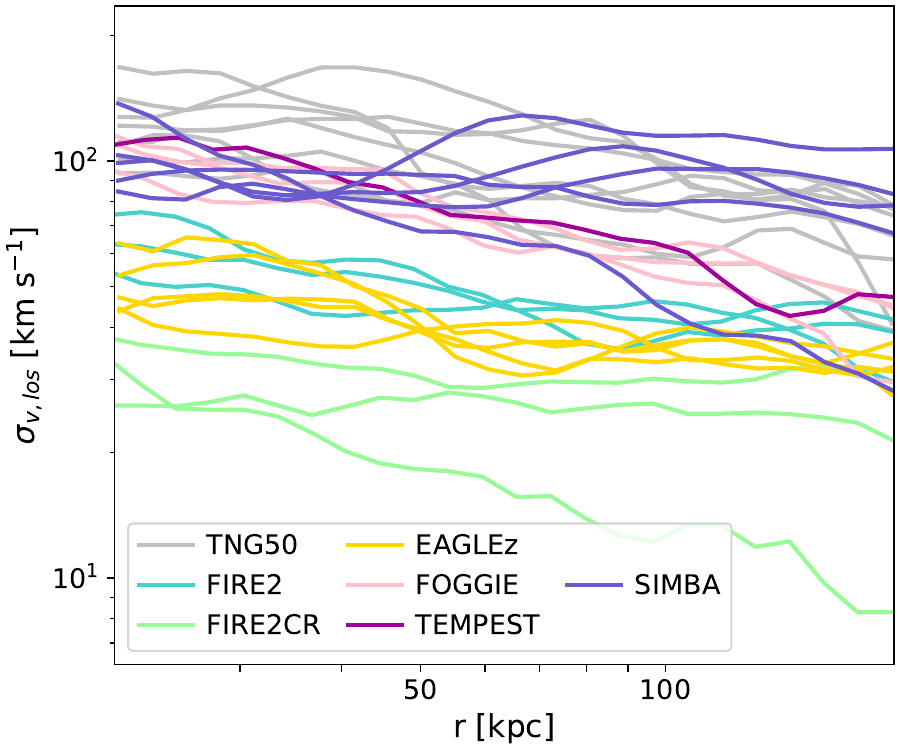}
        \caption{Radial profiles calculated from projections of LOS velocity dispersion ({\sigmav}) weighted by the broadband X-ray emissivity for all galaxies in the sample oriented edge-on. Profiles are calculated in concentric circular annuli centered on the galaxy out to a radius of $\simeq200$ kpc. {\sigmav} does not generally vary by more than a factor of $\sim$a few from galactic radii of 20 to 200 kpc, and the FIRE2CR simulations predict the lowest {\sigmav} on average. } \label{fig:sigmav_edge_1d}
\end{figure} 

Radial profiles derived from edge-on projections of the X-ray line emissivities of the simulated galaxies can therefore vary significantly from one emission line to another as the morphology of each map varies, and comparison of these profiles and their associated XSB maps can provide an important diagnostic of the physical state of the hot CGM for a given galaxy. Even though the O VII, O VIII, and Fe XVII emission lines are separated by only $\lesssim 200$ eV, maps of their XSB can distinguish between different phases of the hot CGM, particularly gas in hotter, often high-velocity outflows versus volume-filling gas at large radii. 

As we have shown, many of the simulations, especially TNG50, FOGGIE, TEMPEST, and Simba, predict highly anisotropic distributions of the hot CGM in XSB (see Figures \ref{fig:xsb_galsamp} and \ref{fig:xsb_lines_maps_profiles}). To contextualize this, we show in Figure \ref{fig:xsb_animation} an animation of the hot, X-ray emitting CGM of both the FOGGIE-B galaxy and the FIRE2-m12f galaxy in XSB, with the young stellar disk ($t_{\text{age}} < 5$ Gyr) additionally plotted for reference. In these examples, the distribution of hot CGM in FOGGIE-B is highly anisotropic, with strong enhancements above and below the disk plane out to large (r $\gtrsim 100$ kpc) radii, which most easily identified when the galaxy is oriented edge-on to the observer. FIRE2-m12f, in contrast, exhibits less XSB azimuthal anisotropy on the same large scales, but has enhanced emission above and below the disk plane on scales $\lesssim$ a few tens of kpc, again most apparent when the galaxy is oriented edge-on. 

\subsection{Velocities} \label{subsec:velocities}
\begin{figure}[htb!]
    \begin{interactive}{animation}{vlos_sigmav_XSB_211.mov}
    \includegraphics[width=0.43\textwidth]{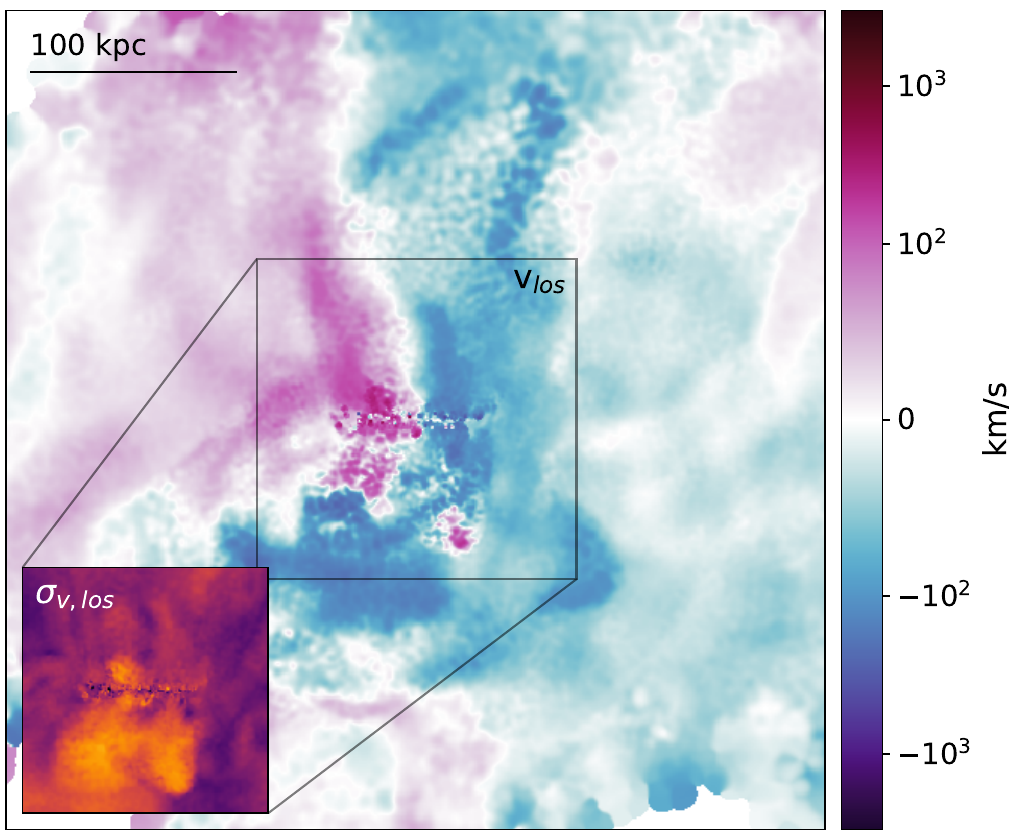}
    \end{interactive}

    \begin{interactive}{animation}{vlos_sigmav_XSB_mCR.mov}
    \includegraphics[width=0.43\textwidth]{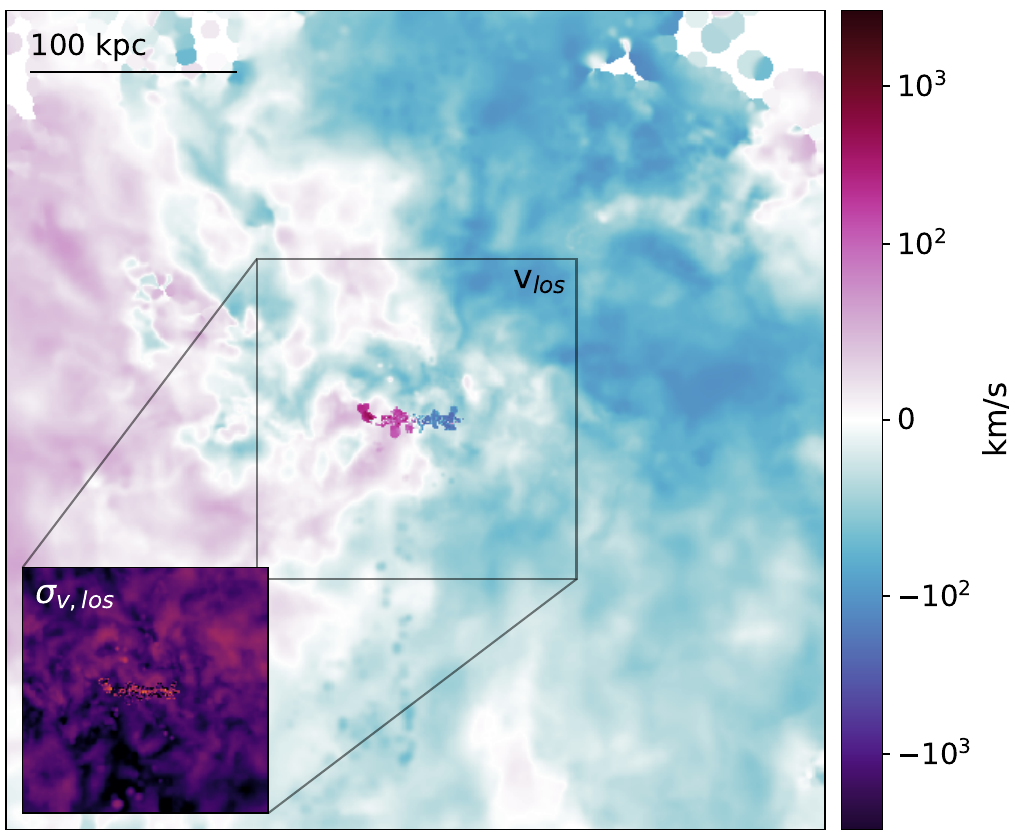}
    \end{interactive}

    \caption{Example animations of the LOS velocity ({\vlos}) and LOS velocity dispersion ({\sigmav}) weighted by the broadband X-ray emissivity as viewed by observers at locations sweeping out $360^{\circ}$ in viewing angle from an edge-on orientation with the galaxies (TNG50-211 and FIRE2CR-m12m). Jointly leveraging maps of {\vlos} and {\sigmav} can help distinguish kinematic contributions from AGN-driven outflows and disk/CGM rotation.} \label{fig:velocity_animation}
\end{figure}

\begin{figure*}[htb!]
    \centering 
        \includegraphics[width=0.95\textwidth]{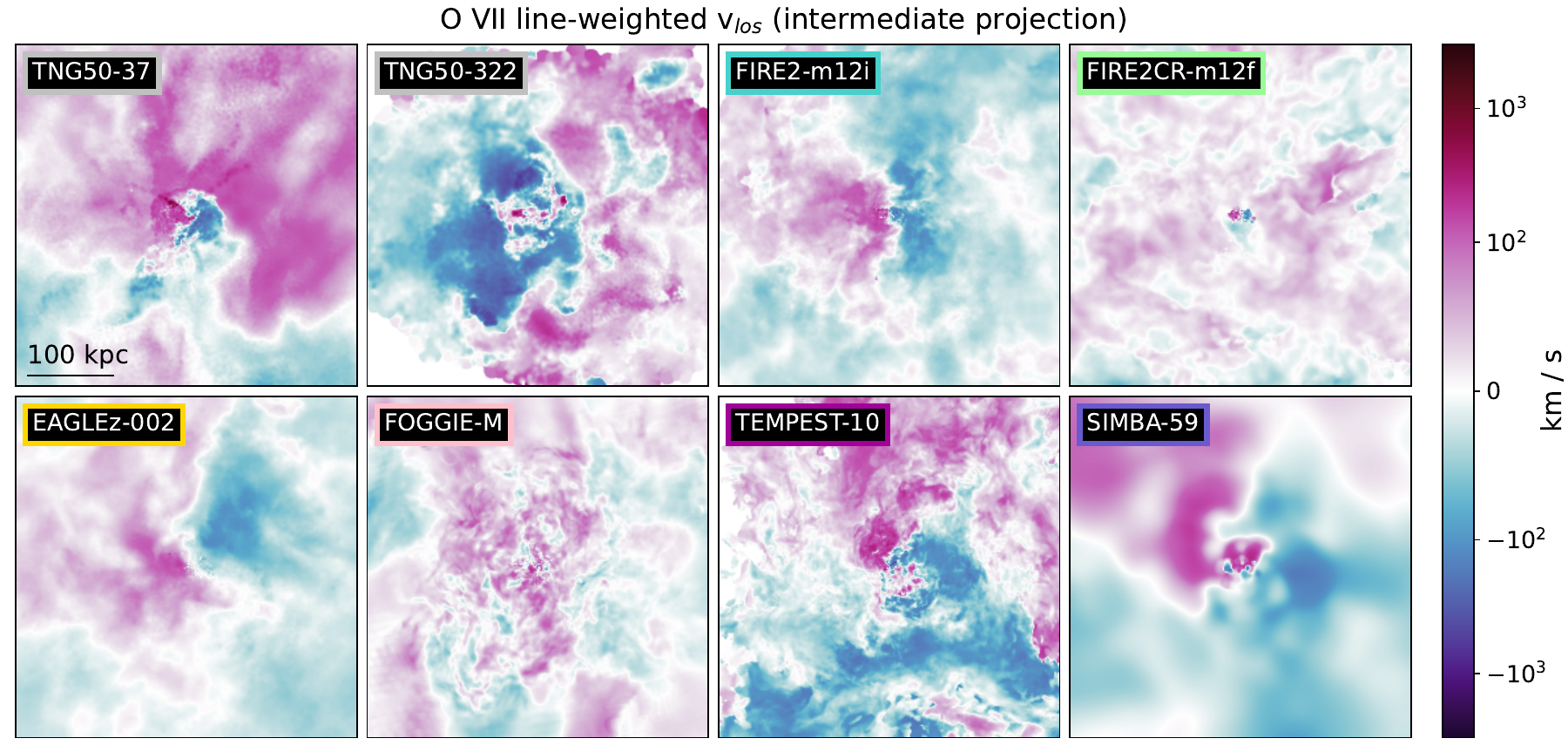}
        
        \vspace{1em}
        \includegraphics[width=0.95\textwidth]{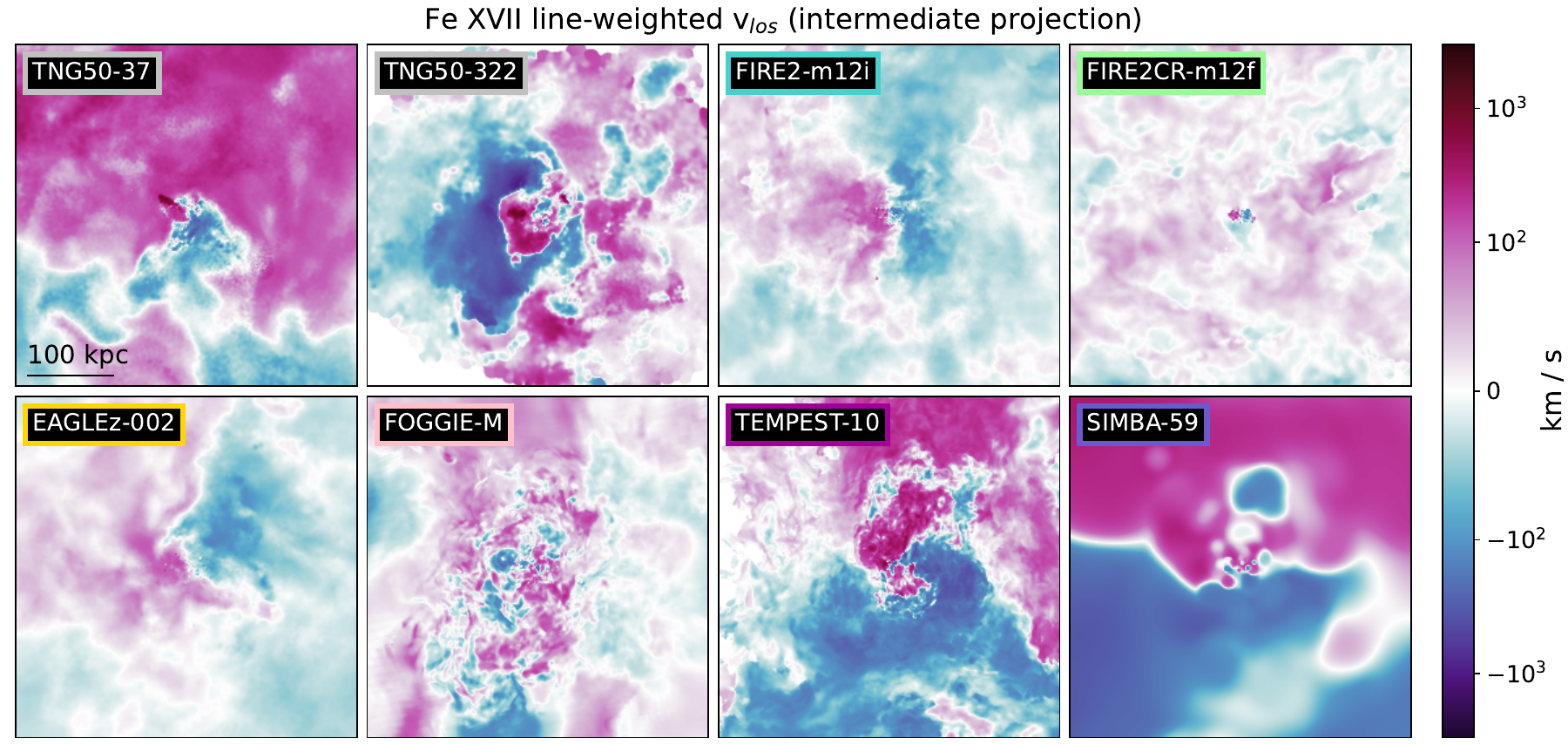}
        \caption{LOS velocity ({\vlos}) weighted by the O VII (\textit{top panels}) and Fe XVII (\textit{bottom panels}) line emissivities for galaxies from each simulation suite (\textit{clockwise, from upper left for each}: TNG (x2), FIRE, FIRE2CR, Simba, TEMPEST, FOGGIE, EAGLE) viewed from an intermediate projection. Here, velocity contributions from both rotation and outflows are visible, and the line weighting influences the observed velocity distribution.}  \label{fig:vlos_xsb_lines_all}
\end{figure*} 

\begin{figure*}[t]
    \centering 
        \includegraphics[width=1\textwidth]{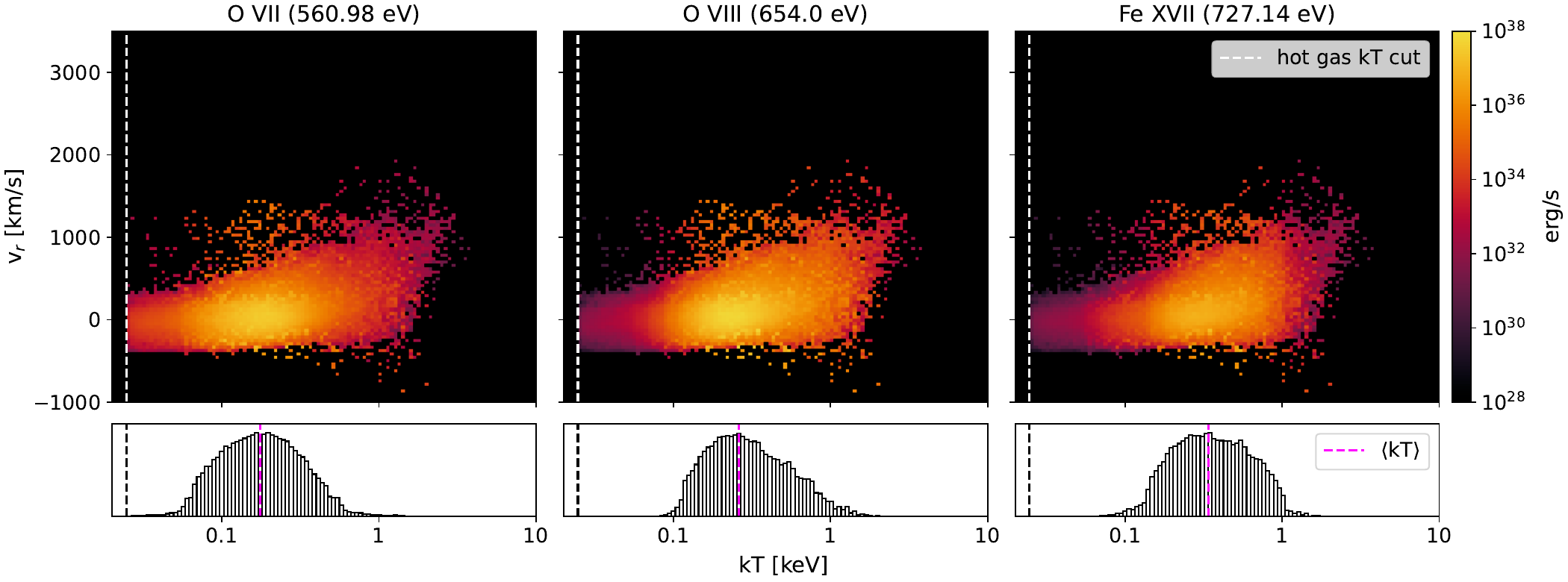}
        \caption{Temperature vs. radial velocity phase diagrams for O VII, O VIII, and Fe XVII line luminosities in the TNG50-31 galaxy. The histograms represent the corresponding distributions collapsed along the temperature axis. For this galaxy—and in general—the luminosity of the highest energy emission line (Fe XVII) peaks at higher temperatures than that of the lower energy emission lines. For galaxies with hot, high-velocity outflows, this corresponds to also peaking at higher radial velocities.}  \label{fig:xline_phases}
\end{figure*} 

\begin{figure*}[t]
    \centering 
        \includegraphics[width=1\textwidth]{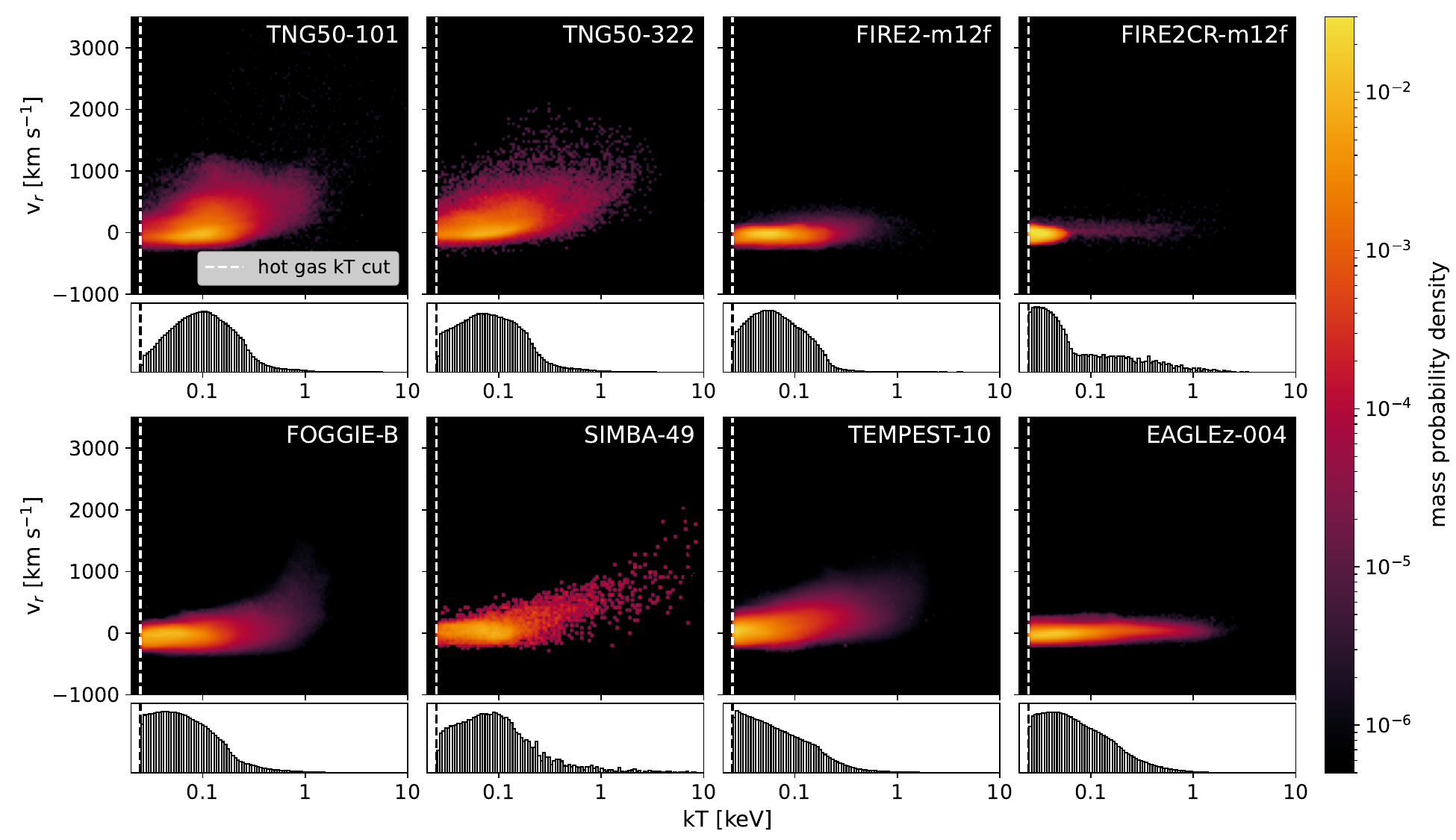}
        
        \vspace{3em}
        
        \includegraphics[width=1\textwidth]{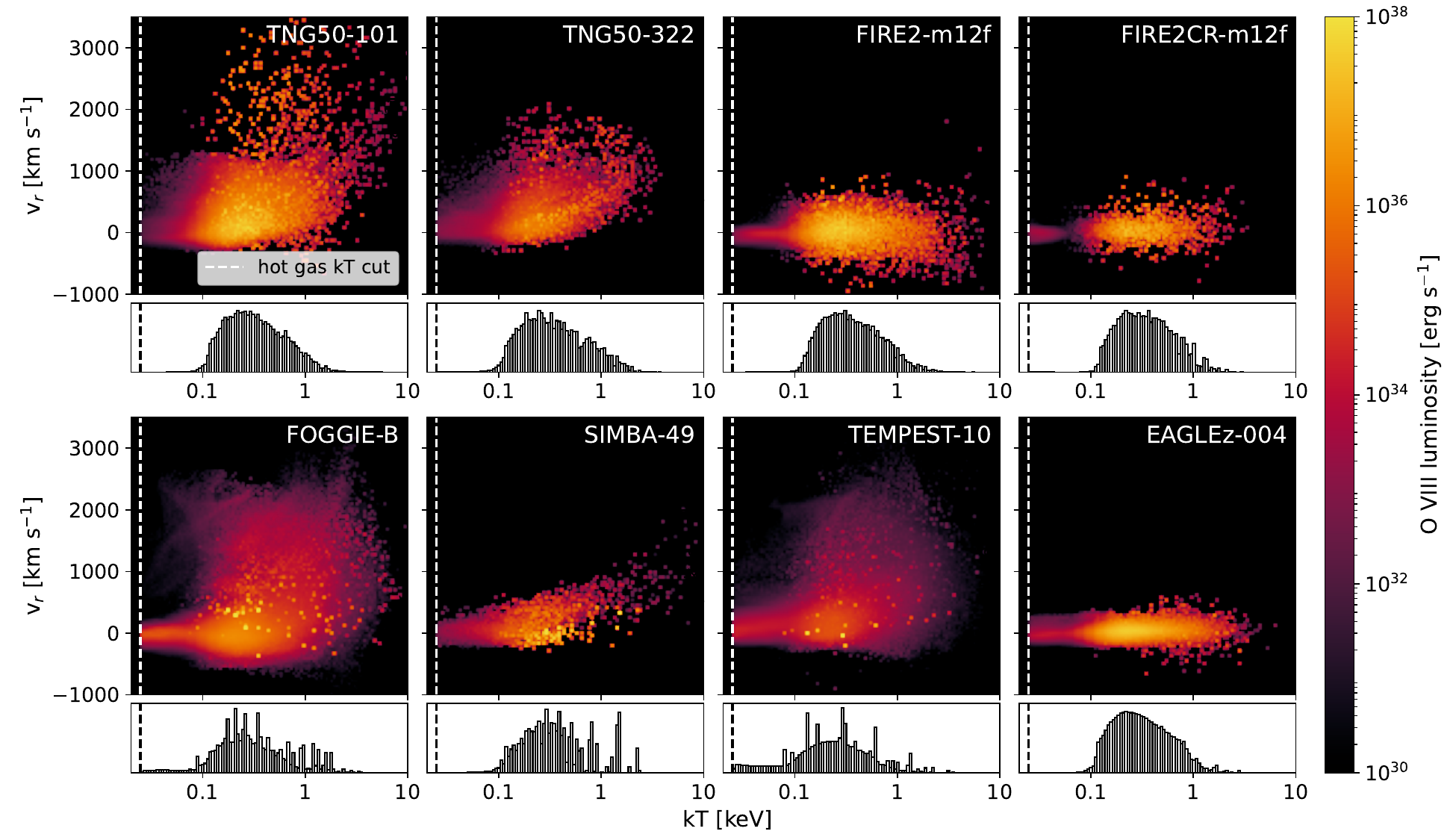}
        \caption{\textit{Top: }mass probability distributions of hot gas for one galaxy from each simulation suite as a function of radial velocity and temperature. \textit{Bottom: }O VIII luminosity distributions for the same. The histograms represent the corresponding distributions collapsed along the temperature axis. There exist high-temperature, high-velocity tails in the mass probability distributions for galaxies that exhibit strong feedback-driven biconical outflow features which are traced strongly by the O VIII luminosity.}  \label{fig:mass_phase_allgals}
\end{figure*} 

\begin{figure*}[htb!]
    \centering 
        \includegraphics[width=0.98\textwidth]{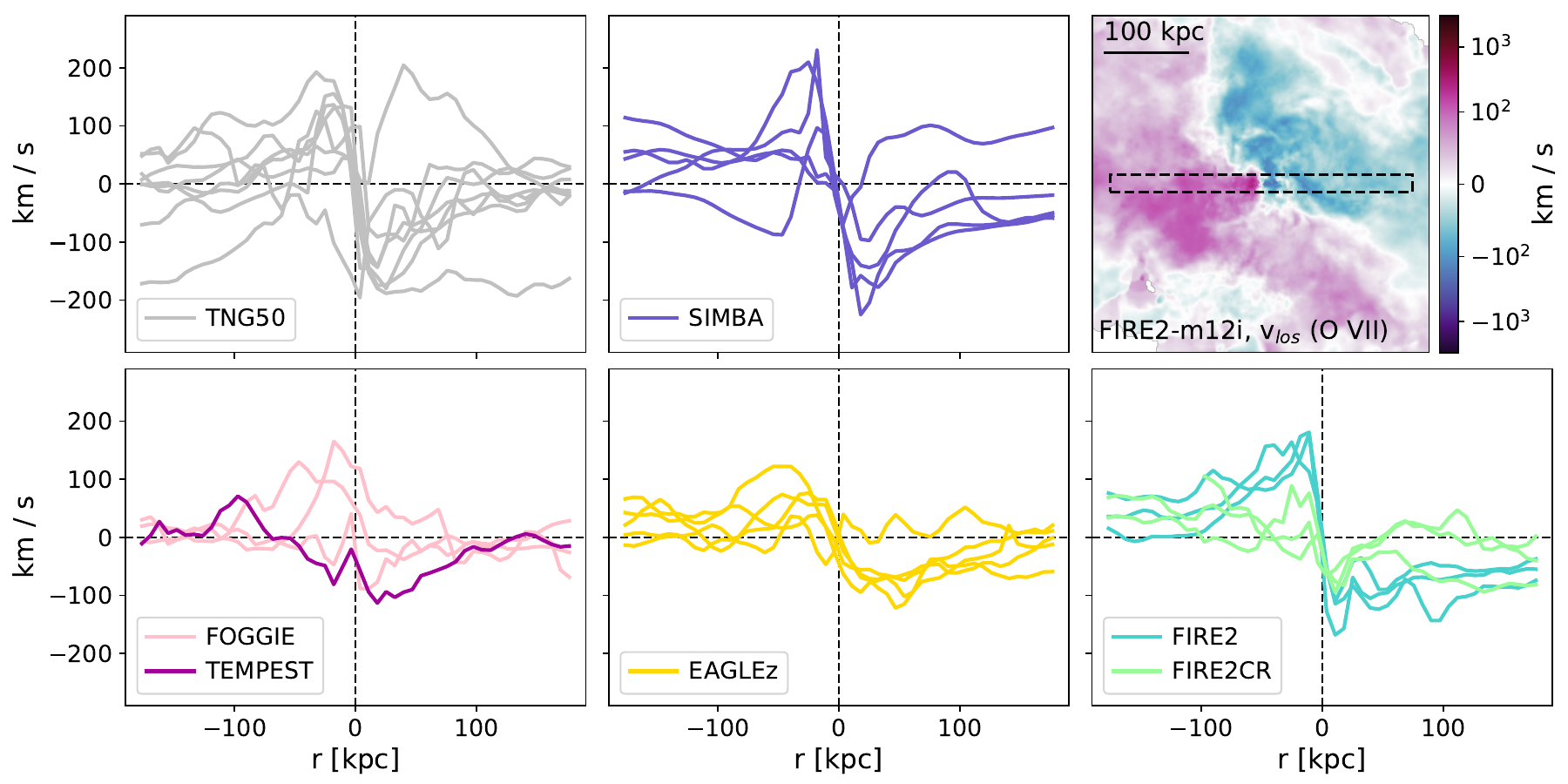}
        \caption{Radial profiles of {\vlos} weighted by the O VII line emissivity derived from edge-on projections (e.g., within the $360 \times 20$ kpc extraction region shown for the FIRE2-m12i galaxy; \textit{top right}) of all galaxies in the sample. This visualization maximizes contributions from rotational motions in the hot CGM.}  \label{fig:vlos_xsb_OVII_prof} \vspace{1em}
\end{figure*} 

The kinematic structure of the CGM can be probed observationally in the soft X-ray band via emission line shifts and broadening \citep[e.g.,][]{Schellenberger2024, ZuHone2024}. X-ray emission line shifts can probe the bulk LOS velocity structure ({\vlos}) of the hot CGM, while measurements of the broadening of these emission lines ({\sigmav}) can provide additional context into the 3D distribution of velocities along the LOS, including turbulence and oppositely directed bulk flows. 

In the simplified case of a galaxy oriented face-on having reasonably symmetric high-velocity biconical outflows driven by galactic feedback processes (e.g., AGN), the system will be observed to have reasonably small bulk velocities ({\vlos}) due to the large positive and negative LOS velocities from the outflows canceling out in projection, but high {\sigmav} across the extent of the outflows in projection \citep[see][]{ZuHone2024}. In this simple case, velocities from the disk rotation and any potentially coupled CGM rotation would not contribute significantly to {\vlos} or {\sigmav}. 

Observing the same simplified galaxy edge-on would maximize the magnitude of {\vlos} on either side of the galactic center from the rotation curve of the galactic disk and any co-rotating CGM, while also minimizing the {\vlos} structure associated with symmetric biconical outflows due to the majority of the outflow velocity being in the plane perpendicular to the LOS, with any LOS outflow velocity components once again canceling out in projection. This orientation would, however, produce regions of enhanced {\sigmav} over the cross-section of the biconical outflows. In the edge-on projection, the shape of the CGM rotation curve and inflowing gas will produce small but non-negligible {\sigmav}. Intermediate orientations (between edge- and face-on) would have contributions to both {\vlos} and {\sigmav} from the outflows and the disk/CGM rotation. Therefore, jointly leveraging the spatial distribution of both {\vlos} and {\sigmav} can provide insight into the 3D velocity structure of feedback on a galaxy-by-galaxy basis. 

Returning to the suite of simulated galaxies, in Figure \ref{fig:vlos_xsb_galsamp}, we show projections of {\vlos} for all galaxies in the sample as observed at an intermediate (midway between edge- and face-on) orientation weighted by the broadband X-ray emissivity. This intermediate orientation allows an observer to simultaneously view component contributions to the LOS velocity from both galactic outflows and the hot CGM bulk angular momentum (e.g., that which typically co-rotates with the galactic disk). The highest {\vlos} magnitudes are generally found in galaxies from the TNG50 and Simba suites, while many of the fiducial FIRE2 galaxies exhibit a large degree of co-rotation between the galactic disk and the CGM on large (r $\sim 100$ kpc) scales.  

To estimate the component contributions of, e.g., high-velocity biconical outflows from a galaxy when the disk is not oriented at an intermediate angle (i.e., the outflows are pointed either parallel or perpendicular to the observer), we can leverage {\sigmav}. In Figure \ref{fig:vlos_sigma_xsb_galsamp}, we show projections of {\sigmav} weighted by the broadband X-ray emissivity for all galaxies in the sample as observed at an edge-on orientation. Outflows above and below the disk plane are traced at high {\sigmav} for many galaxies in the Simba, TNG50, FOGGIE, and TEMPEST suites (see, e.g., SIMBA-40, TNG50-101, TEMPEST-10, or FOGGIE-B). Often, these regions of high {\sigmav} are correlated with XSB enhancements when both quantities are viewed for a galaxy oriented edge-on (compare with Figure \ref{fig:xsb_galsamp}). The velocity dispersion is generally not significantly enhanced in galaxies from the FIRE2, FIRE2CR, or EAGLE NEQ zoom simulations. 

1D radial profiles of {\sigmav} for all simulated galaxies derived from edge-on projections calculated in concentric circular annuli centered on the galaxy out to a radius of $\simeq200$ kpc are shown in Figure \ref{fig:sigmav_edge_1d}. Many of the radial profiles show {\sigmav} to be reasonably flat as a function of galactic radius, generally not varying by more than a factor of $\sim$a few from 20 to 200 kpc. The lowest velocity dispersion is predicted in the FIRE2CR galaxies for all radii examined ($20 \lesssim r \lesssim 200$ kpc). At small radii ($r \lesssim 100$ kpc), the TNG50, Simba, FOGGIE, and TEMPEST simulations predict higher {\sigmav} than any galaxy from the FIRE2, FIRE2CR, and EAGLE simulations. 

To disentangle velocity contributions from feedback-driven outflows and bulk disk/CGM (co-)rotation, we generate illustrative movies of {\vlos} and {\sigmav}. Examples of the TNG50-211 and FIRE2CR-m12m galaxies, mapped in {\vlos} and {\sigmav}, are shown in Figure \ref{fig:velocity_animation}. The viewing angle is initially set to an edge-on orientation relative to the galactic disk, and it sweeps out $360^{\circ}$ through intermediate and face-on orientations. In the case of TNG50-211, when oriented edge-on, {\vlos} traces the global CGM-disk rotation pattern, while {\sigmav} traces the outline of AGN-driven outflows to large radius. In face-on orientations, {\sigmav} is high across the extent of the outflows, and regions of high {\vlos} contributed by asymmetric outflows on either side of the galactic disk can be inferred. At intermediate orientations, contributions from both CGM rotation and outflows can contribute to {\vlos} and {\sigmav}. For FIRE2CR-m12m, the co-rotation of the hot CGM with the galactic disk traced by {\vlos} is apparent in edge-on orientations, while {\sigmav} contains very little enhancement. However, as intermediate orientations are swept out, regions of large-scale enhancement in {\sigmav} associated with outflows from the galactic disk become obvious, and their magnitude is maximized when the galaxy is viewed face-on, along with co-located enhancements in {\vlos}. We discuss this velocity structure further for the FIRE2CR galaxies in Sections \ref{subsec:cylprofs} and \ref{sec:outflows}. For any orientation, models of the 3D kinematics of a galaxy could provide a picture of how feedback processes shape the CGM on large scales. 

In addition to characterizing the kinematic structure of the high X-ray emissivity hot CGM as a whole, future X-ray observatories with $\sim$eV spectral resolution will be able to trace the velocity structure of the hot CGM through individual emission lines \citep{ZuHone2024}. In Figure \ref{fig:vlos_xsb_lines_all}, we show intermediate projections of {\vlos} weighted by the emissivity of the O VII and Fe XVII lines, respectively. The resulting maps often exhibit velocity structures with different magnitudes, and even variable morphologies (see, e.g., TNG50-322 in Figure \ref{fig:vlos_xsb_lines_all}). These soft X-ray emission lines can therefore trace different velocity phases of the hot CGM, even though the emission line energies are only separated by $\lesssim 200$ eV. This is especially true for some of the simulated galaxies that incorporate strong AGN feedback (TNG50, Simba), as well as simulated galaxies with strong signatures of stellar feedback (FOGGIE, TEMPEST). The kinematic differences between {\vlos} distributions weighted by the individual X-ray line emissivities in the FIRE2, FIRE2CR and EAGLE NEQ zoom galaxies are much smaller than for other galaxies. These halos constitute the subset of galaxies with the smallest overall {\sigmav}.

In galaxies with high levels of variation between {\vlos} maps traced by the X-ray emission lines, higher velocity gas is generally traced more clearly by the higher energy emission line(s), Fe XVII (and O VIII). In Figure \ref{fig:xline_phases}, we show phase diagrams of the emission line luminosities as a function of $v_r$ and temperature for one of the TNG50 galaxies (TNG50-31). The luminosity of the highest energy emission line (Fe XVII) peaks at higher temperatures than the luminosity of the lower energy emission lines. In this galaxy, the hot gas near the Fe XVII luminosity peak is associated with higher radial velocities (i.e., gas in outflows). In general, these observable signatures of outflows are evident in the shapes of hot gas mass and emission line (e.g., O VIII) luminosity phase plots as a function of $v_r$ and temperature (Figure \ref{fig:mass_phase_allgals}). Simulations that do not generally produce galaxies with large-scale high-velocity galactic outflows have reasonably flat $v_r$ distributions as a function of temperature for the hot gas mass and O VIII luminosity (e.g., the FIRE2-m12f, FIRE2CR-m12f, and EAGLEz-004 galaxies in Figure \ref{fig:mass_phase_allgals}). Simulations that predict galaxies with high-velocity galactic outflows generally display distributions that sweep up towards higher radial velocities for higher temperatures (see the TNG50-101, TNG50-322, SIMBA-49, FOGGIE-B, and TEMPEST-10 galaxies in Figure \ref{fig:mass_phase_allgals}, and as further discussed in Section \ref{subsec:kT}).  

Individual X-ray emission lines can also be used as tracers of the radial profile of the hot CGM in the disk plane to maximize contributions from the bulk (co-)rotation with the galactic disk. We provide an example of such an application in Figure \ref{fig:vlos_xsb_OVII_prof}. Here, we generate radial profiles of {\vlos} weighted by the O VII line emissivity for each galaxy in the sample from edge-on projections of the corresponding quantity in radial bins out to $180$ kpc in the galactic disk plane and up to $10$ kpc in height from the disk plane. Most of the galaxies exhibit a clear rotation curve structure out to large radii, though galaxies from FOGGIE and TEMPEST produce the least visually identifiable rotation curve structures. At near all radii, the highest magnitude profiles come from galaxies simulated in TNG50 or Simba, though many of these profiles indicate a clear rotation curve structure as well. Several galaxies (largely from TNG50 and Simba) exhibit an apparent counter-rotation of the hot gas in the galactic disk and the large-scale CGM in projection, which is likely associated with the kinematics of the inner regions being dominated by disk rotation and the outer regions being dominated by the kinematics of inflows, outflows, and environment (e.g., the dynamics of nearby satellites). 

\subsection{Cylindrical velocity profiles} \label{subsec:cylprofs}
\begin{figure}[thb!] 
    \centering 
        \includegraphics[width=0.43\textwidth]{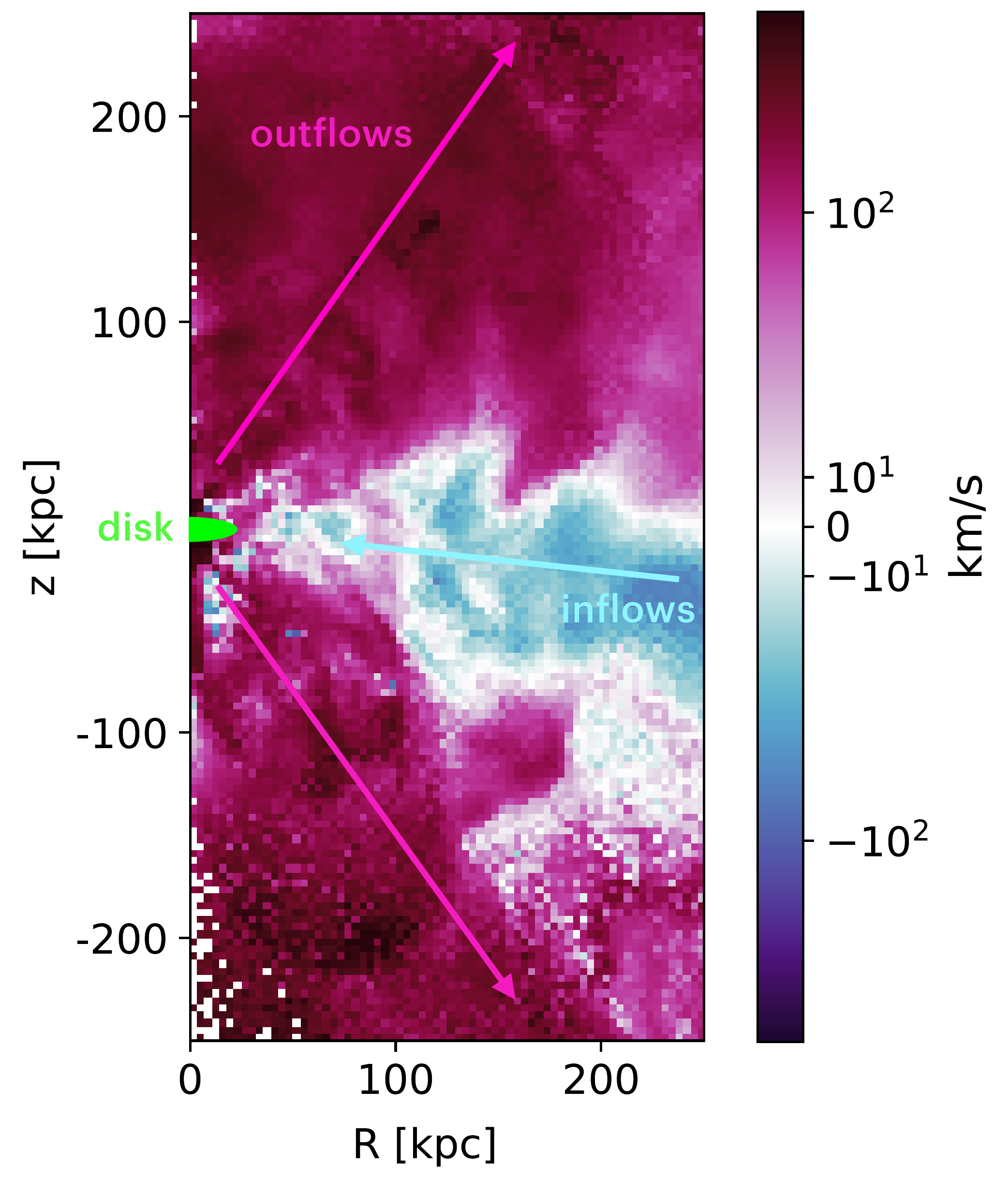}
        \caption{Example cylindrically averaged radial velocity profile weighted by the O VII X-ray line emissivity for TNG50-101. Centered on the galactic disk (green) in an edge-on orientation, positive velocities (pink) represent outward radial flows, and negative velocities (blue) represent inward radial flows.} \label{fig:cyl_profile_example}
\end{figure} 

\begin{figure*}[htb!] 
    \centering 
        \includegraphics[width=1\textwidth]{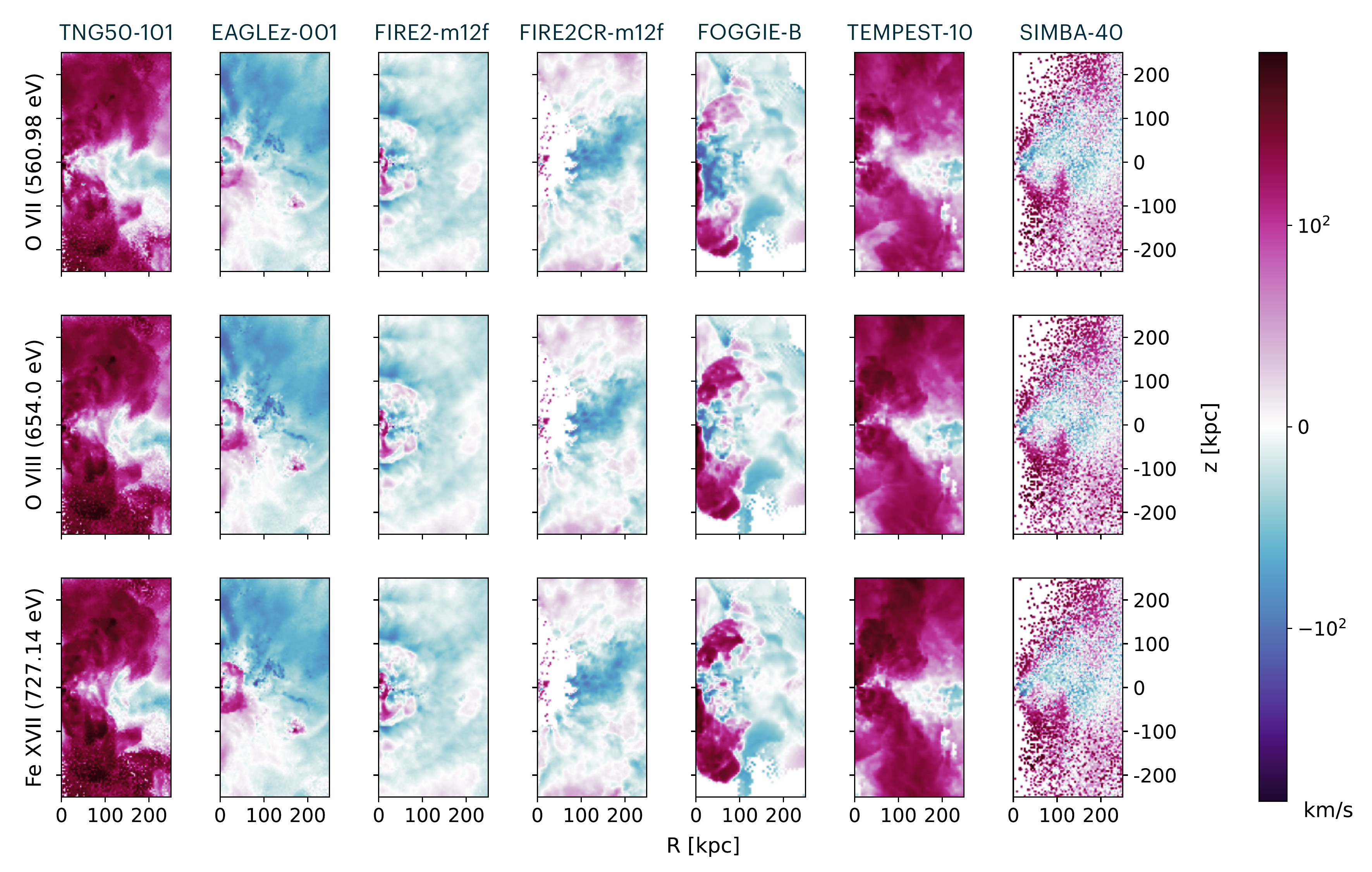}
        \caption{Cylindrically averaged radial velocity profiles weighted by X-ray line emissivities (from top to bottom: O VII, O VIII, Fe XVII) for one galaxy from each simulation suite (from left to right: TNG, EAGLE, FIRE, FIRE2CR, FOGGIE, TEMPEST, SIMBA). Higher outflow velocity magnitudes are traced by higher energy emission lines (e.g., Fe XVII). High-velocity biconical outflows are clearly illustrated for the galaxies from TNG50 and Simba (which incorporate AGN feedback), as well as from TEMPEST and FOGGIE (stellar feedback only). The EAGLE zoom and FIRE galaxies show evidence of smaller-scale outflows above and below the disk. A biconically-oriented outflow is visible in the FIRE2CR galaxy at a somewhat low magnitude due to the relative lack of a resolved hot CGM phase in FIRE2CR runs.} \label{fig:cyl_profs}
\end{figure*} 

In order to clearly isolate inflows from outflows in systems, specifically those that exhibit biconical stellar/AGN-driven outflow geometries, we construct angle-averaged profiles in cylindrical radius $R$ and height $z$ (where the cylinder is aligned with the angular momentum vector of the galaxy's disk) of the spherical radial velocity ($v_r$) for the hot gas, weighted by each X-ray emission line. In Figure \ref{fig:cyl_profile_example}, we provide an illustrative example of such a cylindrical profile traced by the O VII emission line, distinguishing high-velocity outflows above and below the galactic disk plane (positive $v_r$) from inflows concentrated into the disk plane (negative $v_r$), for TNG50-101. Corresponding examples of these cylindrical velocity profiles are shown for one galaxy from each simulation suite in Figure \ref{fig:cyl_profs}. As expected, high-velocity biconical outflows are clearly visible in the galaxies from TNG50 \citep[in agreement with][]{ZuHone2024} and Simba, which incorporate strong AGN feedback. The EAGLE zoom and FIRE galaxy show some evidence of smaller-scale outflows on either side of the disk, though at much lower magnitude than in the aforementioned galaxies. The speed of the outflows are generally higher when traced by the higher energy emission lines relative to lower energy emission lines (compare e.g., the outflow above the FOGGIE-B disk as traced by O VII and Fe XVII in Figure \ref{fig:cyl_profs}), consistent with results from the {\vlos} projections weighted by the X-ray emission lines. 

Biconically-oriented outflows can also manifest in the FIRE2CR m12 galaxies run with CR feedback. This can be seen in Figure \ref{fig:cyl_profs}, where the outflows from FIRE2-m12f are visible only at small distances from the galactic disk, and appear to have a more concentric morphology at large radii. In contrast, FIRE2CR-m12f exhibits fairly slow velocity outflows above and below the disk plane out to large radii, with material inflowing near the disk plane. This behavior is apparent in the mock observables {\vlos} and {\sigmav} for some orientations, in general agreement with the outflow velocity structure predicted for these FIRE2CR galaxies incorporating cosmic ray physics \citep[see][especially their Figure 2]{Hopkins2021}, and qualitatively similar to cosmic-ray driven galactic outflow predictions in other codes \citep[e.g.,][]{Butsky2018,Salem2014}. 

Interestingly, the FOGGIE-B and TEMPEST-10 galaxies also exhibit biconical velocity structures in their high-magnitude outflows. Specifically, TEMPEST-10 produces velocity structures which are nearly indistinguishable from those predicted by some of the TNG50 galaxies (Figure \ref{fig:cyl_profs}), despite the fact that TNG50 incorporates a strong, dual-mode AGN feedback prescription, and TEMPEST-10 (and FOGGIE) incorporate only single-mode models for stellar feedback. We discuss these features in Section \ref{sec:outflows}. 

\subsection{Temperature} \label{subsec:kT}
\begin{figure*}[htb!]
    \centering 
        \includegraphics[width=1\textwidth]{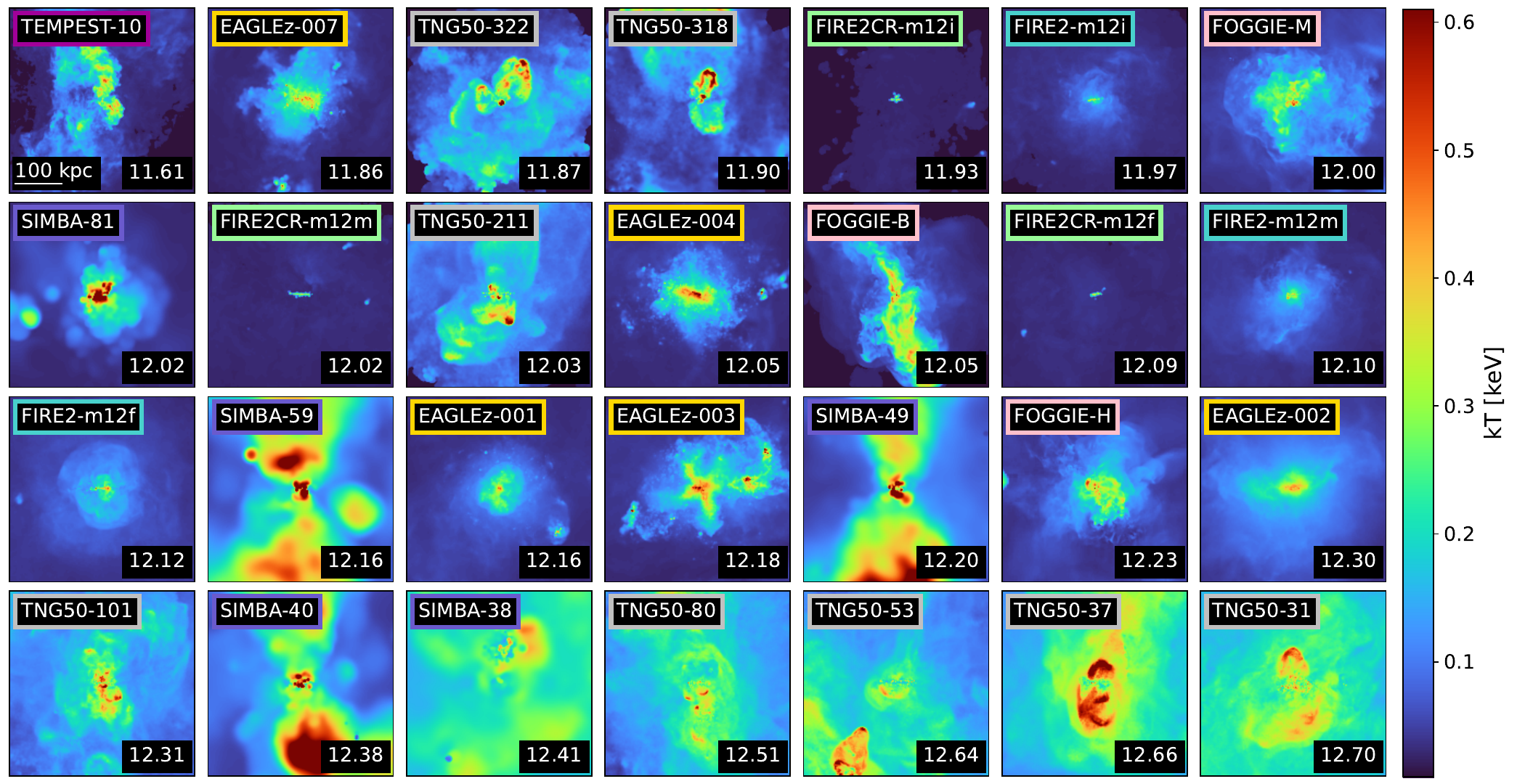}
        \caption{Edge-on projections of kT weighted by the broadband X-ray emissivity for all galaxies in the sample sorted from lowest to highest halo mass (with $\log(M\;[M_{\odot}])$ indicated in each bottom right corner). The temperature structure of hot CGM in galaxies varies significantly as a function of galaxy mass, simulation, and environment.}  \label{fig:kT_xsb_galsamp}
\end{figure*} 

\begin{figure}[thb]
    \begin{interactive}{animation}{hotCGM_prez4_4.mov}
    \includegraphics[width=0.43\textwidth]{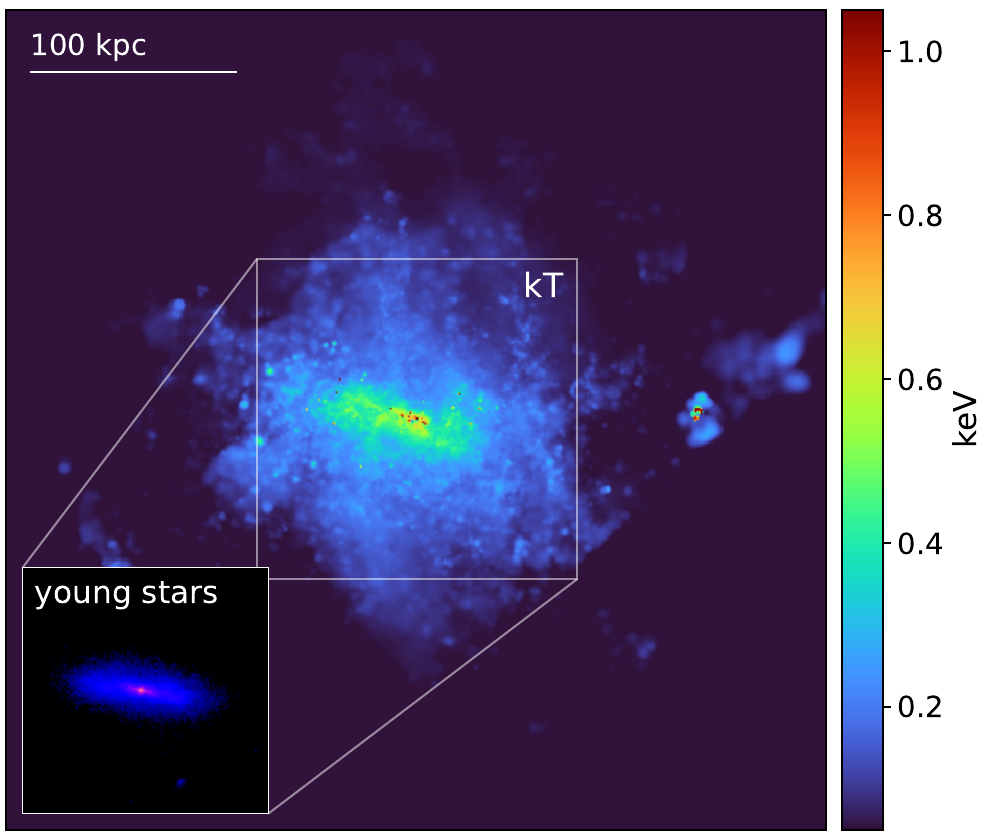}
    \end{interactive}
    \caption{Example animation of kT weighted by the broadband X-ray emissivity and the young stellar disk mass density as viewed by observers at locations sweeping out $360^{\circ}$ in viewing angle from an edge-on orientation with the galaxy (EAGLEz-004). The hot CGM temperature morphology is strongly correlated to the underlying galactic feedback prescription.} \label{fig:kT_animation}
\end{figure}

\begin{figure}[thb] 
    \centering 
        \includegraphics[width=0.47\textwidth]{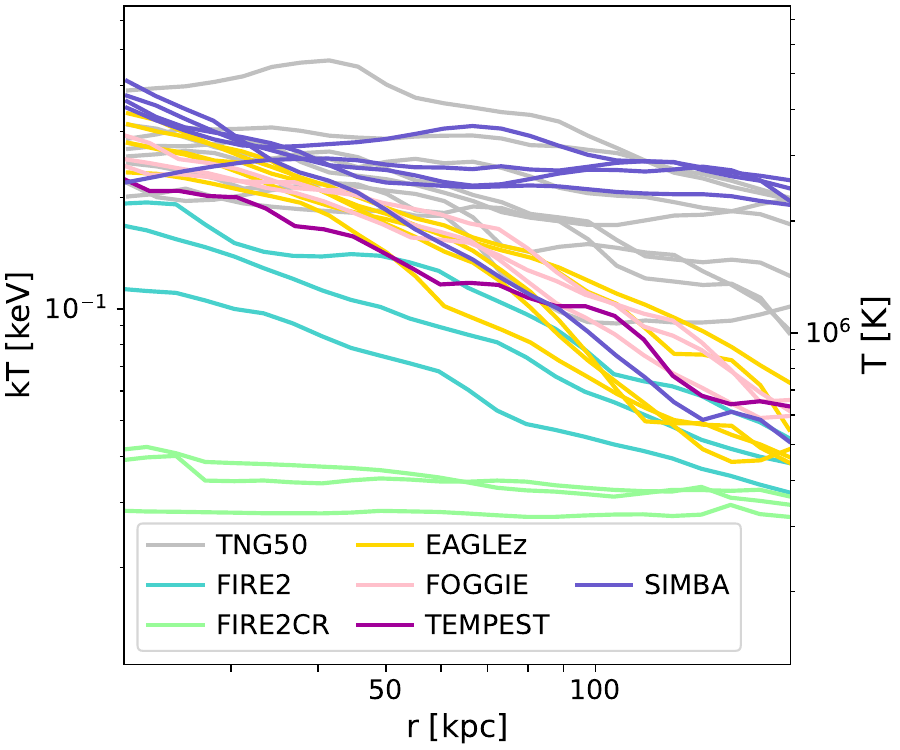}
        \caption{Radial profiles calculated from projections of kT weighted by the broadband X-ray emissivity for all galaxies in the sample oriented edge-on. Profiles are calculated in concentric circular annuli centered on the galaxy out to a radius of $\simeq200$ kpc. The FIRE2CR simulations predict the lowest temperatures at all radii.} \label{fig:kT_edge_1d}
\end{figure} 

In Figure \ref{fig:kT_xsb_galsamp}, we show maps of the gas temperature weighted by the broadband XSB for all simulated galaxies in the sample as viewed by an observer oriented edge-on relative to the galactic disks. The highest temperature regions on large scales in the CGM ($kT \gtrsim 0.6$ keV) occur in the AGN-driven outflows and bubbles of the Simba and TNG50 galaxies (e.g., SIMBA-40 or TNG50-37 in Figure \ref{fig:kT_xsb_galsamp}). The FIRE2CR galaxies exhibit very little hot ($kT \gtrsim 0.1$ keV) CGM out to $\sim$ 200 kpc. While the volume-filling gas is, on average, hotter for higher halo masses, many of the smaller halos can still exhibit feedback-driven hot outflows into the, on average, cooler CGM out to large scales. The TEMPEST-10 galaxy is an example of this: though it is the least massive halo ($M_{200c} \simeq 4.1 \times 10^{11} M_{\odot}$) in our sample, it exhibits hot ($0.2 \lesssim kT \lesssim 0.6$ keV) outflows into the cooler ($kT \lesssim 0.1$ keV) CGM on scales that can extend beyond $\gtrsim 100$ kpc.

The EAGLE NEQ zoom galaxies generally have more centrally concentrated hot gas relative to the galaxies from TNG50 and Simba, which also incorporate AGN feedback physics, albeit with different energy injection prescriptions (see Section \ref{sec:outflows} for a further discussion of these differences and their effects on the physical state of the hot CGM). In Figure \ref{fig:kT_animation}, we provide an animation of the temperature weighted by the X-ray emissivity for EAGLEz-004, with the young stellar disk indicated for orientation reference. In contrast to many other galaxies simulated with AGN feedback physics (in TNG50 and Simba), there is a general absence of hot gas strongly outflowing in biconical orientation or at the edges of feedback-driven bubbles. Few of the galaxies selected from EAGLE NEQ zooms show strong signatures of hot AGN-driven outflows from the galactic disk that extend to large galactic radii, which is unsurprising, given the lack of kinetic energy deposition in the EAGLE NEQ zooms AGN feedback model. Accordingly, \citet{Truong2021} note that galaxies simulated with TNG100 predict temperature enhancements above and below the galactic disk that peak at MW-like halo masses, while EAGLE (Ref-L0100N1504) galaxies only predict higher levels of temperature anisotropy (enhanced along the same axis) for halo masses well below the MW-like masses studied here.

This is further apparent in 1D radial profiles of temperature weighted by the broadband X-ray emissivity and derived from edge-on projections calculated in concentric circular annuli centered on the galaxy out to a radius of $\simeq200$ kpc. (Figure \ref{fig:kT_edge_1d}). The hot CGM temperature for halos simulated in EAGLE zooms fall off much more rapidly than those simulated with TNG50 and Simba. The temperature of the hot gas in the FIRE2CR galaxies remains roughly flat near the lower limit of the hot gas temperature range ($\sim3 \times 10^5$ K) at all galactic radii examined ($20 \lesssim r \lesssim 200$ kpc). This is consistent with analyses of m12 galaxies simulated in FIRE2 and FIRE2CR \citep{chan2022, Ji2020}, where the addition of cosmic ray physics can result in a reduction of hot CGM gas in favor of lower temperature volume-filling gas. These cooler CGM temperatures are a consistent qualitative prediction for MW-like galaxies across many simulations incorporating cosmic ray feedback prescriptions \citep[e.g.,][]{Buck2020,Butsky2018,Salem2016}. Of the remaining simulation suites, at large radius ($r \gtrsim 50$ kpc), the TNG50 and Simba galaxies generally have highest temperatures and shallowest temperature profiles. Radial profiles from the Simba galaxies generally (but notably excepting the lowest mass Simba galaxy, SIMBA-81) remain nearly flat at large radii due to the hot gas in outflows at large radius competing with the falling temperature in the disk plane and outer CGM. 

At smaller radii near the disk-halo interface ($20 \lesssim r \lesssim 30$ kpc), the temperatures of the TNG50, Simba, FOGGIE, TEMPEST, and EAGLE galaxies in projection are similar ($0.2 \lesssim kT \lesssim 0.5$ keV), while the fiducial FIRE2 galaxies are cooler ($0.1 \lesssim kT \lesssim 0.2$ keV). Physically, \citet{chan2022} note that for a sample of fiducial FIRE2 m12 galaxies, hot gas originating from within a high-velocity outflow near a galactic disk will cool and fall back onto the disk, becoming trapped in a small-scale `galactic fountain'. Ultimately, the hot gas is then recycled within $\sim 100$ Myr, rather than reaching the CGM on large scales. This, combined with the lack of hot AGN-driven outflows in the fiducial FIRE2 model, is in agreement with the cooler temperatures seen at smaller radii in Figure \ref{fig:kT_edge_1d}.

\section{Discussion: feedback-driven biconical outflows} \label{sec:outflows}
High-velocity biconical outflows perpendicular to the disk plane that are enhanced in both temperature and XSB are predicted in the X-ray observables for most galaxies we studied within the TNG50 and Simba simulations, as well as for some galaxies within the TEMPEST and FOGGIE simulations. In addition to the X-ray observables, this behavior is apparent in temperature vs. $v_r$ phase space. In Figure \ref{fig:mass_phase_allgals}, we showed the mass and O VIII luminosity profiles in this space for a galaxy from each subsample. Galaxy simulation suites that produce these large-scale biconical outflows in the CGM generally show tails in the distributions extending to high velocity/temperature, while those lacking these outflows are much narrower in velocity space and more symmetric around zero velocity. While present in both profiles, this effect is most obvious when weighted by the O VIII luminosity. Though these outflow structures have been explicitly linked to the kinetic AGN feedback modes in TNG50 \citep{Pillepich2021} and Simba \citep{SIMBA}, the FOGGIE and TEMPEST simulation suites do not model AGN activity, but stellar feedback exclusively.

To investigate a link between the large, X-ray bright outflows and the stellar feedback prescription, we first examined galaxies simulated with exclusively stellar feedback (non-CR). These are the FIRE2, FOGGIE, and TEMPEST galaxies. We visually verified that high velocity, high temperature outflows extending out to large (r $\gtrsim 100$ kpc) radii are not a fundamentally ubiquitous prediction of stellar feedback across all of these simulations by examining projections of the gas temperature at earlier epochs. The FOGGIE and TEMPEST-10 simulations (run with Enzo solely including thermal supernova feedback) do predict stochastic hot, large scale outflows as the galaxy evolves over cosmic time. FOGGIE-B exhibits these outflows at $z=0$, and the presence of such outflows is common at low redshifts in the other two galaxies as well. This behavior is dependent on e.g., the geometry and merger/accretion history of the galaxies. However, we verified that for one of the FIRE2 galaxies, FIRE2-m12i, similar large scale, high velocity outflows with temperatures $\gtrsim10^7$~K up to $z\sim 0$ are not produced. Instead, this galaxy exhibits stochastic galactic outflows over cosmic time that are cooler on average than the FOGGIE and TEMPEST-10 galaxies, do not extend out to large radii, and are not generally oriented biconically relative to the disk center. This is in agreement with studies of characteristic galactic fountains in the fiducial FIRE2 galaxies, where outflows are typically confined to small distances from the galactic disk \citet{chan2022} and with observations that $\sim$MW mass fiducial FIRE2 galaxies at low redshift tend to have more continuous SFR once the inner CGM begins to virialize \citep{Muratov2015,Stern2021}. 

\begin{figure}[thb]
    \centering 
        \includegraphics[width=0.45\textwidth]{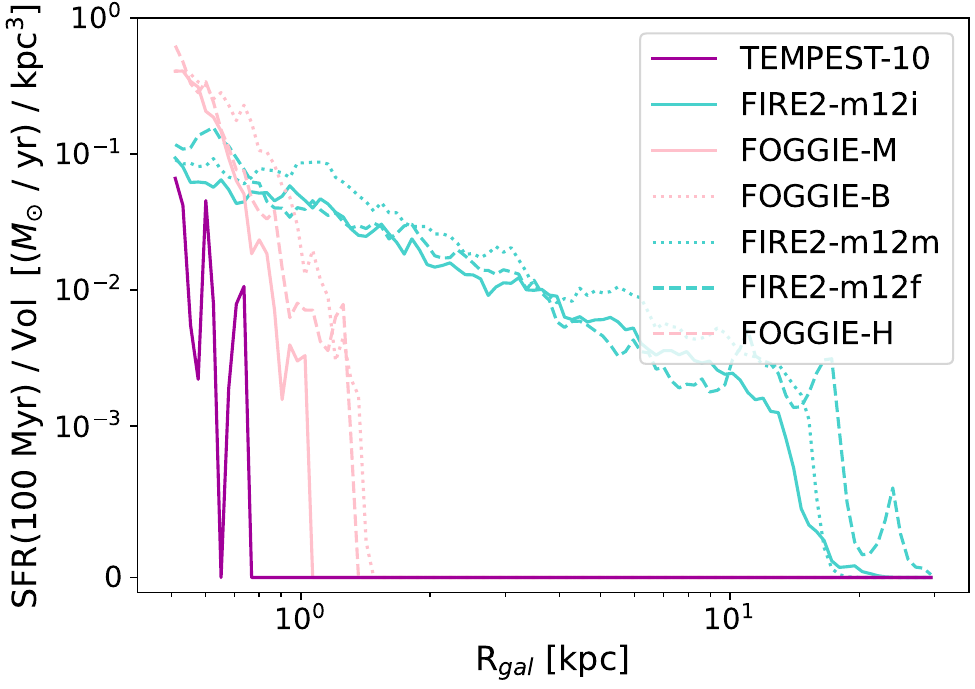}
        \caption{Comparisons of the ``instantaneous'' SFR per volume element as a function of galactic disk radius for galaxies simulated with only stellar (non-CR) feedback, sorted from lowest to highest mass ($M_{200}$). Galaxies from the the FOGGIE and TEMPEST suites, which are simulated in Enzo with simple thermal stellar feedback schemes, exhibit sharply peaked SFR in the central couple kpc of the disk center. The fiducial FIRE2 galaxies, which are simulated in Gizmo with a much more complex stellar feedback model, have extended distributions of SFR out to $\sim$30 kpc.} \label{fig:sfr_stellarsims}
\end{figure}

A key metric for identifying a relationship between large-scale galactic outflows and the underlying feedback physics is the energy injected over a period of time. While the SFR is not itself a measurement of the energy injected by stellar feedback, it is a directly related quantity that we choose to use as a proxy, given this direct relationship and the ease with which we can uniformly estimate it across the three simulations despite differences in energy, mass, and momentum transport into the surrounding medium as stars form and evolve. We therefore estimate the ``instantaneous'' SFR per volume element in the $z=0$ snapshots uniformly across the sample by averaging over 100 Myr the mass of stars with ages $< 100$ Myr in radial bins out to 30 kpc in the galactic disk plane and up to 2 kpc in height from the disk plane (Figure \ref{fig:sfr_stellarsims}). 

Of the galaxies simulated with only stellar feedback physics (non-CR), we find that the distribution of SFR in the galactic disks at $z=0$ is much more centrally peaked for the galaxies simulated in Enzo (FOGGIE/TEMPEST-10), compared to being more radially extended in the fiducial FIRE2 m12 galaxies \citep[in broad agreement with]{Orr2020}. Thus, the distribution of energy injection via stellar feedback being centrally concentrated in the FOGGIE and TEMPEST-10 galaxies is similar to the distribution of energy injection via a central AGN source in other simulation suites. We note that the SFR estimated from older populations of stars (i.e., that associated with earlier epochs) can much more distributed in the disk relative to the instantaneous SFR at $z=0$, though these are still very centrally-peaked. 

Another consideration is the mechanism by which energy and momentum from stellar feedback are imparted into the surrounding medium. For example, the FOGGIE and TEMPEST simulations rely on a thermal energy injection, while the FIRE simulations incorporate additional kinetic components for the transfer of momentum. Analogously, of the simulations that incorporate AGN feedback, the EAGLE NEQ zooms model AGN energy deposition thermally, while TNG50 and Simba include thermal and kinetic energy deposition prescriptions. Of these, the EAGLE NEQ zooms are among the most luminous in the X-ray (see Figs. \ref{fig:xsb_edge_1d} and \ref{fig:xsb_galsamp}), while exhibiting some of the least active signatures of outflows (see Figs. \ref{fig:vlos_sigma_xsb_galsamp}, \ref{fig:sigmav_edge_1d}, and \ref{fig:cyl_profs}). In comparison to the simulations with more active feedback, EAGLE has lower AGN feedback efficiency at late times, which is related to EAGLE's single-mode AGN feedback scheme in contrast to the TNG kinetic-mode feedback that is more prevalent at late times and more massive SMBH \citep{Davies2020}. For the five EAGLE NEQ M4.4 halos we explore here, we see denser hot CGMs that are likely attributable to more inefficient thermal feedback schemes related to SMBH growth. The bipolar outflow lobes seen in EAGLEz-001 (Figure \ref{fig:xsb_lines_maps_profiles}) are an example of active outflows with moderate velocities owing to AGN feedback, though these do not typically reach as large of radii as the outflows launched by TNG50 or Simba (see Fig. \ref{fig:cyl_profs}), nor are they a ubiquitous prediction in the EAGLE NEQ zooms.

Therefore, while it is likely that the form of energy and momentum injection from stellar feedback influences the outflow morphology, it is unlikely that, for the simulations that model exclusively stellar feedback, this is the dominant cause of the extreme biconical outflow differences. The EAGLE NEQ zooms, which lack a kinetic AGN feedback component relative to TNG50 and Simba, produce milder outflows to large galactic radii, while the FOGGIE and TEMPEST-10 simulations, which analogously lack a kinetic stellar feedback component relative to FIRE2, produce much stronger large-scale galactic outflows. We instead speculate that the large, centrally concentrated SFR in the FOGGIE and TEMPEST-10 simulations is a primary driver of the stochastic large scale, hot biconical galactic outflows in these simulations. However, we discuss other physical and numerical considerations that may affect the appearance of hot, fast outflows below. 

\subsection{The effect of spatial resolution} \label{sec:CR_sims}
Another distinguishing factor of the FOGGIE and TEMPEST-10 simulations is the extremely high spatial/mass resolution in the CGM compared to the other simulations in our sample. For instance, the median mass resolution of a hot, X-ray emitting gas cell in TEMPEST-10 is nearly 250 times finer than in the fiducial FIRE2 galaxies, which have the next-highest resolution. Consequently, FOGGIE and TEMPEST are expected to better account for extremes in the thermodynamic properties of the hot CGM gas (e.g., density and temperature peaks) than coarser-resolution simulations. Since the TEMPEST simulations were run for several different resolutions, we performed a basic check of the effects of resolution on the presence of hot, large scale biconical outflows by verifying the presence of such outflows in the lower-resolution TEMPEST-8 simulations.  

The TEMPEST-8 and TEMPEST-10 simulations are identical, except that the forced spatial resolution within the virial radius in TEMPEST-8 is $\sim4 \times$ lower than TEMPEST-10 (2 kpc compared to 0.5 kpc), making the median hot gas cell mass resolution $\sim 100 \times$ lower than in TEMPEST-10. This is $\sim20 \times$ lower than the FOGGIE median hot gas cell mass resolution, and within a factor of a few of the FIRE2 median hot gas cell mass resolution. We visually examined the output mocks for TEMPEST-8, created with the same methods as those in the previous sections for TEMPEST-10, and found that, despite the lower resolution, the presence of hot, large scale biconical outflows in the CGM was also predicted for this simulation, with some unimportant morphological differences. Therefore, it is likely that the presence of large-scale biconical outflows are not predominantly a simulation resolution effect, and instead are more likely an effect of the centrally-concentrated star formation and/or form of stellar feedback energy and momentum deposition. 

\subsection{The effect of incorporating cosmic ray physics} \label{sec:CR_sims}
Though temperatures, velocities, and XSB of the CGM in FIRE2CR galaxies are on average much lower than those of other simulation suites, there does exist some evidence in the cylindrically averaged $v_r$ profiles of biconical outflows above and below the disks out to large radii (see Figure \ref{fig:cyl_profs}). \citet{Hopkins2021} noted that for a sample of FIRE2 m12 galaxies run with and without cosmic ray physics (including the three FIRE2CR galaxies studied here), the addition of cosmic rays in the feedback scheme qualitatively shifted the kinematic structure of the CGM out to to $\gtrsim$Mpc scales from a quasi-spherical turbulence/shock distribution to a biconically oriented outflow/inflow geometry. However, we note that X-ray observations of these outflows would be difficult with future observatories, given the very low density, temperature, and by extension, XSB of the CGM in these cosmic-ray dominated halos (see e.g., \citet{Hopkins2021, Ji2020}; Figures \ref{fig:xsb_edge_1d}, \ref{fig:velocity_animation}, \ref{fig:kT_edge_1d})

\subsection{Caveats} \label{sec:resolution_sims}
A key difference between the galaxies run with solely stellar feedback (non-CR) is that the FOGGIE and TEMPEST-10 simulations are run with the Enzo code, while FIRE2 is run with Gizmo. It is therefore not implausible that some physical manifestation of differences in the numerical implementation of each code could arise, and a true code comparison \citep[in the spirit of, e.g.,][]{agora2021} in future work would be useful. 

It is also worth noting that while \citet{Pillepich2021} did link the vast majority of bubbles/outflows in TNG50 $\sim$MW-mass halos to AGN activity, they did identify some examples of bubbles with complex and non-negligible star formation in the central regions of the host galaxies. Therefore, it is not implausible that the stellar feedback scheme in TNG50 could contribute to the large-scale outflows in some cases. 

\section{Summary and conclusions} \label{sec:summary}
We have analyzed a sample of 28 $\sim$MW-mass halos at $z=0$ from seven cosmological hydrodynamical simulation suites. These simulations are run with a wide variety of galactic feedback physics prescriptions. Our main findings are summarized below. \vspace{0.5em}

The predicted broadband XSB morphology differs drastically between galaxy simulation suites.
\begin{itemize}
    \item  XSB enhancements associated with large-scale outflow and bubble features in the hot CGM are ubiquitous for galaxies simulated with TNG50 and Simba, and often predicted for galaxies in the FOGGIE and TEMPEST simulations. While TNG50 and Simba incorporate both stellar and AGN feedback prescriptions, FOGGIE and TEMPEST include exclusively stellar feedback. 
    \item The FIRE2 and EAGLE NEQ zoom galaxies produce more spherical distributions of hot CGM, despite their different feedback physics (EAGLE with stellar and AGN feedback, and FIRE2 with solely stellar feedback). 
    \item All galaxies simulated without cosmic ray physics exhibit azimuthally averaged radial XSB profiles with generally similar shapes ($\propto r^{-3}$ over the range of $20-200$ kpc), despite the wide range of halo masses, hot CGM spatial/mass resolutions, feedback schemes, and projected XSB morphologies spanned by these simulated galaxies. Scatter about this slope between simulations is likely due to the underlying feedback prescriptions. 
    \item In contrast to the above, FIRE2CR galaxies simulated with cosmic ray physics contain very little hot CGM contributing thermally to the broadband XSB, though additional X-ray emission from cosmic ray inverse-Compton scattering processes (neglected here) could be important. 
    \item The XSB of individual X-ray emission lines can probe different phases of the hot CGM. Lower energy emission lines like O VII probe more volume-filling hot CGM, while higher energy lines like Fe XVII can trace hot, feedback-driven outflows. 
\end{itemize}

 The simulation suites make very different predictions for the hot CGM kinematics.
\begin{itemize}
    \item When weighted by the X-ray emissivity, maps of {\vlos} and {\sigmav} have the power to disentangle bulk hot CGM (co-)rotation relative to the galactic disk, as well as the high-velocity outflows associated with galactic feedback processes. 
    \item High-velocity biconically-oriented outflows extending to large galactic radius are commonly predicted in some simulations with AGN feedback physics (TNG50 and Simba), though the precise mechanism of energy injection by AGN feedback must play a key role in the manifestation of this phenomenon, as they are not frequently seen in the EAGLE NEQ zoom galaxies (having a thermal, but lacking a kinetic, energy injection by AGN). 
    \item These high-velocity outflows can be seen in some galaxies simulated with exclusively stellar feedback physics (FOGGIE and TEMPEST). The radial distribution of SFR (and therefore energy injection) in these galaxies is highly centrally concentrated and reminiscent of an AGN-like energy injection geometry, in contrast to the more radially extended SFR in the fiducial FIRE2 galaxies, which do not show these large-scale galactic outflows, despite the additional kinetic energy injection in these simulations. 
    \item Though the overall XSB of these features is predicted to be low in the FIRE2CR simulations, these galaxies additionally predict similarly-oriented large-scale galactic outflows due to cosmic ray streaming from the galactic disk.
\end{itemize}

Radial profiles of {\sigmav} can provide further insight into the underlying velocity structure of the hot CGM out to large radii. 
\begin{itemize}
    \item Galaxies hosting strong outflows driven by AGN or stellar feedback having higher overall {\sigmav}, especially at smaller radii. 
    \item Individual X-ray emission lines can distinguish the kinematics of different phases of the hot CGM, with higher energy lines (e.g., Fe XVII) often tracing more extreme velocities associated with these feedback-driven outflows. 
\end{itemize}

The temperature of the hot CGM is highly dependent on the nature of underlying feedback physics, and maps of the hot CGM temperature morphology can thus provide a strong constraint on the feedback processes ongoing in galaxies. 
\begin{itemize}
    \item Galaxies simulated with AGN feedback tend to exhibit strong temperature enhancements in large-scale outflows and bubbles of material. This is best illustrated in the TNG50 and Simba simulations, though galaxies simulated in FOGGIE and TEMPEST with a simple prescription for stellar feedback can also produce large-scale outflows with temperature enhancements. 
    \item The fiducial FIRE2 galaxies, with a prescription for stellar feedback exclusively, tend to have cooler CGM on average out to several hundred kpc than simulations with AGN feedback.
    \item Galaxies simulated in FIRE2CR with additional cosmic ray physics have the coolest overall CGM out to large radius (which is also apparent in the lack of strong X-ray emission), consistent with predictions for the cosmic ray-dominated CGM in simulations of MW-like galaxies run with other codes and cosmic ray transport prescriptions. 
    \item The EAGLE NEQ zoom galaxies have temperature distributions that are hot in the center but fall off more rapidly with galactic radius than the other simulations with AGN feedback (TNG50 and Simba). 
\end{itemize}

This combination of XSB, velocity, and temperature observables provides a powerful means to characterize the 3-dimensional geometry of the hot CGM in external galaxies. Characterizing the morphological and kinematic state of the hot CGM in individual galaxies across a population may be vital to distinguishing between different models of galactic feedback physics, most of which are not well constrained by existing observations. Many of the aforementioned X-ray probes, for instance measurements of {\vlos}, are dependent on resolving and disentangling hot CGM X-ray emission lines of external galaxies from those of the Milky Way foreground via high-resolution X-ray imaging spectroscopy. The resolution required to make these measurements is beyond the capabilities of modern CCD instruments, and therefore necessitates a mission with a high spectral resolution imager such as a microcalorimeter.


\begin{acknowledgements}
We thank Rob Crain and Jon Davies for useful discussions on the EAGLE NEQ zooms interpretation. EMS acknowledges support from a National Science Foundation Graduate Research Fellowship (NSF GRFP) under Grant No. DGE‐1745301. Support for JAZ was provided by the {\it Chandra} X-ray Observatory Center, which is operated by the Smithsonian Astrophysical Observatory for and on behalf of NASA under contract NAS8-03060. CL was supported by NASA through the NASA Hubble Fellowship grant \#HST-HF2-51538.001-A awarded by the Space Telescope Science Institute, which is operated by the Association of Universities for Research in Astronomy, Inc., for NASA, under contract NAS5-26555. 
\end{acknowledgements}

\software{yt \citep{turk2011}, pyXSIM \citep{pyxsim}, mpi4py \citep{dalcin2021}, Astropy \citep{astropy2013, astropy2018}, NumPy \citep{vanderwalt2011, numpy}, Matplotlib \citep{matplotlib}}
    
\newpage
\bibliography{refs}{}

\begin{thebibliography}{}
\expandafter\ifx\csname natexlab\endcsname\relax\def\natexlab#1{#1}\fi
\providecommand{\url}[1]{\href{#1}{#1}}
\providecommand{\dodoi}[1]{doi:~\href{http://doi.org/#1}{\nolinkurl{#1}}}
\providecommand{\doeprint}[1]{\href{http://ascl.net/#1}{\nolinkurl{http://ascl.net/#1}}}
\providecommand{\doarXiv}[1]{\href{https://arxiv.org/abs/#1}{\nolinkurl{https://arxiv.org/abs/#1}}}

\bibitem[{{Anderson} \& {Bregman}(2011)}]{Anderson2011}
{Anderson}, M.~E., \& {Bregman}, J.~N. 2011, \apj, 737, 22,
  \dodoi{10.1088/0004-637X/737/1/22}

\bibitem[{{Anderson} {et~al.}(2013){Anderson}, {Bregman}, \&
  {Dai}}]{Anderson2013}
{Anderson}, M.~E., {Bregman}, J.~N., \& {Dai}, X. 2013, \apj, 762, 106,
  \dodoi{10.1088/0004-637X/762/2/106}

\bibitem[{{Anderson} {et~al.}(2015){Anderson}, {Gaspari}, {White}, {Wang}, \&
  {Dai}}]{Anderson2015}
{Anderson}, M.~E., {Gaspari}, M., {White}, S. D.~M., {Wang}, W., \& {Dai}, X.
  2015, \mnras, 449, 3806, \dodoi{10.1093/mnras/stv437}

\bibitem[{{Astropy Collaboration} {et~al.}(2013){Astropy Collaboration},
  {Robitaille}, {Tollerud}, {Greenfield}, {Droettboom}, {Bray}, {Aldcroft},
  {Davis}, {Ginsburg}, {Price-Whelan}, {Kerzendorf}, {Conley}, {Crighton},
  {Barbary}, {Muna}, {Ferguson}, {Grollier}, {Parikh}, {Nair}, {Unther},
  {Deil}, {Woillez}, {Conseil}, {Kramer}, {Turner}, {Singer}, {Fox}, {Weaver},
  {Zabalza}, {Edwards}, {Azalee Bostroem}, {Burke}, {Casey}, {Crawford},
  {Dencheva}, {Ely}, {Jenness}, {Labrie}, {Lim}, {Pierfederici}, {Pontzen},
  {Ptak}, {Refsdal}, {Servillat}, \& {Streicher}}]{astropy2013}
{Astropy Collaboration}, {Robitaille}, T.~P., {Tollerud}, E.~J., {et~al.} 2013,
  \aap, 558, A33, \dodoi{10.1051/0004-6361/201322068}

\bibitem[{{Astropy Collaboration} {et~al.}(2018){Astropy Collaboration},
  {Price-Whelan}, {Sip{\H{o}}cz}, {G{\"u}nther}, {Lim}, {Crawford}, {Conseil},
  {Shupe}, {Craig}, {Dencheva}, {Ginsburg}, {VanderPlas}, {Bradley},
  {P{\'e}rez-Su{\'a}rez}, {de Val-Borro}, {Aldcroft}, {Cruz}, {Robitaille},
  {Tollerud}, {Ardelean}, {Babej}, {Bach}, {Bachetti}, {Bakanov}, {Bamford},
  {Barentsen}, {Barmby}, {Baumbach}, {Berry}, {Biscani}, {Boquien}, {Bostroem},
  {Bouma}, {Brammer}, {Bray}, {Breytenbach}, {Buddelmeijer}, {Burke},
  {Calderone}, {Cano Rodr{\'\i}guez}, {Cara}, {Cardoso}, {Cheedella}, {Copin},
  {Corrales}, {Crichton}, {D'Avella}, {Deil}, {Depagne}, {Dietrich}, {Donath},
  {Droettboom}, {Earl}, {Erben}, {Fabbro}, {Ferreira}, {Finethy}, {Fox},
  {Garrison}, {Gibbons}, {Goldstein}, {Gommers}, {Greco}, {Greenfield},
  {Groener}, {Grollier}, {Hagen}, {Hirst}, {Homeier}, {Horton}, {Hosseinzadeh},
  {Hu}, {Hunkeler}, {Ivezi{\'c}}, {Jain}, {Jenness}, {Kanarek}, {Kendrew},
  {Kern}, {Kerzendorf}, {Khvalko}, {King}, {Kirkby}, {Kulkarni}, {Kumar},
  {Lee}, {Lenz}, {Littlefair}, {Ma}, {Macleod}, {Mastropietro}, {McCully},
  {Montagnac}, {Morris}, {Mueller}, {Mumford}, {Muna}, {Murphy}, {Nelson},
  {Nguyen}, {Ninan}, {N{\"o}the}, {Ogaz}, {Oh}, {Parejko}, {Parley}, {Pascual},
  {Patil}, {Patil}, {Plunkett}, {Prochaska}, {Rastogi}, {Reddy Janga},
  {Sabater}, {Sakurikar}, {Seifert}, {Sherbert}, {Sherwood-Taylor}, {Shih},
  {Sick}, {Silbiger}, {Singanamalla}, {Singer}, {Sladen}, {Sooley},
  {Sornarajah}, {Streicher}, {Teuben}, {Thomas}, {Tremblay}, {Turner},
  {Terr{\'o}n}, {van Kerkwijk}, {de la Vega}, {Watkins}, {Weaver}, {Whitmore},
  {Woillez}, {Zabalza}, \& {Astropy Contributors}}]{astropy2018}
{Astropy Collaboration}, {Price-Whelan}, A.~M., {Sip{\H{o}}cz}, B.~M., {et~al.}
  2018, \aj, 156, 123, \dodoi{10.3847/1538-3881/aabc4f}

\bibitem[{{Bandler} {et~al.}(2019){Bandler}, {Chervenak}, {Datesman},
  {Devasia}, {DiPirro}, {Sakai}, {Smith}, {Stevenson}, {Yoon}, {Bennett},
  {Mates}, {Swetz}, {Ullom}, {Irwin}, {Eckart}, {Figueroa-Feliciano},
  {McCammon}, {Ryu}, {Olson}, \& {Zeiger}}]{Bandler2019}
{Bandler}, S.~R., {Chervenak}, J.~A., {Datesman}, A.~M., {et~al.} 2019, Journal
  of Astronomical Telescopes, Instruments, and Systems, 5, 021017,
  \dodoi{10.1117/1.JATIS.5.2.021017}

\bibitem[{{Barret} {et~al.}(2018){Barret}, {Lam Trong}, {den Herder}, {Piro},
  {Cappi}, {Houvelin}, {Kelley}, {Mas-Hesse}, {Mitsuda}, {Paltani}, {Rauw},
  {Rozanska}, {Wilms}, {Bandler}, {Barbera}, {Barcons}, {Bozzo}, {Ceballos},
  {Charles}, {Costantini}, {Decourchelle}, {den Hartog}, {Duband}, {Duval},
  {Fiore}, {Gatti}, {Goldwurm}, {Jackson}, {Jonker}, {Kilbourne}, {Macculi},
  {Mendez}, {Molendi}, {Orleanski}, {Pajot}, {Pointecouteau}, {Porter},
  {Pratt}, {Pr{\^e}le}, {Ravera}, {Sato}, {Schaye}, {Shinozaki}, {Thibert},
  {Valenziano}, {Valette}, {Vink}, {Webb}, {Wise}, {Yamasaki}, {Douchin},
  {Mesnager}, {Pontet}, {Pradines}, {Branduardi-Raymont}, {Bulbul}, {Dadina},
  {Ettori}, {Finoguenov}, {Fukazawa}, {Janiuk}, {Kaastra}, {Mazzotta},
  {Miller}, {Miniutti}, {Naze}, {Nicastro}, {Scioritino}, {Simonescu},
  {Torrejon}, {Frezouls}, {Geoffray}, {Peille}, {Aicardi}, {Andr{\'e}},
  {Daniel}, {Cl{\'e}net}, {Etcheverry}, {Gloaguen}, {Hervet}, {Jolly}, {Ledot},
  {Paillet}, {Schmisser}, {Vella}, {Damery}, {Boyce}, {Dipirro}, {Lotti},
  {Schwander}, {Smith}, {Van Leeuwen}, {van Weers}, {Clerc}, {Cobo}, {Dauser},
  {Kirsch}, {Cucchetti}, {Eckart}, {Ferrando}, \& {Natalucci}}]{athena2018}
{Barret}, D., {Lam Trong}, T., {den Herder}, J.-W., {et~al.} 2018, in Society
  of Photo-Optical Instrumentation Engineers (SPIE) Conference Series, Vol.
  10699, Space Telescopes and Instrumentation 2018: Ultraviolet to Gamma Ray,
  ed. J.-W.~A. {den Herder}, S.~{Nikzad}, \& K.~{Nakazawa}, 106991G,
  \dodoi{10.1117/12.2312409}

\bibitem[{{Bertone} {et~al.}(2013){Bertone}, {Aguirre}, \&
  {Schaye}}]{Bertone2013}
{Bertone}, S., {Aguirre}, A., \& {Schaye}, J. 2013, \mnras, 430, 3292,
  \dodoi{10.1093/mnras/stt131}

\bibitem[{{Bogd{\'a}n} {et~al.}(2017){Bogd{\'a}n}, {Bourdin}, {Forman},
  {Kraft}, {Vogelsberger}, {Hernquist}, \& {Springel}}]{bogdan2017}
{Bogd{\'a}n}, {\'A}., {Bourdin}, H., {Forman}, W.~R., {et~al.} 2017, \apj, 850,
  98, \dodoi{10.3847/1538-4357/aa9523}

\bibitem[{{Bogd{\'a}n} {et~al.}(2013){Bogd{\'a}n}, {Forman}, {Vogelsberger},
  {Bourdin}, {Sijacki}, {Mazzotta}, {Kraft}, {Jones}, {Gilfanov}, {Churazov},
  \& {David}}]{Bogdan2013}
{Bogd{\'a}n}, {\'A}., {Forman}, W.~R., {Vogelsberger}, M., {et~al.} 2013, \apj,
  772, 97, \dodoi{10.1088/0004-637X/772/2/97}

\bibitem[{{Bregman} {et~al.}(2022){Bregman}, {Hodges-Kluck}, {Qu}, {Pratt},
  {Li}, \& {Yun}}]{Bregman2022}
{Bregman}, J.~N., {Hodges-Kluck}, E., {Qu}, Z., {et~al.} 2022, \apj, 928, 14,
  \dodoi{10.3847/1538-4357/ac51de}

\bibitem[{{Bryan} {et~al.}(2014){Bryan}, {Norman}, {O'Shea}, {Abel}, {Wise},
  {Turk}, {Reynolds}, {Collins}, {Wang}, {Skillman}, {Smith}, {Harkness},
  {Bordner}, {Kim}, {Kuhlen}, {Xu}, {Goldbaum}, {Hummels}, {Kritsuk}, {Tasker},
  {Skory}, {Simpson}, {Hahn}, {Oishi}, {So}, {Zhao}, {Cen}, {Li}, \& {Enzo
  Collaboration}}]{enzo}
{Bryan}, G.~L., {Norman}, M.~L., {O'Shea}, B.~W., {et~al.} 2014, \apjs, 211,
  19, \dodoi{10.1088/0067-0049/211/2/19}

\bibitem[{{Buck} {et~al.}(2020){Buck}, {Pfrommer}, {Pakmor}, {Grand}, \&
  {Springel}}]{Buck2020}
{Buck}, T., {Pfrommer}, C., {Pakmor}, R., {Grand}, R. J.~J., \& {Springel}, V.
  2020, \mnras, 497, 1712, \dodoi{10.1093/mnras/staa1960}

\bibitem[{{Burchett} {et~al.}(2021){Burchett}, {Rubin}, {Prochaska}, {Coil},
  {Vaught}, \& {Hennawi}}]{Burchett2021}
{Burchett}, J.~N., {Rubin}, K. H.~R., {Prochaska}, J.~X., {et~al.} 2021, \apj,
  909, 151, \dodoi{10.3847/1538-4357/abd4e0}

\bibitem[{{Burchett} {et~al.}(2016){Burchett}, {Tripp}, {Bordoloi}, {Werk},
  {Prochaska}, {Tumlinson}, {Willmer}, {O'Meara}, \& {Katz}}]{Burchett2016}
{Burchett}, J.~N., {Tripp}, T.~M., {Bordoloi}, R., {et~al.} 2016, \apj, 832,
  124, \dodoi{10.3847/0004-637X/832/2/124}

\bibitem[{{Burchett} {et~al.}(2019){Burchett}, {Tripp}, {Prochaska}, {Werk},
  {Tumlinson}, {Howk}, {Willmer}, {Lehner}, {Meiring}, {Bowen}, {Bordoloi},
  {Peeples}, {Jenkins}, {O'Meara}, {Tejos}, \& {Katz}}]{Burchett2019}
{Burchett}, J.~N., {Tripp}, T.~M., {Prochaska}, J.~X., {et~al.} 2019, \apjl,
  877, L20, \dodoi{10.3847/2041-8213/ab1f7f}

\bibitem[{{Butsky} \& {Quinn}(2018)}]{Butsky2018}
{Butsky}, I.~S., \& {Quinn}, T.~R. 2018, \apj, 868, 108,
  \dodoi{10.3847/1538-4357/aaeac2}

\bibitem[{{Cai} {et~al.}(2018){Cai}, {Hamden}, {Matuszewski}, {Prochaska},
  {Li}, {Cantalupo}, {Arrigoni Battaia}, {Martin}, {Neill}, {O'Sullivan},
  {Wang}, {Moore}, \& {Morrissey}}]{cai2018}
{Cai}, Z., {Hamden}, E., {Matuszewski}, M., {et~al.} 2018, \apjl, 861, L3,
  \dodoi{10.3847/2041-8213/aacce6}

\bibitem[{{Chadayammuri} {et~al.}(2022){Chadayammuri}, {Bogd{\'a}n},
  {Oppenheimer}, {Kraft}, {Forman}, \& {Jones}}]{Chadayammuri2022}
{Chadayammuri}, U., {Bogd{\'a}n}, {\'A}., {Oppenheimer}, B.~D., {et~al.} 2022,
  \apjl, 936, L15, \dodoi{10.3847/2041-8213/ac8936}

\bibitem[{{Chan} {et~al.}(2022){Chan}, {Kere{\v{s}}}, {Gurvich}, {Hopkins},
  {Trapp}, {Ji}, \& {Faucher-Gigu{\`e}re}}]{chan2022}
{Chan}, T.~K., {Kere{\v{s}}}, D., {Gurvich}, A.~B., {et~al.} 2022, \mnras, 517,
  597, \dodoi{10.1093/mnras/stac2236}

\bibitem[{{Chan} {et~al.}(2019){Chan}, {Kere{\v{s}}}, {Hopkins}, {Quataert},
  {Su}, {Hayward}, \& {Faucher-Gigu{\`e}re}}]{Chan2019}
{Chan}, T.~K., {Kere{\v{s}}}, D., {Hopkins}, P.~F., {et~al.} 2019, \mnras, 488,
  3716, \dodoi{10.1093/mnras/stz1895}

\bibitem[{{Churazov} {et~al.}(2001){Churazov}, {Haehnelt}, {Kotov}, \&
  {Sunyaev}}]{Churazov2001}
{Churazov}, E., {Haehnelt}, M., {Kotov}, O., \& {Sunyaev}, R. 2001, \mnras,
  323, 93, \dodoi{10.1046/j.1365-8711.2001.04090.x}

\bibitem[{{Churchill} {et~al.}(2013){Churchill}, {Trujillo-Gomez}, {Nielsen},
  \& {Kacprzak}}]{Churchill2013}
{Churchill}, C.~W., {Trujillo-Gomez}, S., {Nielsen}, N.~M., \& {Kacprzak},
  G.~G. 2013, \apj, 779, 87, \dodoi{10.1088/0004-637X/779/1/87}

\bibitem[{{Comparat} {et~al.}(2022){Comparat}, {Truong}, {Merloni},
  {Pillepich}, {Ponti}, {Driver}, {Bellstedt}, {Liske}, {Aird}, {Br{\"u}ggen},
  {Bulbul}, {Davies}, {Villalba}, {Georgakakis}, {Haberl}, {Liu}, {Maitra},
  {Nandra}, {Popesso}, {Predehl}, {Robotham}, {Salvato}, {Thorne}, \&
  {Zhang}}]{Comparat2022}
{Comparat}, J., {Truong}, N., {Merloni}, A., {et~al.} 2022, \aap, 666, A156,
  \dodoi{10.1051/0004-6361/202243101}

\bibitem[{{Connor} \& {Ravi}(2022)}]{Connor2022}
{Connor}, L., \& {Ravi}, V. 2022, Nature Astronomy, 6, 1035,
  \dodoi{10.1038/s41550-022-01719-7}

\bibitem[{{Crain} \& {van de Voort}(2023)}]{Crain2023}
{Crain}, R.~A., \& {van de Voort}, F. 2023, \araa, 61, 473,
  \dodoi{10.1146/annurev-astro-041923-043618}

\bibitem[{{Crain} {et~al.}(2015){Crain}, {Schaye}, {Bower}, {Furlong},
  {Schaller}, {Theuns}, {Dalla Vecchia}, {Frenk}, {McCarthy}, {Helly},
  {Jenkins}, {Rosas-Guevara}, {White}, \& {Trayford}}]{Crain2015}
{Crain}, R.~A., {Schaye}, J., {Bower}, R.~G., {et~al.} 2015, \mnras, 450, 1937,
  \dodoi{10.1093/mnras/stv725}

\bibitem[{{Dalcin} \& {Fang}(2021)}]{dalcin2021}
{Dalcin}, L., \& {Fang}, Y.-L.~L. 2021, Computing in Science and Engineering,
  23, 47, \dodoi{10.1109/MCSE.2021.3083216}

\bibitem[{{Das} {et~al.}(2020){Das}, {Mathur}, \& {Gupta}}]{Das2020}
{Das}, S., {Mathur}, S., \& {Gupta}, A. 2020, \apj, 897, 63,
  \dodoi{10.3847/1538-4357/ab93d2}

\bibitem[{{Das} {et~al.}(2019){Das}, {Mathur}, {Gupta}, {Nicastro}, {Krongold},
  \& {Null}}]{Das2019}
{Das}, S., {Mathur}, S., {Gupta}, A., {et~al.} 2019, \apj, 885, 108,
  \dodoi{10.3847/1538-4357/ab48df}

\bibitem[{{Dav{\'e}} {et~al.}(2019){Dav{\'e}}, {Angl{\'e}s-Alc{\'a}zar},
  {Narayanan}, {Li}, {Rafieferantsoa}, \& {Appleby}}]{SIMBA}
{Dav{\'e}}, R., {Angl{\'e}s-Alc{\'a}zar}, D., {Narayanan}, D., {et~al.} 2019,
  \mnras, 486, 2827, \dodoi{10.1093/mnras/stz937}

\bibitem[{{Davies} {et~al.}(2020){Davies}, {Crain}, {Oppenheimer}, \&
  {Schaye}}]{Davies2020}
{Davies}, J.~J., {Crain}, R.~A., {Oppenheimer}, B.~D., \& {Schaye}, J. 2020,
  \mnras, 491, 4462, \dodoi{10.1093/mnras/stz3201}

\bibitem[{{El-Badry} {et~al.}(2018){El-Badry}, {Quataert}, {Wetzel}, {Hopkins},
  {Weisz}, {Chan}, {Fitts}, {Boylan-Kolchin}, {Kere{\v{s}}},
  {Faucher-Gigu{\`e}re}, \& {Garrison-Kimmel}}]{El-Badry2018}
{El-Badry}, K., {Quataert}, E., {Wetzel}, A., {et~al.} 2018, \mnras, 473, 1930,
  \dodoi{10.1093/mnras/stx2482}

\bibitem[{{Faucher-Gigu{\`e}re} \& {Oh}(2023)}]{FGrev2023}
{Faucher-Gigu{\`e}re}, C.-A., \& {Oh}, S.~P. 2023, \araa, 61, 131,
  \dodoi{10.1146/annurev-astro-052920-125203}

\bibitem[{{Ferland} {et~al.}(2017){Ferland}, {Chatzikos}, {Guzm{\'a}n},
  {Lykins}, {van Hoof}, {Williams}, {Abel}, {Badnell}, {Keenan}, {Porter}, \&
  {Stancil}}]{Ferland2017}
{Ferland}, G.~J., {Chatzikos}, M., {Guzm{\'a}n}, F., {et~al.} 2017, \rmxaa, 53,
  385, \dodoi{10.48550/arXiv.1705.10877}

\bibitem[{{Foster} {et~al.}(2012){Foster}, {Ji}, {Smith}, \&
  {Brickhouse}}]{Foster2012}
{Foster}, A.~R., {Ji}, L., {Smith}, R.~K., \& {Brickhouse}, N.~S. 2012, \apj,
  756, 128, \dodoi{10.1088/0004-637X/756/2/128}

\bibitem[{{Garrison-Kimmel} {et~al.}(2018){Garrison-Kimmel}, {Hopkins},
  {Wetzel}, {El-Badry}, {Sanderson}, {Bullock}, {Ma}, {van de Voort}, {Hafen},
  {Faucher-Gigu{\`e}re}, {Hayward}, {Quataert}, {Kere{\v{s}}}, \&
  {Boylan-Kolchin}}]{Garrison-Kimmel2018}
{Garrison-Kimmel}, S., {Hopkins}, P.~F., {Wetzel}, A., {et~al.} 2018, \mnras,
  481, 4133, \dodoi{10.1093/mnras/sty2513}

\bibitem[{{Harris} {et~al.}(2020){Harris}, {Millman}, {van der Walt},
  {Gommers}, {Virtanen}, {Cournapeau}, {Wieser}, {Taylor}, {Berg}, {Smith},
  {Kern}, {Picus}, {Hoyer}, {van Kerkwijk}, {Brett}, {Haldane}, {del R{\'\i}o},
  {Wiebe}, {Peterson}, {G{\'e}rard-Marchant}, {Sheppard}, {Reddy}, {Weckesser},
  {Abbasi}, {Gohlke}, \& {Oliphant}}]{numpy}
{Harris}, C.~R., {Millman}, K.~J., {van der Walt}, S.~J., {et~al.} 2020, \nat,
  585, 357, \dodoi{10.1038/s41586-020-2649-2}

\bibitem[{{Hayes} {et~al.}(2016){Hayes}, {Melinder}, {{\"O}stlin}, {Scarlata},
  {Lehnert}, \& {Mannerstr{\"o}m-Jansson}}]{Hayes2016}
{Hayes}, M., {Melinder}, J., {{\"O}stlin}, G., {et~al.} 2016, \apj, 828, 49,
  \dodoi{10.3847/0004-637X/828/1/49}

\bibitem[{{Hopkins}(2015)}]{gizmo}
{Hopkins}, P.~F. 2015, \mnras, 450, 53, \dodoi{10.1093/mnras/stv195}

\bibitem[{{Hopkins}(2017)}]{gizmoPR}
---. 2017, arXiv e-prints, arXiv:1712.01294, \dodoi{10.48550/arXiv.1712.01294}

\bibitem[{{Hopkins} {et~al.}(2021){Hopkins}, {Chan}, {Ji}, {Hummels},
  {Kere{\v{s}}}, {Quataert}, \& {Faucher-Gigu{\`e}re}}]{Hopkins2021}
{Hopkins}, P.~F., {Chan}, T.~K., {Ji}, S., {et~al.} 2021, \mnras, 501, 3640,
  \dodoi{10.1093/mnras/staa3690}

\bibitem[{{Hopkins} {et~al.}(2014){Hopkins}, {Kere{\v{s}}}, {O{\~n}orbe},
  {Faucher-Gigu{\`e}re}, {Quataert}, {Murray}, \& {Bullock}}]{FIRE1}
{Hopkins}, P.~F., {Kere{\v{s}}}, D., {O{\~n}orbe}, J., {et~al.} 2014, \mnras,
  445, 581, \dodoi{10.1093/mnras/stu1738}

\bibitem[{{Hopkins} {et~al.}(2025){Hopkins}, {Quataert}, {Ponnada}, \&
  {Silich}}]{Hopkins2025}
{Hopkins}, P.~F., {Quataert}, E., {Ponnada}, S.~B., \& {Silich}, E. 2025, arXiv
  e-prints, arXiv:2501.18696.
\newblock \doarXiv{2501.18696}

\bibitem[{{Hopkins} \& {Raives}(2016)}]{hopkins2016}
{Hopkins}, P.~F., \& {Raives}, M.~J. 2016, \mnras, 455, 51,
  \dodoi{10.1093/mnras/stv2180}

\bibitem[{{Hopkins} {et~al.}(2018){Hopkins}, {Wetzel}, {Kere{\v{s}}},
  {Faucher-Gigu{\`e}re}, {Quataert}, {Boylan-Kolchin}, {Murray}, {Hayward},
  {Garrison-Kimmel}, {Hummels}, {Feldmann}, {Torrey}, {Ma},
  {Angl{\'e}s-Alc{\'a}zar}, {Su}, {Orr}, {Schmitz}, {Escala}, {Sanderson},
  {Grudi{\'c}}, {Hafen}, {Kim}, {Fitts}, {Bullock}, {Wheeler}, {Chan},
  {Elbert}, \& {Narayanan}}]{FIRE2}
{Hopkins}, P.~F., {Wetzel}, A., {Kere{\v{s}}}, D., {et~al.} 2018, \mnras, 480,
  800, \dodoi{10.1093/mnras/sty1690}

\bibitem[{{Hopkins} {et~al.}(2020){Hopkins}, {Chan}, {Garrison-Kimmel}, {Ji},
  {Su}, {Hummels}, {Kere{\v{s}}}, {Quataert}, \&
  {Faucher-Gigu{\`e}re}}]{FIRE2CR}
{Hopkins}, P.~F., {Chan}, T.~K., {Garrison-Kimmel}, S., {et~al.} 2020, \mnras,
  492, 3465, \dodoi{10.1093/mnras/stz3321}

\bibitem[{{Hummels} {et~al.}(2013){Hummels}, {Bryan}, {Smith}, \&
  {Turk}}]{Hummels2013}
{Hummels}, C.~B., {Bryan}, G.~L., {Smith}, B.~D., \& {Turk}, M.~J. 2013,
  \mnras, 430, 1548, \dodoi{10.1093/mnras/sts702}

\bibitem[{{Hummels} {et~al.}(2019){Hummels}, {Smith}, {Hopkins}, {O'Shea},
  {Silvia}, {Werk}, {Lehner}, {Wise}, {Collins}, \& {Butsky}}]{Hummels2019}
{Hummels}, C.~B., {Smith}, B.~D., {Hopkins}, P.~F., {et~al.} 2019, \apj, 882,
  156, \dodoi{10.3847/1538-4357/ab378f}

\bibitem[{{Humphrey} {et~al.}(2011){Humphrey}, {Buote}, {Canizares}, {Fabian},
  \& {Miller}}]{Humphrey2011}
{Humphrey}, P.~J., {Buote}, D.~A., {Canizares}, C.~R., {Fabian}, A.~C., \&
  {Miller}, J.~M. 2011, \apj, 729, 53, \dodoi{10.1088/0004-637X/729/1/53}

\bibitem[{{Hunter}(2007)}]{matplotlib}
{Hunter}, J.~D. 2007, Computing in Science and Engineering, 9, 90,
  \dodoi{10.1109/MCSE.2007.55}

\bibitem[{{Ji} {et~al.}(2019){Ji}, {Oh}, \& {Masterson}}]{Ji2019}
{Ji}, S., {Oh}, S.~P., \& {Masterson}, P. 2019, \mnras, 487, 737,
  \dodoi{10.1093/mnras/stz1248}

\bibitem[{{Ji} {et~al.}(2020){Ji}, {Chan}, {Hummels}, {Hopkins}, {Stern},
  {Kere{\v{s}}}, {Quataert}, {Faucher-Gigu{\`e}re}, \& {Murray}}]{Ji2020}
{Ji}, S., {Chan}, T.~K., {Hummels}, C.~B., {et~al.} 2020, \mnras, 496, 4221,
  \dodoi{10.1093/mnras/staa1849}

\bibitem[{{Khabibullin} \& {Churazov}(2019)}]{Khabibullin2019}
{Khabibullin}, I., \& {Churazov}, E. 2019, \mnras, 482, 4972,
  \dodoi{10.1093/mnras/sty2992}

\bibitem[{{Kraft} {et~al.}(2022){Kraft}, {Markevitch}, {Kilbourne}, {Adams},
  {Akamatsu}, {Ayromlou}, {Bandler}, {Barbera}, {Bennett}, {Bhardwaj}, {Biffi},
  {Bodewits}, {Bogdan}, {Bonamente}, {Borgani}, {Branduardi-Raymont},
  {Bregman}, {Burchett}, {Cann}, {Carter}, {Chakraborty}, {Churazov}, {Crain},
  {Cumbee}, {Dave}, {DiPirro}, {Dolag}, {Bertrand Doriese}, {Drake}, {Dunn},
  {Eckart}, {Eckert}, {Ettori}, {Forman}, {Galeazzi}, {Gall}, {Gatuzz}, {Hell},
  {Hodges-Kluck}, {Jackman}, {Jahromi}, {Jennings}, {Jones}, {Kaaret},
  {Kavanagh}, {Kelley}, {Khabibullin}, {Kim}, {Koutroumpa}, {Kovacs}, {Kuntz},
  {Lau}, {Lee}, {Leutenegger}, {Lin}, {Lisse}, {Lo Cicero}, {Lovisari},
  {McCammon}, {McEntee}, {Mernier}, {Miller}, {Nagai}, {Negro}, {Nelson},
  {Ness}, {Nulsen}, {Ogorzalek}, {Oppenheimer}, {Oskinova}, {Patnaude},
  {Pfeifle}, {Pillepich}, {Plucinsky}, {Pooley}, {Porter}, {Randall}, {Rasia},
  {Raymond}, {Ruszkowski}, {Sakai}, {Sarkar}, {Sasaki}, {Sato},
  {Schellenberger}, {Schaye}, {Simionescu}, {Smith}, {Steiner}, {Stern}, {Su},
  {Sun}, {Tremblay}, {Truong}, {Tutt}, {Ursino}, {Veilleux}, {Vikhlinin},
  {Vladutescu-Zopp}, {Vogelsberger}, {Walker}, {Weaver}, {Weigt}, {Werk},
  {Werner}, {Wolk}, {Zhang}, {Zhang}, {Zhuravleva}, \& {ZuHone}}]{Kraft2022}
{Kraft}, R., {Markevitch}, M., {Kilbourne}, C., {et~al.} 2022, arXiv e-prints,
  arXiv:2211.09827, \dodoi{10.48550/arXiv.2211.09827}

\bibitem[{{Li} {et~al.}(2018){Li}, {Bregman}, {Wang}, {Crain}, \&
  {Anderson}}]{Li2018}
{Li}, J.-T., {Bregman}, J.~N., {Wang}, Q.~D., {Crain}, R.~A., \& {Anderson},
  M.~E. 2018, \apjl, 855, L24, \dodoi{10.3847/2041-8213/aab2af}

\bibitem[{{Li} {et~al.}(2017){Li}, {Bregman}, {Wang}, {Crain}, {Anderson}, \&
  {Zhang}}]{Li2017}
{Li}, J.-T., {Bregman}, J.~N., {Wang}, Q.~D., {et~al.} 2017, \apjs, 233, 20,
  \dodoi{10.3847/1538-4365/aa96fc}

\bibitem[{{Lu} {et~al.}(2025){Lu}, {Kere{\v{s}}}, {Hopkins}, {Ponnada},
  {Faucher-Gigu{\'e}re}, \& {Hummels}}]{Lu2025}
{Lu}, Y.~S., {Kere{\v{s}}}, D., {Hopkins}, P.~F., {et~al.} 2025, arXiv
  e-prints, arXiv:2505.13597, \dodoi{10.48550/arXiv.2505.13597}

\bibitem[{{Muratov} {et~al.}(2015){Muratov}, {Kere{\v{s}}},
  {Faucher-Gigu{\`e}re}, {Hopkins}, {Quataert}, \& {Murray}}]{Muratov2015}
{Muratov}, A.~L., {Kere{\v{s}}}, D., {Faucher-Gigu{\`e}re}, C.-A., {et~al.}
  2015, \mnras, 454, 2691, \dodoi{10.1093/mnras/stv2126}

\bibitem[{{Nelson} {et~al.}(2019){Nelson}, {Pillepich}, {Springel}, {Pakmor},
  {Weinberger}, {Genel}, {Torrey}, {Vogelsberger}, {Marinacci}, \&
  {Hernquist}}]{tng1}
{Nelson}, D., {Pillepich}, A., {Springel}, V., {et~al.} 2019, \mnras, 490,
  3234, \dodoi{10.1093/mnras/stz2306}

\bibitem[{{Nielsen} {et~al.}(2013){Nielsen}, {Churchill}, {Kacprzak}, \&
  {Murphy}}]{Nielsen2013}
{Nielsen}, N.~M., {Churchill}, C.~W., {Kacprzak}, G.~G., \& {Murphy}, M.~T.
  2013, \apj, 776, 114, \dodoi{10.1088/0004-637X/776/2/114}

\bibitem[{{Oppenheimer} \& {Schaye}(2013)}]{Oppenheimer2013}
{Oppenheimer}, B.~D., \& {Schaye}, J. 2013, \mnras, 434, 1043,
  \dodoi{10.1093/mnras/stt1043}

\bibitem[{{Oppenheimer} {et~al.}(2016){Oppenheimer}, {Crain}, {Schaye},
  {Rahmati}, {Richings}, {Trayford}, {Tumlinson}, {Bower}, {Schaller}, \&
  {Theuns}}]{Oppenheimer2016}
{Oppenheimer}, B.~D., {Crain}, R.~A., {Schaye}, J., {et~al.} 2016, \mnras, 460,
  2157, \dodoi{10.1093/mnras/stw1066}

\bibitem[{{Orr} {et~al.}(2020){Orr}, {Hayward}, {Medling}, {Gurvich},
  {Hopkins}, {Murray}, {Pineda}, {Faucher-Gigu{\`e}re}, {Kere{\v{s}}},
  {Wetzel}, \& {Su}}]{Orr2020}
{Orr}, M.~E., {Hayward}, C.~C., {Medling}, A.~M., {et~al.} 2020, \mnras, 496,
  1620, \dodoi{10.1093/mnras/staa1619}

\bibitem[{{Peeples} {et~al.}(2019){Peeples}, {Corlies}, {Tumlinson}, {O'Shea},
  {Lehner}, {O'Meara}, {Howk}, {Earl}, {Smith}, {Wise}, \&
  {Hummels}}]{Peeples2019}
{Peeples}, M.~S., {Corlies}, L., {Tumlinson}, J., {et~al.} 2019, \apj, 873,
  129, \dodoi{10.3847/1538-4357/ab0654}

\bibitem[{{Pillepich} {et~al.}(2021){Pillepich}, {Nelson}, {Truong},
  {Weinberger}, {Martin-Navarro}, {Springel}, {Faber}, \&
  {Hernquist}}]{Pillepich2021}
{Pillepich}, A., {Nelson}, D., {Truong}, N., {et~al.} 2021, \mnras, 508, 4667,
  \dodoi{10.1093/mnras/stab2779}

\bibitem[{{Pillepich} {et~al.}(2018){Pillepich}, {Springel}, {Nelson}, {Genel},
  {Naiman}, {Pakmor}, {Hernquist}, {Torrey}, {Vogelsberger}, {Weinberger}, \&
  {Marinacci}}]{Pillepich2018}
{Pillepich}, A., {Springel}, V., {Nelson}, D., {et~al.} 2018, \mnras, 473,
  4077, \dodoi{10.1093/mnras/stx2656}

\bibitem[{{Pillepich} {et~al.}(2019){Pillepich}, {Nelson}, {Springel},
  {Pakmor}, {Torrey}, {Weinberger}, {Vogelsberger}, {Marinacci}, {Genel}, {van
  der Wel}, \& {Hernquist}}]{tng2}
{Pillepich}, A., {Nelson}, D., {Springel}, V., {et~al.} 2019, \mnras, 490,
  3196, \dodoi{10.1093/mnras/stz2338}

\bibitem[{{Pillepich} {et~al.}(2024){Pillepich}, {Sotillo-Ramos}, {Ramesh},
  {Nelson}, {Engler}, {Rodriguez-Gomez}, {Fournier}, {Donnari}, {Springel}, \&
  {Hernquist}}]{Pillepich2024}
{Pillepich}, A., {Sotillo-Ramos}, D., {Ramesh}, R., {et~al.} 2024, \mnras, 535,
  1721, \dodoi{10.1093/mnras/stae2165}

\bibitem[{{Planck Collaboration} {et~al.}(2014){Planck Collaboration}, {Ade},
  {Aghanim}, {Armitage-Caplan}, {Arnaud}, {Ashdown}, {Atrio-Barandela},
  {Aumont}, {Baccigalupi}, {Banday}, {Barreiro}, {Bartlett}, {Battaner},
  {Benabed}, {Beno{\^\i}t}, {Benoit-L{\'e}vy}, {Bernard}, {Bersanelli},
  {Bielewicz}, {Bobin}, {Bock}, {Bonaldi}, {Bond}, {Borrill}, {Bouchet},
  {Bridges}, {Bucher}, {Burigana}, {Butler}, {Calabrese}, {Cappellini},
  {Cardoso}, {Catalano}, {Challinor}, {Chamballu}, {Chary}, {Chen}, {Chiang},
  {Chiang}, {Christensen}, {Church}, {Clements}, {Colombi}, {Colombo},
  {Couchot}, {Coulais}, {Crill}, {Curto}, {Cuttaia}, {Danese}, {Davies},
  {Davis}, {de Bernardis}, {de Rosa}, {de Zotti}, {Delabrouille}, {Delouis},
  {D{\'e}sert}, {Dickinson}, {Diego}, {Dolag}, {Dole}, {Donzelli}, {Dor{\'e}},
  {Douspis}, {Dunkley}, {Dupac}, {Efstathiou}, {Elsner}, {En{\ss}lin},
  {Eriksen}, {Finelli}, {Forni}, {Frailis}, {Fraisse}, {Franceschi}, {Gaier},
  {Galeotta}, {Galli}, {Ganga}, {Giard}, {Giardino}, {Giraud-H{\'e}raud},
  {Gjerl{\o}w}, {Gonz{\'a}lez-Nuevo}, {G{\'o}rski}, {Gratton}, {Gregorio},
  {Gruppuso}, {Gudmundsson}, {Haissinski}, {Hamann}, {Hansen}, {Hanson},
  {Harrison}, {Henrot-Versill{\'e}}, {Hern{\'a}ndez-Monteagudo}, {Herranz},
  {Hildebrandt}, {Hivon}, {Hobson}, {Holmes}, {Hornstrup}, {Hou}, {Hovest},
  {Huffenberger}, {Jaffe}, {Jaffe}, {Jewell}, {Jones}, {Juvela},
  {Keih{\"a}nen}, {Keskitalo}, {Kisner}, {Kneissl}, {Knoche}, {Knox}, {Kunz},
  {Kurki-Suonio}, {Lagache}, {L{\"a}hteenm{\"a}ki}, {Lamarre}, {Lasenby},
  {Lattanzi}, {Laureijs}, {Lawrence}, {Leach}, {Leahy}, {Leonardi},
  {Le{\'o}n-Tavares}, {Lesgourgues}, {Lewis}, {Liguori}, {Lilje},
  {Linden-V{\o}rnle}, {L{\'o}pez-Caniego}, {Lubin}, {Mac{\'\i}as-P{\'e}rez},
  {Maffei}, {Maino}, {Mandolesi}, {Maris}, {Marshall}, {Martin},
  {Mart{\'\i}nez-Gonz{\'a}lez}, {Masi}, {Massardi}, {Matarrese}, {Matthai},
  {Mazzotta}, {Meinhold}, {Melchiorri}, {Melin}, {Mendes}, {Menegoni},
  {Mennella}, {Migliaccio}, {Millea}, {Mitra}, {Miville-Desch{\^e}nes},
  {Moneti}, {Montier}, {Morgante}, {Mortlock}, {Moss}, {Munshi}, {Murphy},
  {Naselsky}, {Nati}, {Natoli}, {Netterfield}, {N{\o}rgaard-Nielsen},
  {Noviello}, {Novikov}, {Novikov}, {O'Dwyer}, {Osborne}, {Oxborrow}, {Paci},
  {Pagano}, {Pajot}, {Paladini}, {Paoletti}, {Partridge}, {Pasian},
  {Patanchon}, {Pearson}, {Pearson}, {Peiris}, {Perdereau}, {Perotto},
  {Perrotta}, {Pettorino}, {Piacentini}, {Piat}, {Pierpaoli}, {Pietrobon},
  {Plaszczynski}, {Platania}, \& {Pointecouteau}}]{planck2014}
{Planck Collaboration}, {Ade}, P.~A.~R., {Aghanim}, N., {et~al.} 2014, \aap,
  571, A16, \dodoi{10.1051/0004-6361/201321591}

\bibitem[{{Planck Collaboration} {et~al.}(2016){Planck Collaboration}, {Ade},
  {Aghanim}, {Arnaud}, {Ashdown}, {Aumont}, {Baccigalupi}, {Banday},
  {Barreiro}, {Bartlett}, {Bartolo}, {Battaner}, {Battye}, {Benabed},
  {Beno{\^\i}t}, {Benoit-L{\'e}vy}, {Bernard}, {Bersanelli}, {Bielewicz},
  {Bock}, {Bonaldi}, {Bonavera}, {Bond}, {Borrill}, {Bouchet}, {Boulanger},
  {Bucher}, {Burigana}, {Butler}, {Calabrese}, {Cardoso}, {Catalano},
  {Challinor}, {Chamballu}, {Chary}, {Chiang}, {Chluba}, {Christensen},
  {Church}, {Clements}, {Colombi}, {Colombo}, {Combet}, {Coulais}, {Crill},
  {Curto}, {Cuttaia}, {Danese}, {Davies}, {Davis}, {de Bernardis}, {de Rosa},
  {de Zotti}, {Delabrouille}, {D{\'e}sert}, {Di Valentino}, {Dickinson},
  {Diego}, {Dolag}, {Dole}, {Donzelli}, {Dor{\'e}}, {Douspis}, {Ducout},
  {Dunkley}, {Dupac}, {Efstathiou}, {Elsner}, {En{\ss}lin}, {Eriksen},
  {Farhang}, {Fergusson}, {Finelli}, {Forni}, {Frailis}, {Fraisse},
  {Franceschi}, {Frejsel}, {Galeotta}, {Galli}, {Ganga}, {Gauthier}, {Gerbino},
  {Ghosh}, {Giard}, {Giraud-H{\'e}raud}, {Giusarma}, {Gjerl{\o}w},
  {Gonz{\'a}lez-Nuevo}, {G{\'o}rski}, {Gratton}, {Gregorio}, {Gruppuso},
  {Gudmundsson}, {Hamann}, {Hansen}, {Hanson}, {Harrison}, {Helou},
  {Henrot-Versill{\'e}}, {Hern{\'a}ndez-Monteagudo}, {Herranz}, {Hildebrandt},
  {Hivon}, {Hobson}, {Holmes}, {Hornstrup}, {Hovest}, {Huang}, {Huffenberger},
  {Hurier}, {Jaffe}, {Jaffe}, {Jones}, {Juvela}, {Keih{\"a}nen}, {Keskitalo},
  {Kisner}, {Kneissl}, {Knoche}, {Knox}, {Kunz}, {Kurki-Suonio}, {Lagache},
  {L{\"a}hteenm{\"a}ki}, {Lamarre}, {Lasenby}, {Lattanzi}, {Lawrence}, {Leahy},
  {Leonardi}, {Lesgourgues}, {Levrier}, {Lewis}, {Liguori}, {Lilje},
  {Linden-V{\o}rnle}, {L{\'o}pez-Caniego}, {Lubin}, {Mac{\'\i}as-P{\'e}rez},
  {Maggio}, {Maino}, {Mandolesi}, {Mangilli}, {Marchini}, {Maris}, {Martin},
  {Martinelli}, {Mart{\'\i}nez-Gonz{\'a}lez}, {Masi}, {Matarrese}, {McGehee},
  {Meinhold}, {Melchiorri}, {Melin}, {Mendes}, {Mennella}, {Migliaccio},
  {Millea}, {Mitra}, {Miville-Desch{\^e}nes}, {Moneti}, {Montier}, {Morgante},
  {Mortlock}, {Moss}, {Munshi}, {Murphy}, {Naselsky}, {Nati}, {Natoli},
  {Netterfield}, {N{\o}rgaard-Nielsen}, {Noviello}, {Novikov}, {Novikov},
  {Oxborrow}, {Paci}, {Pagano}, {Pajot}, {Paladini}, {Paoletti}, {Partridge},
  {Pasian}, {Patanchon}, {Pearson}, {Perdereau}, {Perotto}, {Perrotta},
  {Pettorino}, {Piacentini}, {Piat}, {Pierpaoli}, {Pietrobon}, {Plaszczynski},
  {Pointecouteau}, {Polenta}, {Popa}, {Pratt}, {Pr{\'e}zeau}, {Prunet},
  {Puget}, {Rachen}, {Reach}, {Rebolo}, {Reinecke}, {Remazeilles}, {Renault},
  {Renzi}, {Ristorcelli}, {Rocha}, {Rosset}, {Rossetti}, {Roudier},
  {Rouill{\'e} d'Orfeuil}, {Rowan-Robinson}, {Rubi{\~n}o-Mart{\'\i}n},
  {Rusholme}, {Said}, {Salvatelli}, {Salvati}, {Sandri}, {Santos},
  {Savelainen}, {Savini}, {Scott}, {Seiffert}, {Serra}, {Shellard}, {Spencer},
  {Spinelli}, {Stolyarov}, {Stompor}, {Sudiwala}, {Sunyaev}, {Sutton},
  {Suur-Uski}, {Sygnet}, {Tauber}, {Terenzi}, {Toffolatti}, {Tomasi},
  {Tristram}, {Trombetti}, {Tucci}, {Tuovinen}, {T{\"u}rler}, {Umana},
  {Valenziano}, {Valiviita}, {Van Tent}, {Vielva}, {Villa}, {Wade}, {Wandelt},
  {Wehus}, {White}, {White}, {Wilkinson}, {Yvon}, {Zacchei}, \&
  {Zonca}}]{Planck2016}
---. 2016, \aap, 594, A13, \dodoi{10.1051/0004-6361/201525830}

\bibitem[{Ponnada {et~al.}(2022)Ponnada, Panopoulou, Butsky, Hopkins, Loebman,
  Hummels, Ji, Wetzel, Faucher-Giguère, \& Hayward}]{ponnada_magnetic_2022}
Ponnada, S.~B., Panopoulou, G.~V., Butsky, I.~S., {et~al.} 2022, Monthly
  Notices of the Royal Astronomical Society, 516, 4417,
  \dodoi{10.1093/mnras/stac2448}

\bibitem[{{Porter} {et~al.}(2024){Porter}, {Kilbourne}, {Chiao}, {Cumbee},
  {Eckart}, {Fujimoto}, {Ishisaki}, {Kanemaru}, {Kelley}, {Leutenegger},
  {Maeda}, {Mizumoto}, {Sato}, {Sawada}, {Sneiderman}, {Takei}, {Tsujimoto},
  {Uchida}, {Watanabe}, \& {Yamada}}]{Porter2024}
{Porter}, F.~S., {Kilbourne}, C.~A., {Chiao}, M., {et~al.} 2024, in Society of
  Photo-Optical Instrumentation Engineers (SPIE) Conference Series, Vol. 13093,
  Space Telescopes and Instrumentation 2024: Ultraviolet to Gamma Ray, ed.
  J.-W.~A. {den Herder}, S.~{Nikzad}, \& K.~{Nakazawa}, 130931K,
  \dodoi{10.1117/12.3018882}

\bibitem[{{Ravi}(2019)}]{Ravi2019}
{Ravi}, V. 2019, \apj, 872, 88, \dodoi{10.3847/1538-4357/aafb30}

\bibitem[{{Roca-F{\`a}brega} {et~al.}(2021){Roca-F{\`a}brega}, {Kim},
  {Hausammann}, {Nagamine}, {Lupi}, {Powell}, {Shimizu}, {Ceverino}, {Primack},
  {Quinn}, {Revaz}, {Vel{\'a}zquez}, {Abel}, {Buehlmann}, {Dekel}, {Dong},
  {Hahn}, {Hummels}, {Kim}, {Smith}, {Strawn}, {Teyssier}, {Turk}, \& {AGORA
  Collaboration}}]{agora2021}
{Roca-F{\`a}brega}, S., {Kim}, J.-H., {Hausammann}, L., {et~al.} 2021, \apj,
  917, 64, \dodoi{10.3847/1538-4357/ac088a}

\bibitem[{{Rupke} {et~al.}(2019){Rupke}, {Coil}, {Geach}, {Tremonti},
  {Diamond-Stanic}, {George}, {Hickox}, {Kepley}, {Leung}, {Moustakas},
  {Rudnick}, \& {Sell}}]{Rupke2019}
{Rupke}, D. S.~N., {Coil}, A., {Geach}, J.~E., {et~al.} 2019, \nat, 574, 643,
  \dodoi{10.1038/s41586-019-1686-1}

\bibitem[{{Salem} \& {Bryan}(2014)}]{Salem2014}
{Salem}, M., \& {Bryan}, G.~L. 2014, \mnras, 437, 3312,
  \dodoi{10.1093/mnras/stt2121}

\bibitem[{{Salem} {et~al.}(2016){Salem}, {Bryan}, \& {Corlies}}]{Salem2016}
{Salem}, M., {Bryan}, G.~L., \& {Corlies}, L. 2016, \mnras, 456, 582,
  \dodoi{10.1093/mnras/stv2641}

\bibitem[{{Schaye} {et~al.}(2015){Schaye}, {Crain}, {Bower}, {Furlong},
  {Schaller}, {Theuns}, {Dalla Vecchia}, {Frenk}, {McCarthy}, {Helly},
  {Jenkins}, {Rosas-Guevara}, {White}, {Baes}, {Booth}, {Camps}, {Navarro},
  {Qu}, {Rahmati}, {Sawala}, {Thomas}, \& {Trayford}}]{Schaye2015}
{Schaye}, J., {Crain}, R.~A., {Bower}, R.~G., {et~al.} 2015, \mnras, 446, 521,
  \dodoi{10.1093/mnras/stu2058}

\bibitem[{{Schellenberger} {et~al.}(2024){Schellenberger}, {Bogd{\'a}n},
  {ZuHone}, {Oppenheimer}, {Truong}, {Khabibullin}, {Jennings}, {Pillepich},
  {Burchett}, {Carr}, {Chakraborty}, {Crain}, {Forman}, {Jones}, {Kilbourne},
  {Kraft}, {Markevitch}, {Nagai}, {Nelson}, {Ogorzalek}, {Randall}, {Sarkar},
  {Schaye}, {Veilleux}, {Vogelsberger}, {Wang}, \&
  {Zhuravleva}}]{Schellenberger2024}
{Schellenberger}, G., {Bogd{\'a}n}, {\'A}., {ZuHone}, J.~A., {et~al.} 2024,
  \apj, 969, 85, \dodoi{10.3847/1538-4357/ad4548}

\bibitem[{{Smith} {et~al.}(2001){Smith}, {Brickhouse}, {Liedahl}, \&
  {Raymond}}]{Smith2001}
{Smith}, R.~K., {Brickhouse}, N.~S., {Liedahl}, D.~A., \& {Raymond}, J.~C.
  2001, \apjl, 556, L91, \dodoi{10.1086/322992}

\bibitem[{{Springel}(2010)}]{arepo}
{Springel}, V. 2010, \mnras, 401, 791, \dodoi{10.1111/j.1365-2966.2009.15715.x}

\bibitem[{{Steidel} {et~al.}(2000){Steidel}, {Adelberger}, {Shapley},
  {Pettini}, {Dickinson}, \& {Giavalisco}}]{steidel2000}
{Steidel}, C.~C., {Adelberger}, K.~L., {Shapley}, A.~E., {et~al.} 2000, \apj,
  532, 170, \dodoi{10.1086/308568}

\bibitem[{{Stern} {et~al.}(2021){Stern}, {Faucher-Gigu{\`e}re}, {Fielding},
  {Quataert}, {Hafen}, {Gurvich}, {Ma}, {Byrne}, {El-Badry},
  {Angl{\'e}s-Alc{\'a}zar}, {Chan}, {Feldmann}, {Kere{\v{s}}}, {Wetzel},
  {Murray}, \& {Hopkins}}]{Stern2021}
{Stern}, J., {Faucher-Gigu{\`e}re}, C.-A., {Fielding}, D., {et~al.} 2021, \apj,
  911, 88, \dodoi{10.3847/1538-4357/abd776}

\bibitem[{{Stocke} {et~al.}(2013){Stocke}, {Keeney}, {Danforth}, {Shull},
  {Froning}, {Green}, {Penton}, \& {Savage}}]{Stocke2013}
{Stocke}, J.~T., {Keeney}, B.~A., {Danforth}, C.~W., {et~al.} 2013, \apj, 763,
  148, \dodoi{10.1088/0004-637X/763/2/148}

\bibitem[{{Su} {et~al.}(2017){Su}, {Hopkins}, {Hayward}, {Faucher-Gigu{\`e}re},
  {Kere{\v{s}}}, {Ma}, \& {Robles}}]{Su2017}
{Su}, K.-Y., {Hopkins}, P.~F., {Hayward}, C.~C., {et~al.} 2017, \mnras, 471,
  144, \dodoi{10.1093/mnras/stx1463}

\bibitem[{{Suresh} {et~al.}(2015){Suresh}, {Bird}, {Vogelsberger}, {Genel},
  {Torrey}, {Sijacki}, {Springel}, \& {Hernquist}}]{Suresh2015}
{Suresh}, J., {Bird}, S., {Vogelsberger}, M., {et~al.} 2015, \mnras, 448, 895,
  \dodoi{10.1093/mnras/stu2762}

\bibitem[{{Truong} {et~al.}(2021){Truong}, {Pillepich}, {Nelson}, {Werner}, \&
  {Hernquist}}]{Truong2021}
{Truong}, N., {Pillepich}, A., {Nelson}, D., {Werner}, N., \& {Hernquist}, L.
  2021, \mnras, 508, 1563, \dodoi{10.1093/mnras/stab2638}

\bibitem[{{Truong} {et~al.}(2023){Truong}, {Pillepich}, {Nelson}, {Bogd{\'a}n},
  {Schellenberger}, {Chakraborty}, {Forman}, {Kraft}, {Markevitch},
  {Ogorzalek}, {Oppenheimer}, {Sarkar}, {Veilleux}, {Vogelsberger}, {Wang},
  {Werner}, {Zhuravleva}, \& {Zuhone}}]{Truong2023}
{Truong}, N., {Pillepich}, A., {Nelson}, D., {et~al.} 2023, \mnras, 525, 1976,
  \dodoi{10.1093/mnras/stad2216}

\bibitem[{{Tumlinson} {et~al.}(2017){Tumlinson}, {Peeples}, \&
  {Werk}}]{Tumlinson2017}
{Tumlinson}, J., {Peeples}, M.~S., \& {Werk}, J.~K. 2017, \araa, 55, 389,
  \dodoi{10.1146/annurev-astro-091916-055240}

\bibitem[{{Tumlinson} {et~al.}(2011){Tumlinson}, {Thom}, {Werk}, {Prochaska},
  {Tripp}, {Weinberg}, {Peeples}, {O'Meara}, {Oppenheimer}, {Meiring}, {Katz},
  {Dav{\'e}}, {Ford}, \& {Sembach}}]{Tumlinson2011}
{Tumlinson}, J., {Thom}, C., {Werk}, J.~K., {et~al.} 2011, Science, 334, 948,
  \dodoi{10.1126/science.1209840}

\bibitem[{{Tumlinson} {et~al.}(2013){Tumlinson}, {Thom}, {Werk}, {Prochaska},
  {Tripp}, {Katz}, {Dav{\'e}}, {Oppenheimer}, {Meiring}, {Ford}, {O'Meara},
  {Peeples}, {Sembach}, \& {Weinberg}}]{Tumlinson2013}
---. 2013, \apj, 777, 59, \dodoi{10.1088/0004-637X/777/1/59}

\bibitem[{{Turk} {et~al.}(2011){Turk}, {Smith}, {Oishi}, {Skory}, {Skillman},
  {Abel}, \& {Norman}}]{turk2011}
{Turk}, M.~J., {Smith}, B.~D., {Oishi}, J.~S., {et~al.} 2011, \apjs, 192, 9,
  \dodoi{10.1088/0067-0049/192/1/9}

\bibitem[{{van de Voort} {et~al.}(2019){van de Voort}, {Springel}, {Mandelker},
  {van den Bosch}, \& {Pakmor}}]{vandeVoort2019}
{van de Voort}, F., {Springel}, V., {Mandelker}, N., {van den Bosch}, F.~C., \&
  {Pakmor}, R. 2019, \mnras, 482, L85, \dodoi{10.1093/mnrasl/sly190}

\bibitem[{{van der Walt} {et~al.}(2011){van der Walt}, {Colbert}, \&
  {Varoquaux}}]{vanderwalt2011}
{van der Walt}, S., {Colbert}, S.~C., \& {Varoquaux}, G. 2011, Computing in
  Science and Engineering, 13, 22, \dodoi{10.1109/MCSE.2011.37}

\bibitem[{{Werk} {et~al.}(2013){Werk}, {Prochaska}, {Thom}, {Tumlinson},
  {Tripp}, {O'Meara}, \& {Peeples}}]{Werk2013}
{Werk}, J.~K., {Prochaska}, J.~X., {Thom}, C., {et~al.} 2013, \apjs, 204, 17,
  \dodoi{10.1088/0067-0049/204/2/17}

\bibitem[{{Werk} {et~al.}(2014){Werk}, {Prochaska}, {Tumlinson}, {Peeples},
  {Tripp}, {Fox}, {Lehner}, {Thom}, {O'Meara}, {Ford}, {Bordoloi}, {Katz},
  {Tejos}, {Oppenheimer}, {Dav{\'e}}, \& {Weinberg}}]{Werk2014}
{Werk}, J.~K., {Prochaska}, J.~X., {Tumlinson}, J., {et~al.} 2014, \apj, 792,
  8, \dodoi{10.1088/0004-637X/792/1/8}

\bibitem[{{Werk} {et~al.}(2016){Werk}, {Prochaska}, {Cantalupo}, {Fox},
  {Oppenheimer}, {Tumlinson}, {Tripp}, {Lehner}, \& {McQuinn}}]{Werk2016}
{Werk}, J.~K., {Prochaska}, J.~X., {Cantalupo}, S., {et~al.} 2016, \apj, 833,
  54, \dodoi{10.3847/1538-4357/833/1/54}

\bibitem[{{Wijers} {et~al.}(2020){Wijers}, {Schaye}, \&
  {Oppenheimer}}]{Wijers2020}
{Wijers}, N.~A., {Schaye}, J., \& {Oppenheimer}, B.~D. 2020, \mnras, 498, 574,
  \dodoi{10.1093/mnras/staa2456}

\bibitem[{{Wright} {et~al.}(2024){Wright}, {Tumlinson}, {Peeples}, {O'Shea},
  {Lochhaas}, {Corlies}, {Smith}, {Binh}, {Augustin}, \& {Simons}}]{Wright2024}
{Wright}, A.~C., {Tumlinson}, J., {Peeples}, M.~S., {et~al.} 2024, \apj, 970,
  70, \dodoi{10.3847/1538-4357/ad49a3}

\bibitem[{{Wu} \& {McQuinn}(2023)}]{Wu2023}
{Wu}, X., \& {McQuinn}, M. 2023, \apj, 945, 87,
  \dodoi{10.3847/1538-4357/acbc7d}

\bibitem[{{Zhang} {et~al.}(2024){Zhang}, {Comparat}, {Ponti}, {Merloni},
  {Nandra}, {Haberl}, {Locatelli}, {Zhang}, {Sanders}, {Zheng}, {Liu},
  {Popesso}, {Liu}, {Truong}, {Pillepich}, {Predehl}, {Salvato}, {Shreeram},
  {Yeung}, \& {Ni}}]{Zhang2024}
{Zhang}, Y., {Comparat}, J., {Ponti}, G., {et~al.} 2024, arXiv e-prints,
  arXiv:2401.17308, \dodoi{10.48550/arXiv.2401.17308}

\bibitem[{{ZuHone} \& {Hallman}(2016)}]{pyxsim}
{ZuHone}, J.~A., \& {Hallman}, E.~J. 2016, {pyXSIM: Synthetic X-ray
  observations generator}.
\newblock \doeprint{1608.002}

\bibitem[{{ZuHone} {et~al.}(2024){ZuHone}, {Schellenberger}, {Ogorza{\l}ek},
  {Oppenheimer}, {Stern}, {Bogd{\'a}n}, {Truong}, {Markevitch}, {Pillepich},
  {Nelson}, {Burchett}, {Khabibullin}, {Kilbourne}, {Kraft}, {Nulsen},
  {Veilleux}, {Vogelsberger}, {Wang}, \& {Zhuravleva}}]{ZuHone2024}
{ZuHone}, J.~A., {Schellenberger}, G., {Ogorza{\l}ek}, A., {et~al.} 2024, \apj,
  967, 49, \dodoi{10.3847/1538-4357/ad36c1}

\end{thebibliography}
\bibliographystyle{aasjournal}

\appendix 
\restartappendixnumbering
\section{The need for photoionization in X-ray emissivity calculations} \label{appendix:photoion}

\begin{figure*}[h]
    \centering 
        \includegraphics[width=1\textwidth]{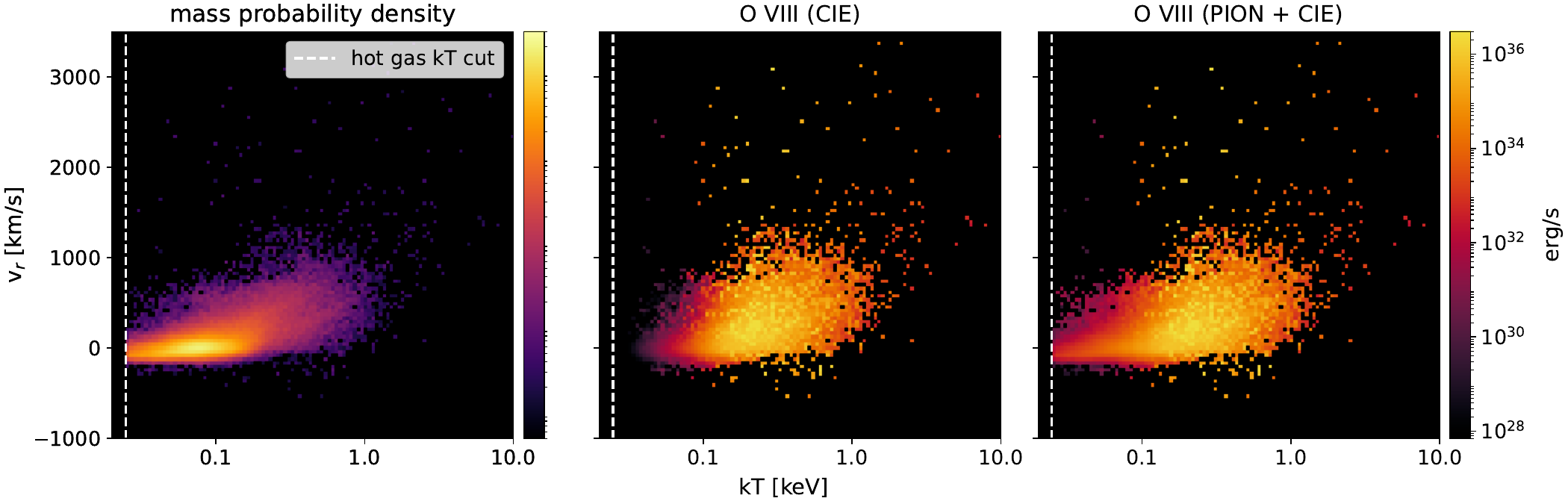}
        \caption{Comparison of the mass and O VIII luminosity profiles in temperature vs $v_r$ space for emission models with CIE and CIE$+$photoionization in the TNG50-211 galaxy.} \label{fig:CIE_PION_comps_TNG}
\end{figure*}

\begin{figure*}[h]
    \centering 
        \includegraphics[width=1\textwidth]{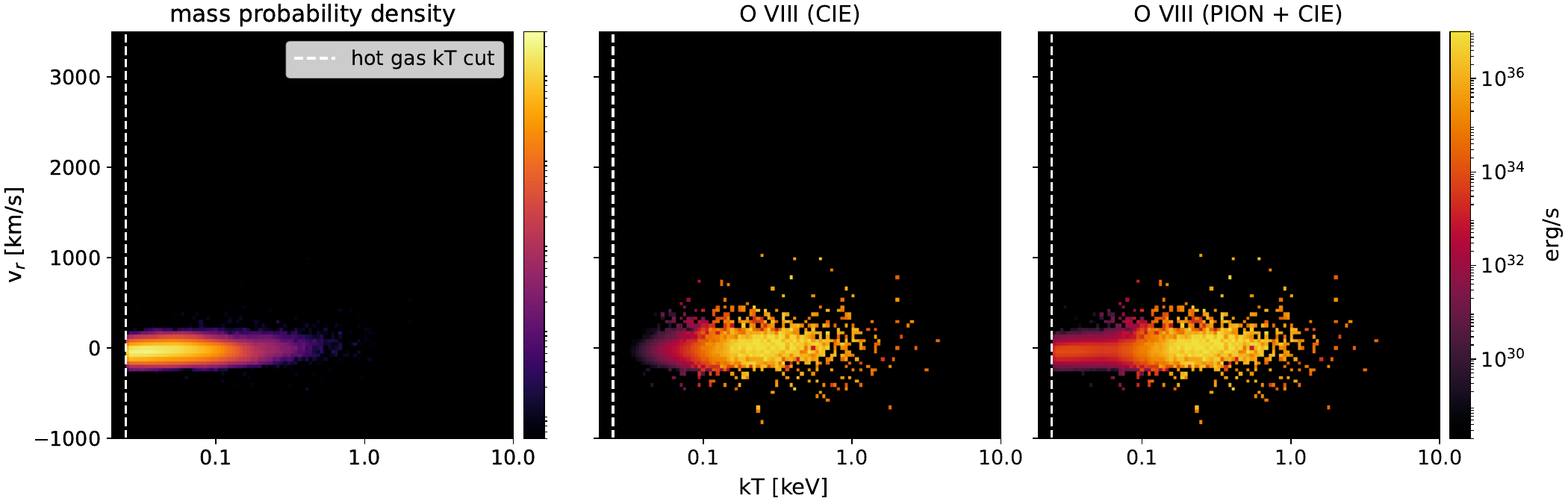}
        \caption{Comparison of the mass and O VIII luminosity profiles in temperature vs $v_r$ space for emission models with CIE and CIE$+$photoionization in the FIRE2-m12i galaxy.}  \label{fig:CIE_PION_comps_FIRE}
\end{figure*}

In the low-density warm/hot CGM, photoionization by the cosmic UV/X-ray background radiation fields can contribute non-negligibly relative to collisional ionization, additionally shifting ions to higher states than exclusively CIE \citep{Churazov2001, Khabibullin2019, Wijers2020}. We verified the need for the inclusion of photoionization in our X-ray emissivity calculations by comparing the temperature vs $v_r$ phase diagrams of the mass and O VIII emission line luminosity generated for two $\sim$MW-mass halos with two different source models. The first assumes pure CIE (via APEC), and the other is a \texttt{cloudy}-based model that includes additional photoionization effects below $\sim1$ keV (Figures \ref{fig:CIE_PION_comps_TNG} and \ref{fig:CIE_PION_comps_FIRE}). 

Thus, we see that the CIE model does not sufficiently account for all of the hot gas mass expected to be emitting at low temperatures, especially below $\sim 0.1$ keV. The additional photoionization contribution to the X-ray emissivity in the CIE$+$photoionization model better incorporates this low-temperature gas.

\newpage
\section{Additional X-ray observable projections} \label{appendix:moreprojs}
We provide additional projections of the X-ray observables outlined in this study (XSB, {\vlos}, {\sigmav}, and \textit{kT}) for all galaxies in our sample viewed from orientations that were not highlighted in the main text. In each figure, the galaxies are sorted from lowest to highest halo mass (with $\log(M\;[M_{\odot}])$ indicated in each bottom right corner). \vspace{-0.5em}
\begin{figure*}[thb!]
    \centering 
        \includegraphics[width=1\textwidth]{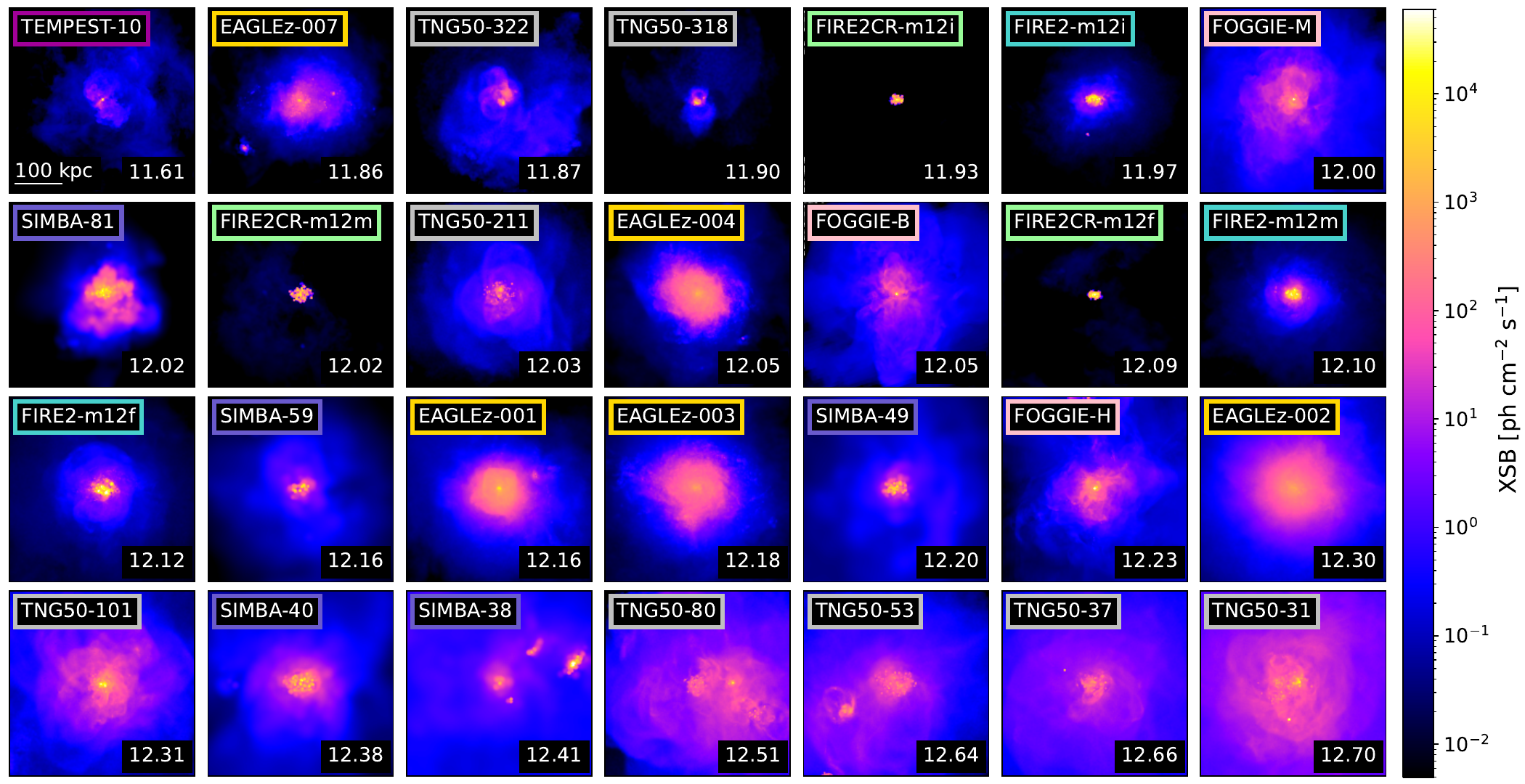}
        \caption{Intermediate (midway between edge- and face-on) projections of broadband X-ray surface brightness (XSB). \vspace{-1em}}  \label{fig:XSB_mid_projs}
\end{figure*}
\begin{figure*}[thb!]
    \centering 
        \includegraphics[width=1\textwidth]{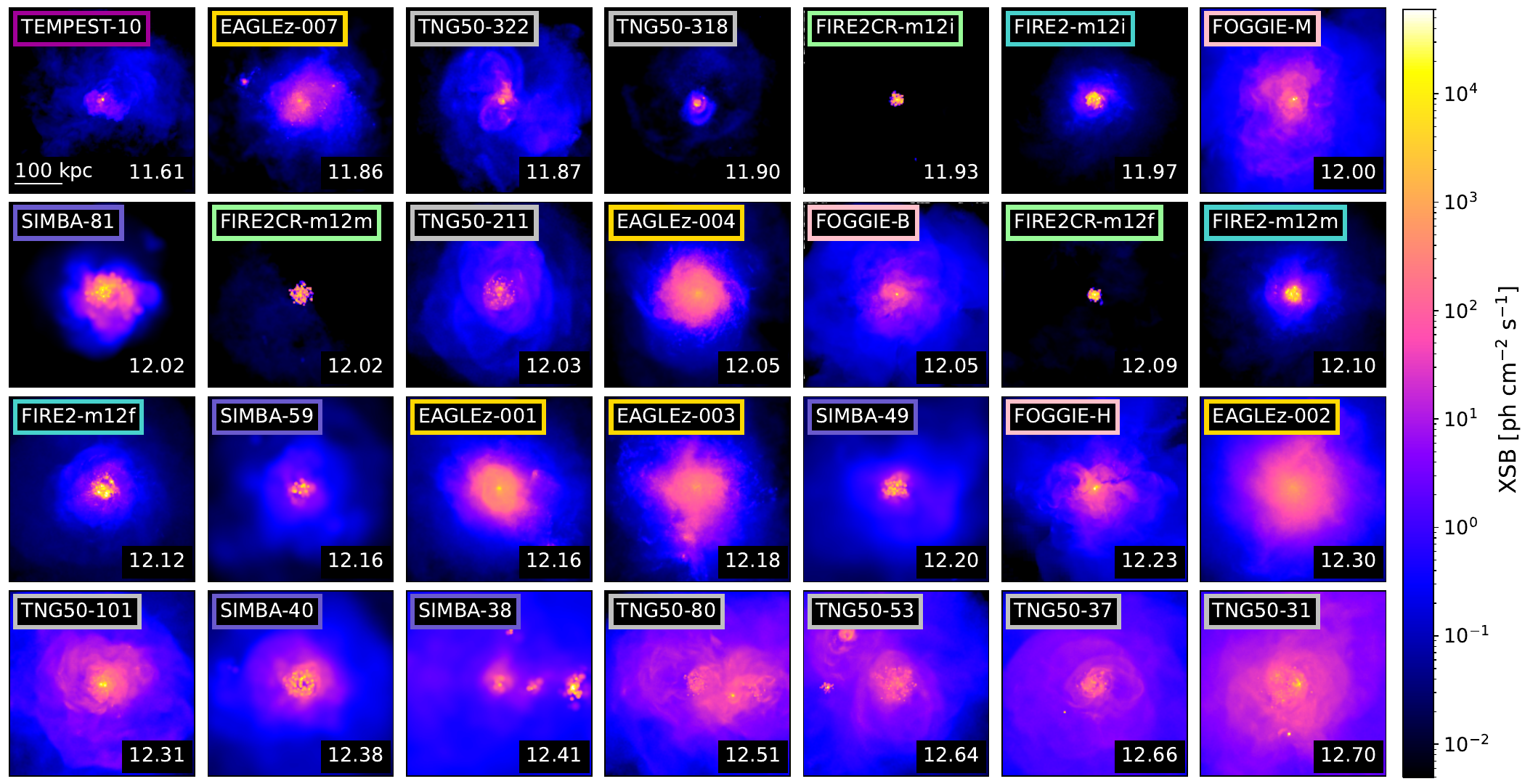}\vspace{-1em}
        \caption{Face-on projections of broadband X-ray surface brightness (XSB). \vspace{-1em}}  \label{fig:XSB_face_projs}
\end{figure*}

\begin{figure*}[thb!]
    \centering 
        \includegraphics[width=1\textwidth]{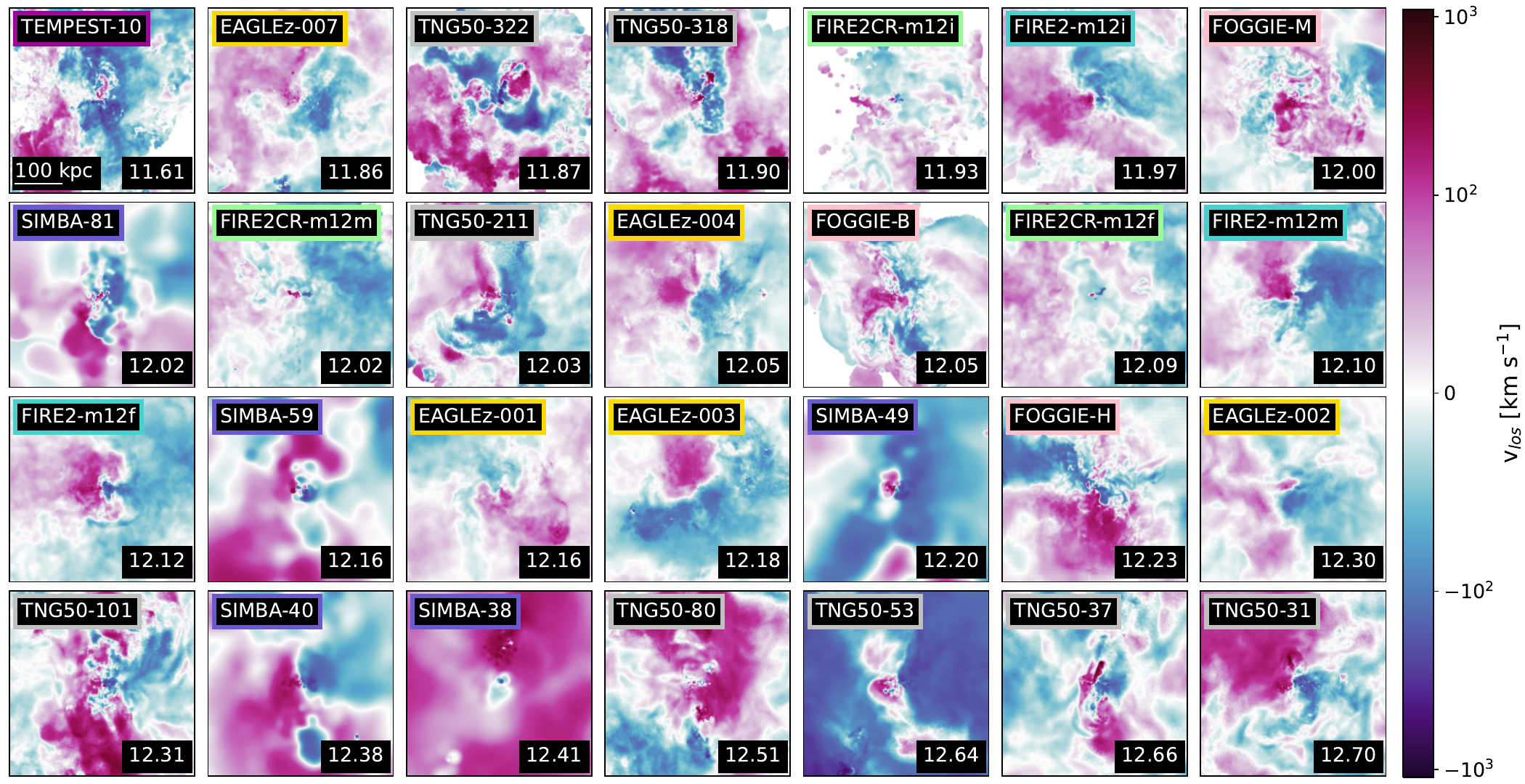}\vspace{-0.5em}
        \caption{Edge-on projections of LOS velocity ({\vlos}) weighted by the broadband X-ray emissivity. \vspace{-1em}}  \label{fig:vlos_edge_projs}
\end{figure*}
\begin{figure*}[thb!]
    \centering 
        \includegraphics[width=1\textwidth]{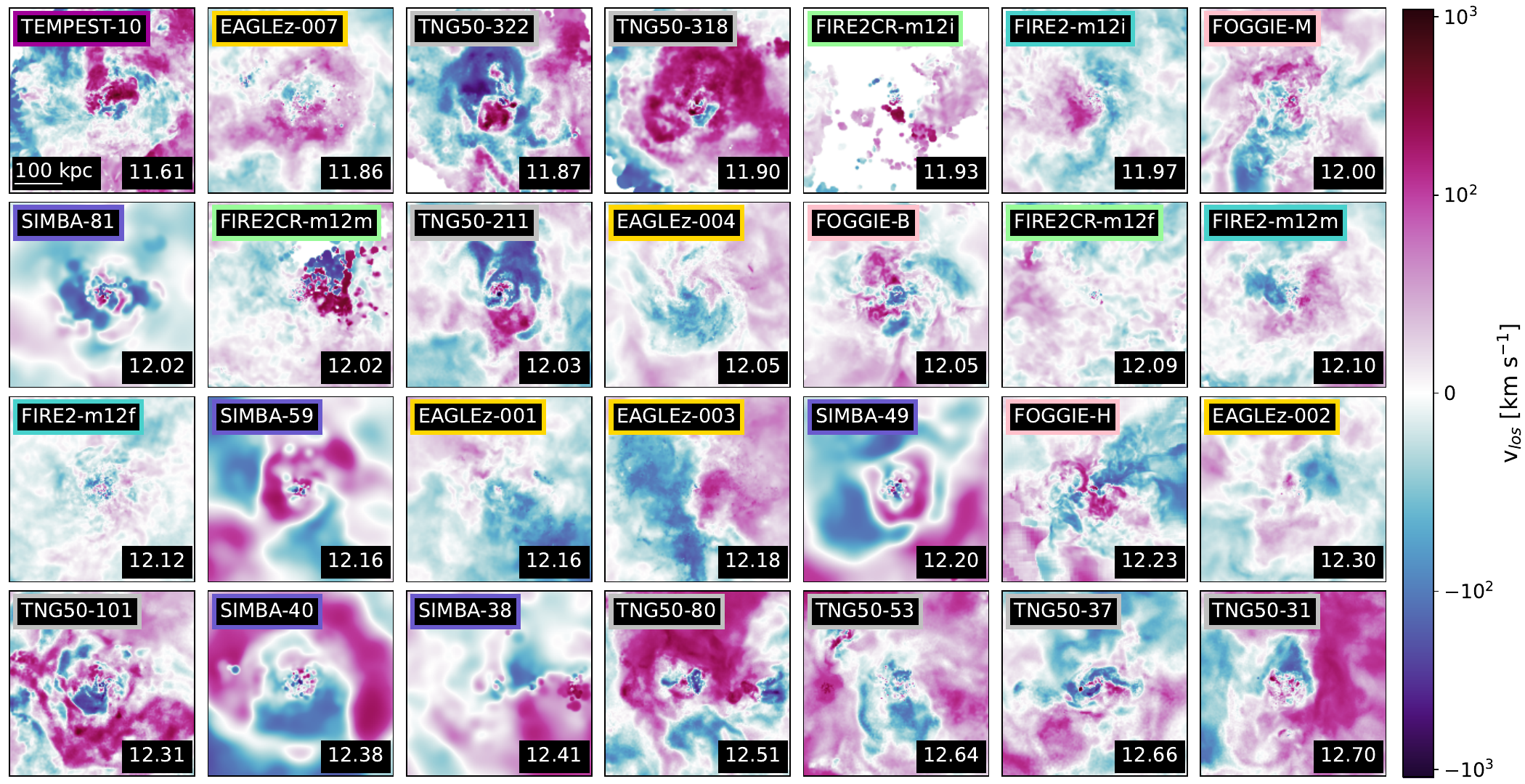}
        \caption{Face-on projections of LOS velocity ({\vlos}) weighted by the broadband X-ray emissivity. \vspace{-1em}}  \label{fig:vlos_face_projs}
\end{figure*}

\begin{figure*}[thb!]
    \centering 
        \includegraphics[width=1\textwidth]{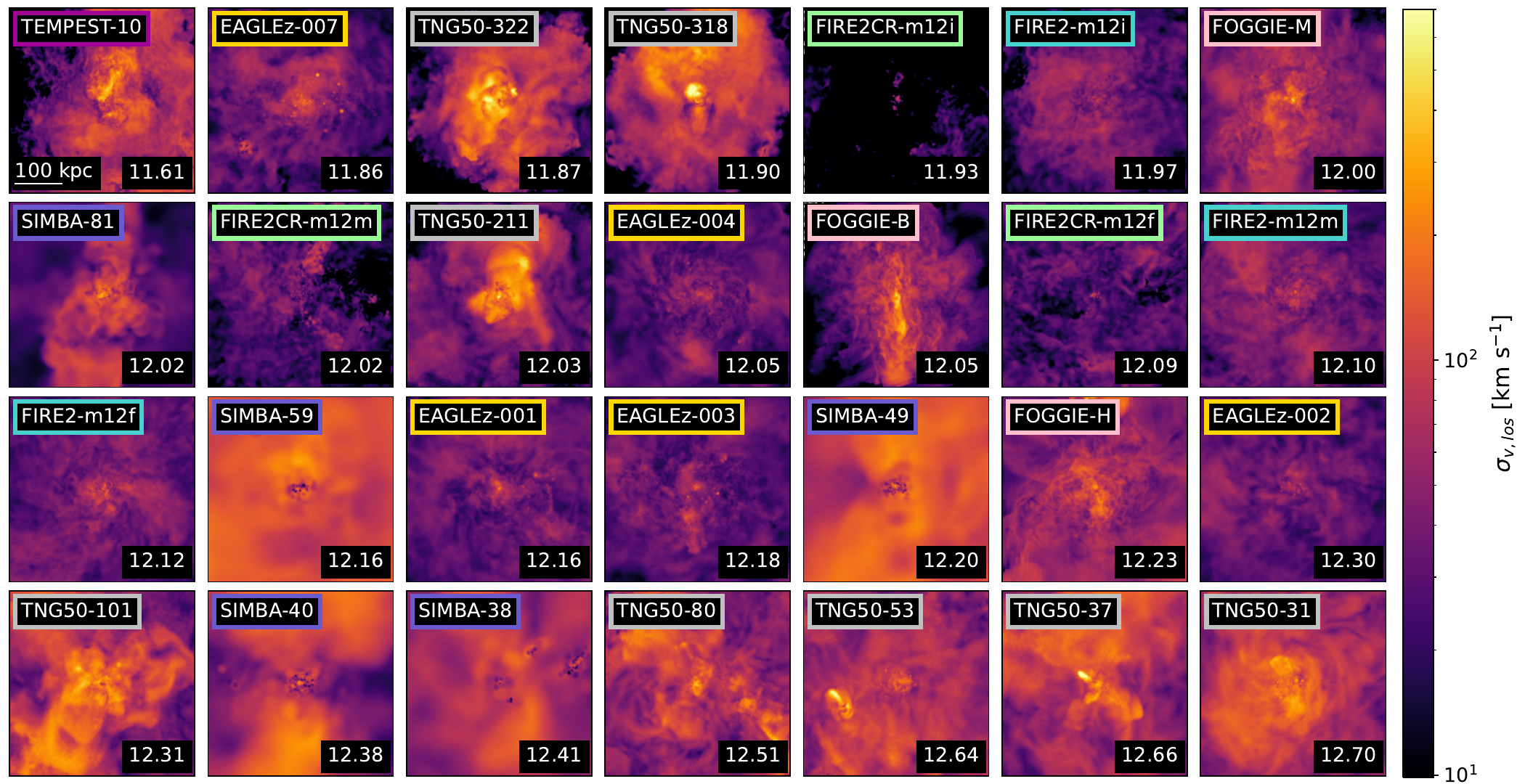}
        \caption{Intermediate (midway between edge- and face-on) projections of LOS velocity dispersion ({\sigmav}) weighted by the broadband X-ray emissivity. \vspace{-1em}} \label{fig:sigmav_mid_projs} 
\end{figure*} 
\begin{figure*}[thb!]
    \centering 
        \includegraphics[width=1\textwidth]{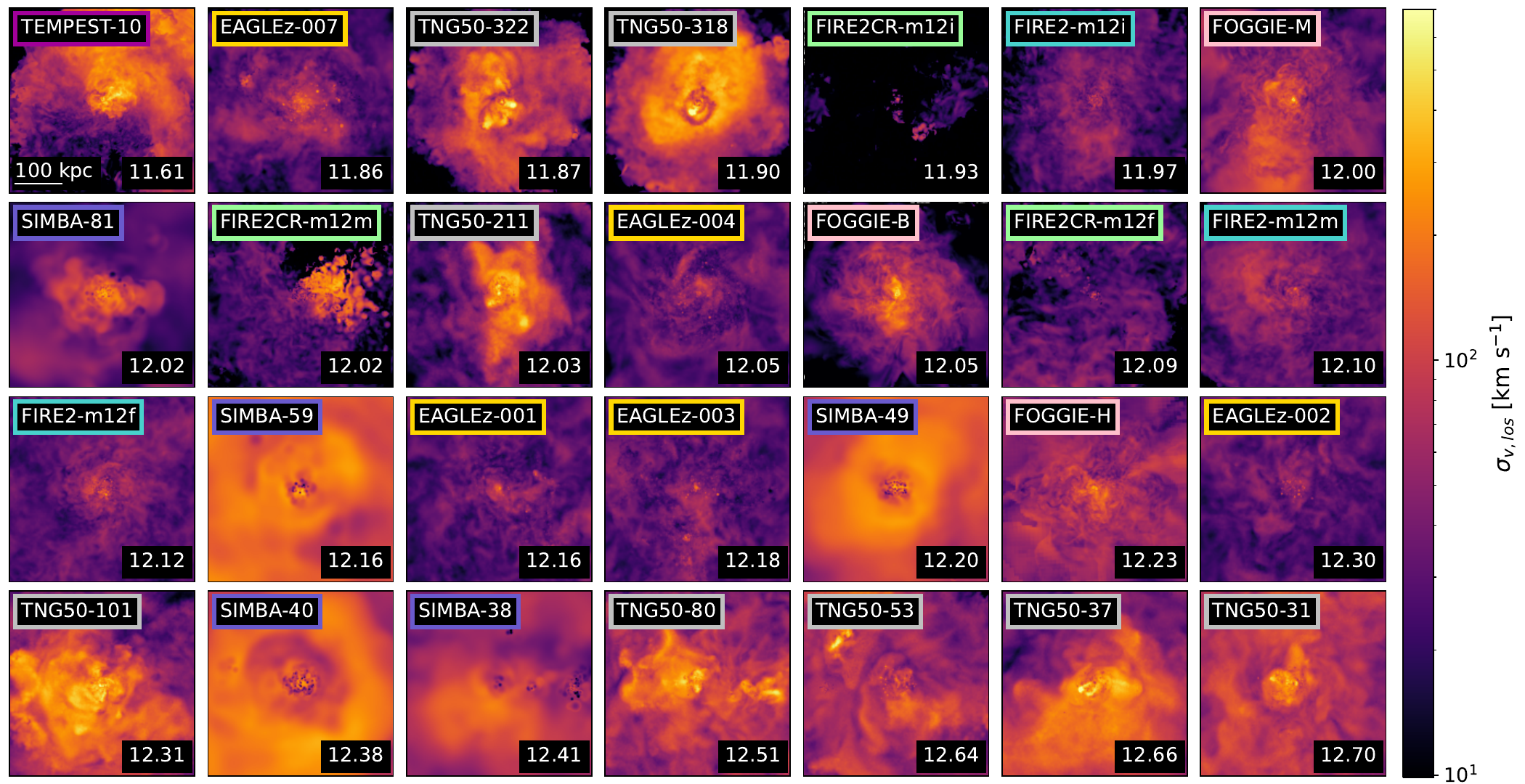}
        \caption{Face-on projections of LOS velocity dispersion ({\sigmav}) weighted by the broadband X-ray emissivity. \vspace{-1em}} \label{fig:sigmav_face_projs} 
\end{figure*} 

\begin{figure*}[htb!]
    \centering 
        \includegraphics[width=1\textwidth]{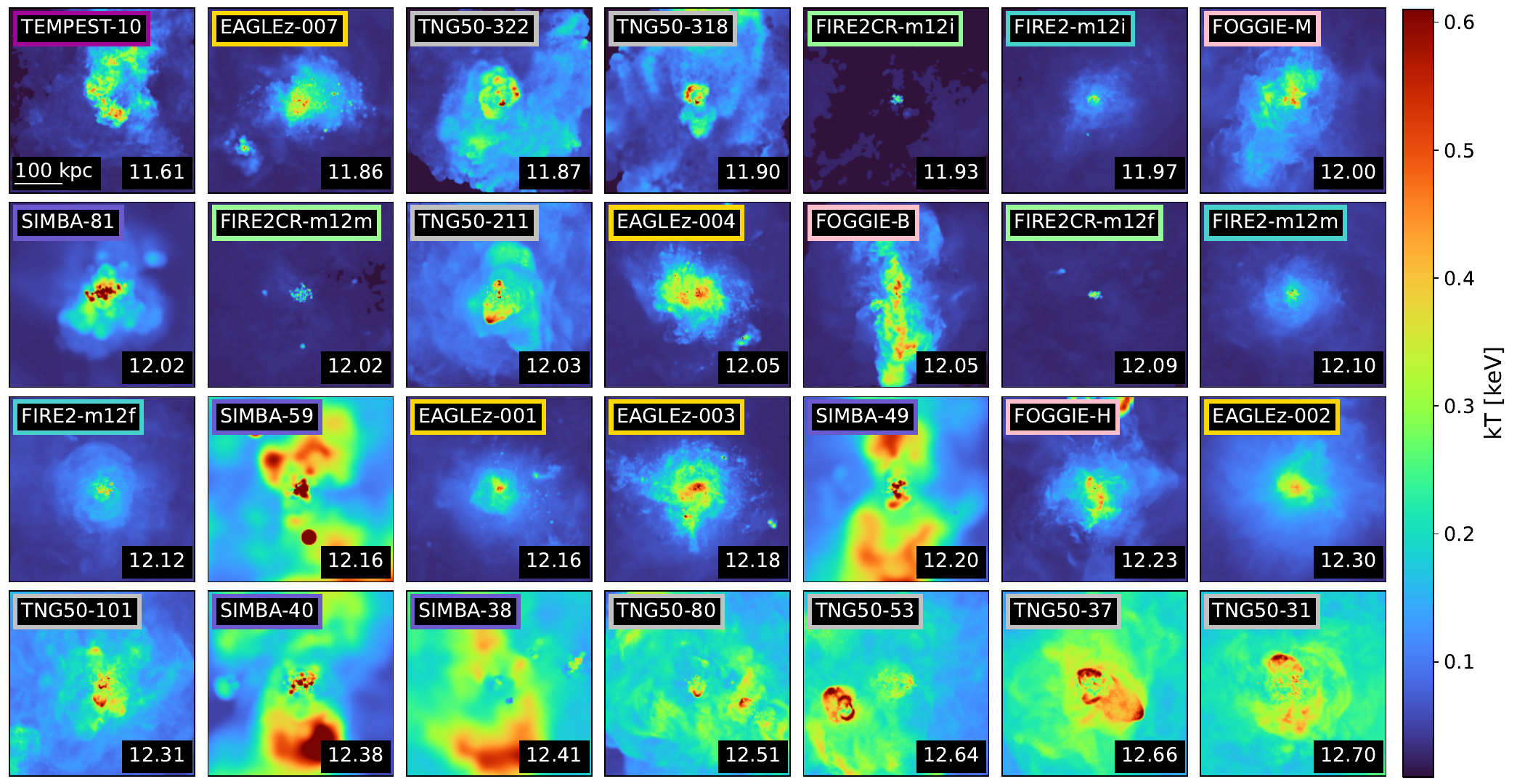}
        \caption{Intermediate (midway between edge- and face-on) projections of kT weighted by the broadband X-ray emissivity. \vspace{-1em}}  \label{fig:kT_mid_projs}
\end{figure*} 
\begin{figure*}[htb!]
    \centering 
        \includegraphics[width=1\textwidth]{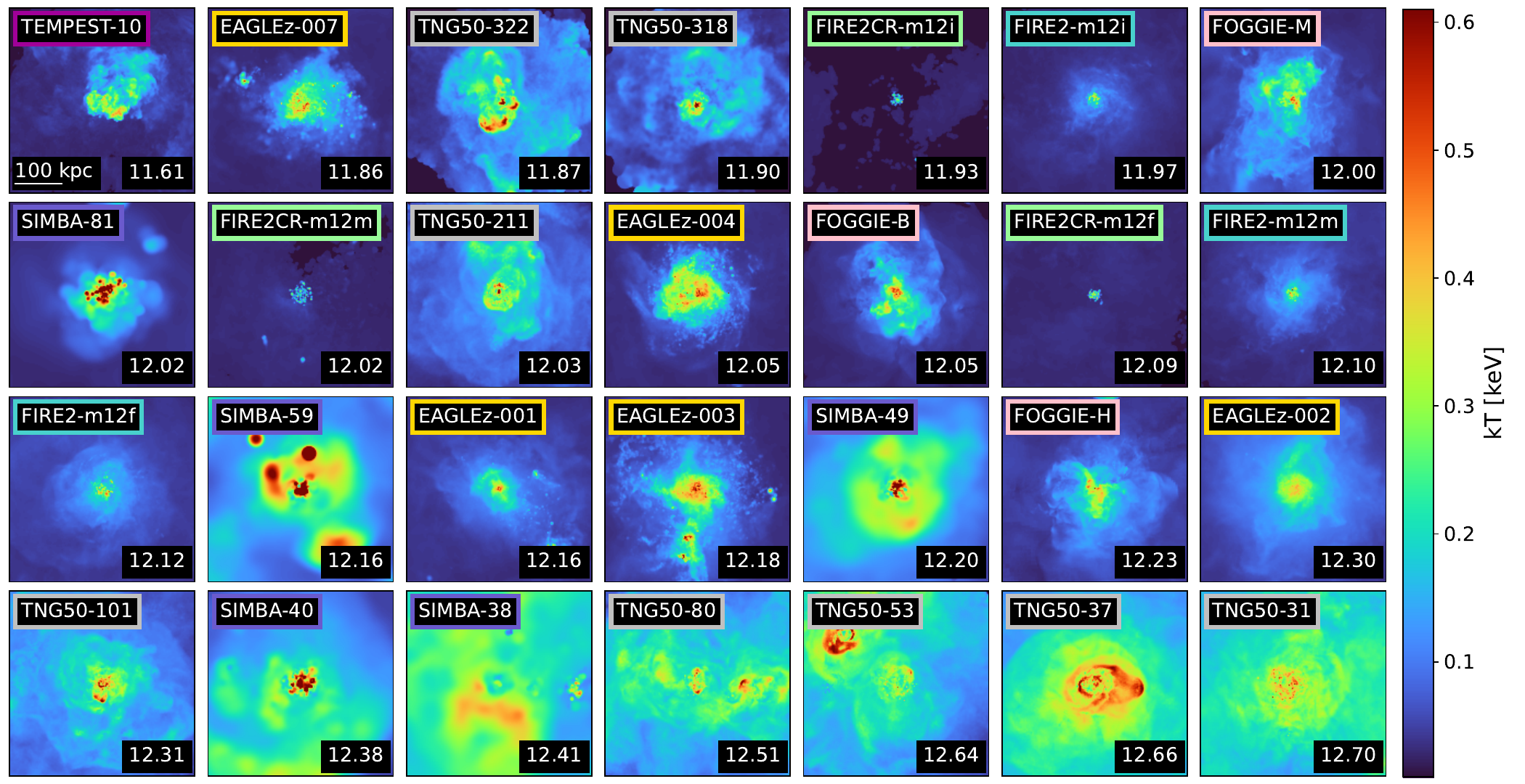}
        \caption{Face-on projections of kT weighted by the broadband X-ray emissivity. \vspace{-1em}}  \label{fig:kT_face_projs}
\end{figure*}

\end{document}